\documentclass[prd, reprint, showpacs, nofootinbib, raggedbottom, floatfix]{revtex4-1}
\usepackage{amsmath, mathtools, graphicx, color, latexsym, array, setspace, bigints, dcolumn, bm, natbib, hyperref}
\usepackage[inline]{enumitem}
\usepackage[retainorgcmds]{IEEEtrantools}
\usepackage[explicit]{titlesec}
\usepackage[caption=false]{subfig}

\newcolumntype{S}{>{\centering\arraybackslash}m{2.1cm}}
\newcolumntype{N}{>{\centering\arraybackslash}m{2.1cm}}
\newcolumntype{B}{>{\centering\arraybackslash}m{2.1cm}}
\newcolumntype{T}{>{\centering\arraybackslash}m{2cm}}
\newcolumntype{d}[1]{D{.}{.}{#1} }
\allowdisplaybreaks

\newcommand{\Tensb}[1]{\mathbf{#1}}

\newcommand*{\lin}{\mathrm{lin}}
\newcommand*{\tol}{\varepsilon_{tol}}
\newcommand*{\enumref}[1]{(\ref{#1}}

\newcommand*{\LinEl}{s}
\newcommand*{\FriedScale}{a}
\newcommand*{\CosmoConst}{\Lambda}

\newcommand*{\LattTime}{\tau}
\newcommand*{\LattScale}{a}

\newcommand*{\Figuref}[1]{Figure \ref{#1}}
\newcommand*{\FigRef}[1]{Fig.\ \ref{#1}}
\newcommand*{\tabref}[1]{Table \ref{#1}}
\newcommand*{\secref}[1]{Section \ref{#1}}
\newcommand*{\appendref}[1]{Appendix \ref{#1}}

\DeclareMathOperator{\Tr}{Tr}

\renewcommand{\thesection}{\Roman{section}}
\titleformat{\section}[block]
  {\normalfont\bfseries}{\thesection.}{1em}{\centering \MakeUppercase{#1}}
\renewcommand{\thesubsection}{\Alph{subsection}}
\titleformat{\subsection}[block]
  {\normalfont\bfseries}{\thesubsection.}{1em}{\centering {#1}}

\begin{document}

\title{\bf The Lindquist-Wheeler formulation of lattice universes}
\author{Rex G \surname{Liu}}
\email{R.Liu@damtp.cam.ac.uk}
\affiliation{Trinity College, Cambridge CB2 1TQ, UK,}
\affiliation{DAMTP, CMS, Wilberforce Road, Cambridge CB3 0WA, UK.}

\begin{abstract}
This paper examines the properties of `lattice universes' wherein point masses are arranged in a regular lattice on space-like hypersurfaces; open, flat, and closed universes are considered.  The universes are modelled using the Lindquist-Wheeler (LW) approximation scheme, which approximates the space-time in each lattice cell by Schwarz\-schild geometry.  Extending Lindquist and Wheeler's work, we derive cosmological scale factors describing the evolution of all three types of universes, and we use these scale factors to show that the universes' dynamics strongly resemble those of Friedmann-Lema\^itre-Robertson-Walker (FLRW) universes.  In particular, we use the scale factors to make more salient the resemblance between Clifton and Ferreira's Friedmann-like equations for the LW models and the actual Friedmann equations of FLRW space-times.  Cosmological redshifts for such universes are then determined numerically, using a modification of Clifton and Ferreira's approach; the redshifts are found to closely resemble their FLRW counterparts, though with certain differences attributable to the `lumpiness' in the underlying matter content.  Most notably, the LW redshifts can differ from their FLRW counterparts by as much as 30\%, even though they increase linearly with FLRW redshifts, and they exhibit a non-zero integrated Sachs-Wolfe effect, something which would not be possible in matter-dominated FLRW universes without a cosmological constant.
\end{abstract}



\maketitle

\section{Introduction}

\begin{figure}[htbp]
\input{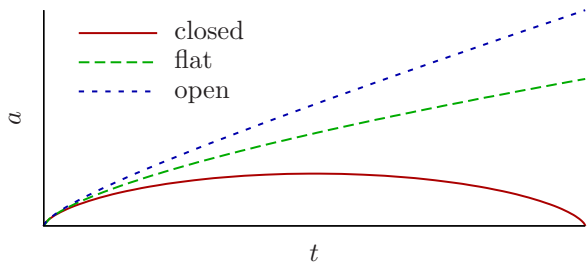}
\caption{\label{cosmo-graphs}Plots of $\FriedScale$ versus $t$ for open, flat, and closed dust-filled universes.}
\end{figure}

Modern cosmology is founded upon the so-called \emph{Copernican principle}, which posits that the universe `looks' on average to be the same regardless of where one is in the universe or in which direction one looks.  More formally, the Copernican principle states that Cauchy surfaces of the universe can be admitted that are homogeneous and isotropic, and this symmetry can be expressed mathematically by writing the universe's metric in the form
\begin{equation}
d\LinEl^2 = - dt^2 + \FriedScale^2(t) \left[\frac{dr^2}{1- k r^2} + r^2 \big(d\theta^2 + \sin^2\theta\, d\phi^2 \big) \right],
\label{FLRW-metric}
\end{equation}
where $\FriedScale(t)$ is the time-dependent conformal scale factor, and $k$ is the curvature constant.  The sign of $k$ determines whether 3-spaces of constant $t$ will be open, flat, or closed, with $k<0$ corresponding to open universes, $k=0$ to flat universes, and $k>0$ to closed universes; one can always re-scale $\FriedScale(t)$ and $k$ such that $k = +1$, $0$, or $-1$ as appropriate, in which case $\FriedScale(t)$ becomes the radius of curvature for the open and closed universes.  This family of metrics is known as the Friedmann-Lema\^itre-Robertson-Walker (FLRW) metric.  For the metric to satisfy the Einstein field equations of general relativity, $a(t)$ must obey the Friedmann equations
\begin{IEEEeqnarray}{rCl}
\left( \frac{\dot{\FriedScale}}{\FriedScale} \right)^2 &=& \frac{1}{3} \Big(8\pi\rho + \CosmoConst \Big) - \frac{k}{\FriedScale^2},
\label{Friedmann1}\\
\frac{\ddot{\FriedScale}}{\FriedScale} \quad &=& -\frac{4\pi}{3} \Big(\rho + 3p \Big) + \frac{\CosmoConst}{3},
\label{Friedmann2}
\end{IEEEeqnarray}
where $\CosmoConst$ is the cosmological constant, and $\rho$ and $p$ are the energy density and pressure of the fluid filling the space.  For dust-filled $\CosmoConst=0$ universes, the relationship between $\FriedScale$ and $t$ is given by
\begin{IEEEeqnarray}{lr}
\begin{aligned}
\FriedScale &= \displaystyle \frac{\FriedScale_0}{2}(1 - \cos \eta)\\[3mm]
t &= \displaystyle \frac{\FriedScale_0}{2}(\eta - \sin \eta)
\end{aligned}
& \qquad \text{for } k>0,
\label{closed_a}\\[5mm]
\begin{aligned}
\FriedScale &= \displaystyle \left(\frac{9}{4}\FriedScale_0\right)^{1/3} t^{2/3}
\end{aligned}
& \qquad \text{for } k=0,
\label{flat_a}\\[5mm]
\begin{aligned}
\FriedScale &= \displaystyle \frac{\FriedScale_0}{2}(\cosh \eta - 1)\\[3mm]
t &= \displaystyle \frac{\FriedScale_0}{2}(\sinh \eta - \eta)
\end{aligned}
& \qquad \text{for } k<0,
\label{open_a}
\end{IEEEeqnarray}
as depicted in \FigRef{cosmo-graphs}, where 
\begin{equation}
\FriedScale_0 = \frac{8\pi\rho_0}{3},
\label{FLRW_a0}
\end{equation}
and $\rho_0$ is the energy density when $\FriedScale=1$.  In the case of $k>0$, the factor $\FriedScale_0$ also corresponds to the maximum of $\FriedScale$.

These FLRW models have had great success in explaining much of the universe's behaviour, including most notably the Hubble expansion of the universe, the cosmic microwave background (CMB), and baryon acoustic oscillations.  Indeed, the underlying assumption of homogeneity and isotropy appears well-supported by precision measurements showing the CMB to be isotropic to within one part in 100,000.  Yet recent measurements of redshifts from Type Ia supernovae (SN1a), when fitted to FLRW models, have led to the conclusion that the universe's expansion is actually accelerating, and to account for this acceleration, cosmologists have had to modify the standard models by introducing some exotic matter, known generally as \emph{dark energy}.  Two of the most-favoured dark energy models are an FLRW model with non-zero cosmological constant $\CosmoConst$, which describes a negative-pressure fluid acting repulsively under gravity and permeating the entire universe, and a scalar field theory called \emph{quintessence}.  However attempts to directly observe any exotic matter and determine its nature have so far proven unsuccessful, and some cosmologists have begun to seek other explanations for the SN1a observations, questioning whether there is any need for exotic matter after all.

There has been an intense debate recently over the relative importance of inhomogeneities in influencing cosmological observations \citep{EllisBuchert, Ellis2011, CELU}.  In spite of the wide success of FLRW models, observations clearly show that the late, matter-dominated universe is not homogeneous and isotropic except at the coarsest of scales.  Instead, matter is distributed predominantly in clusters and superclusters of galaxies with large voids in between.  Standard FLRW models ignore such inhomogeneities completely, with proponents arguing that the inhomogeneities' effects should be small and hence unimportant.  Yet others have argued the converse, proposing that inhomogeneities may have non-trivial implications on both the large-scale dynamics of the universe and on cosmological observations themselves; indeed, inhomogeneities may even explain the SN1a data as an apparent acceleration without any need for exotic matter; this apparent acceleration and analogous effects arise as a result of fitting data from an inhomogeneous universe onto a homogeneous model.

Therefore because of the possibly important implications of inhomogeneities to cosmology, there has been much recent effort in trying to quantify their effects.  As the actual universe's structure is far too complicated to model directly, much effort has been focused on studying the effects of inhomogeneities in toy models.  Additionally, there has been much debate on whether perturbed FLRW models can adequately capture the inhomogeneities' non-linear structure \citep{KMR, ClarksonMaartens, ClarksonUmeh}.  Thus a non-perturbative approach is necessary to unambiguously understand inhomogeneities.

Perhaps the simplest inhomogeneous models that can account for all known cosmological observations are the Lema\^itre-Tolman-Bondi (LTB) models  \citep{MHE, CS, Tomita1, Tomita2, AAG, Moffat, Mansouri, GH, CFL, CGH, ABNV, BW, FLSC}.  In such models, the observer is located at the centre of a Hubble-scale under-dense region in a spherically-symmetric, dust-dominated universe.  These models are therefore isotropic but not homogeneous.  What is most noteworthy is that these models can successfully generate the SN1a redshift data without requiring any exotic matter.

FLRW-based inhomogeneous models have also been constructed using the `Swiss Cheese' methodology developed by Einstein and Straus \citep{EinsteinStraus, *EinsteinStraus-err}.  In these models, co-moving spheres are excised from FLRW space-times and replaced by either Schwarzschild space-time or LTB space-time \citep{Kantowski, BiswasNotari, MKMR, BTT, MKM, CliftonZuntz}.  The resulting models are still exact solutions to the Einstein field equations.  By construction, the fluid region surrounding the excised spheres is still described by the FLRW metric; therefore these models would still be dynamically identical to FLRW.  However, there is no \emph{a priori} reason to believe their optical properties should be the same; nevertheless, studies have shown that they are generally similar though not identical to FLRW \citep{BiswasNotari, MKMR, BTT, MKM, CliftonZuntz}.

In contrast to the above approaches, we shall consider a completely different space-time that is not in any way based on FLRW nor includes any continuous fluid in its matter content.  Rather, we shall consider a space-time where the matter content consists solely of discrete point masses with vacuum in between, as this would provide a more realistic description of the late universe's matter content.  However, for simplicity, our space-time will still possess a high degree of symmetry, though not to the extent of the FLRW universes.  Specifically, we shall consider the so-called \emph{lattice universes} where the matter content on each Cauchy surface consists of identical point masses arranged into a regular lattice.  We shall focus on lattices constructed by tessellating a 3-space of constant curvature with identical regular polyhedral cells; the possible lattices obtainable from such a tessellation are summarised in \appendref{App_Latt}.  To `construct' a lattice universe, one of the lattices in \appendref{App_Latt} is selected, and all mass in each cell is `concentrated' into a point at its centre.  The metric consistent with such a matter distribution is expected to be invariant under the same symmetry transformations that leave the lattice invariant, symmetries which include discrete translation symmetries, discrete rotational symmetries, and reflection symmetries at the cell boundaries.  In other words, the lattice universe should have a metric of the form
\begin{equation}
d\LinEl^2 = -dt^2 + \gamma^{(3)}_{ab}\left(t,\Tensb{x}\right)\,dx^a dx^b,
\end{equation}
where $\Tensb{\gamma}^{(3)}_{ab}\left(t,\Tensb{x}\right)$ is the 3-dimensional metric for constant $t$ hypersurfaces, and Latin indices $a,b = 1, 2, 3$ denote spatial co-ordinates only; the spatial metric $\Tensb{\gamma}^{(3)}_{ab}\left(t,\Tensb{x}\right)$ at constant $t$ would possess the lattice symmetries.  Effectively, the Copernican symmetries of FLRW universes have been reduced to just these symmetries.  Yet these lattice universes still satisfy the Copernican principle in a coarse-grained manner, in contrast to LTB models where homogeneity of the universe is completely surrendered.

There has been some progress towards determining the 3-metric $\Tensb{\gamma}^{(3)}\left(t,\Tensb{x}\right)$.  For the closed universe, the 3-metric $\Tensb{\gamma}^{(3)}\left(t=0,\Tensb{x}\right)$ on the time-symmetric hypersurface was first determined by Wheeler for the 5-cell universe \citep{Wheeler1982} and then more recently by Clifton \emph{et al.}\ for all closed lattices \citep{CRT}; this work has been further generalised by Korzy{\'n}ski who examined closed universes with an arbitrary number of identical masses, not necessarily arranged in a lattice \citep{Korzynski}.  The time evolution of this initial data has been investigated numerically by Bentivegna and Korzy{\'n}ski \citep{BentivegnaKorzynski2012} and analytically by Clifton \emph{et al.}\ \citep{CGRT, CGR}; however Clifton \emph{et al.}\ only considered the evolution of certain highly-symmetric curves on the initial surface, so the complete, analytic 4-metric of the closed universe is still lacking.  The flat lattice universe has been modelled perturbatively by Bruneton and Larena \citep{BrunetonLarena1} and numerically by various authors \citep{YATN, YON, BentivegnaKorzynski2013}; using their perturbative model, Bruneton and Larena have also investigated the flat universe's optical properties \citep{BrunetonLarena2}.  Yet in spite of the progress in determining $\Tensb{\gamma}^{(3)}\left(t,\Tensb{x}\right)$, a complete, non-perturbative solution for any lattice universe remains elusive.  Moreover, similar analyses of the open universe, analytical or numerical, have yet to be performed; and the optical properties of the closed universe have yet to be examined as well.

Since the complete 4-metric for any lattice universe is still unknown, constructing a non-perturbative approximation of it instead may help provide further insights into the universe's properties.  Such an approximation may, for instance, offer qualitative insights into the universe's dynamics as well as into the behaviour of photons as they propagate along any arbitrary direction; indeed, it would be difficult to propagate photons through any universe without having a full 4-metric, so for this reason, an analytic approximation of the metric would be especially invaluable.  In fact, the lattice universe was first studied using one such approximation.  In 1957, Lindquist and Wheeler (LW) \citep{LW, *LW-err, Houches} devised a construction wherein each polyhedral lattice cell was approximated by a spherical cell with Schwarzschild geometry inside.  For this approximation to work though, the masses were required to be of such magnitude and separation that the geometry around each could be reasonably approximated by Schwarzschild geometry; for example, should the masses get too close together, the deviation around each mass from Schwarzschild geometry would become too great such that the LW approximation would break down.  The original construction, however, was restricted to approximating closed universes.  Yet even in this limited case, Lindquist and Wheeler could obtain a scale factor that very closely resembled the scale factor of closed dust-filled FLRW universes.  Indeed, the LW scale factor would asymptotically approach the FLRW scale factor as the total number of masses in the universe increased while the universe's total mass was held constant.

More recently, Clifton and Ferreira (CF) \citep{CF, *CF-err} have re-visited the LW construction and extended it in several notable ways.  First, they were able to generalise the construction so that flat and open universes could be modelled as well.  Secondly, while Lindquist and Wheeler had defined a `cosmological time' co-ordinate in the neighbourhood of the cell boundaries, Clifton and Ferreira were able to extend the co-ordinate to cover the entire space-time globally.  This opened the way to a definition of co-moving observers for lattice universes, thereby allowing quantities such as cosmological redshifts to be defined.  Finally, Clifton and Ferreira were able to define a set of boundary conditions that photons must obey when crossing from one cell into the next, thereby allowing photons to be propagated across the entire LW model.  With their extension, Clifton and Ferreira then modelled the flat lattice universe and found that its evolution and redshifts generally agreed well with those of the flat dust-filled FLRW universe but that its angular diameters and luminosity distances differed \citep{CFO}.

In this paper, we shall further explore the properties of the lattice universe using the LW approximation with the CF extensions.  We shall consider all three types of universes, closed, flat, as well as open.  First, we shall extend the LW evolution equation for the closed universe's scale factor and derive analogous expressions for the flat and open universes.  These scale factors are related to the radius of the cell boundary by a constant scaling, and in all three cases, the scale factors have the same functional form as their FLRW counterparts in \eqref{closed_a}--\eqref{open_a}.  Secondly, as Clifton and Ferreira have shown, the LW cell boundary radius satisfies a Friedmann-like equation for all three types of lattice universes; they have also shown that a similar equation holds for the closed universe's scale factor.  By using the LW scale factors we derived rather than cell radii, we can make the analogy between the CF Friedmann equation and the FLRW Friedmann equation more salient; specifically, we shall demonstrate that the CF Friedmann equation takes a form essentially identical to \eqref{Friedmann1} for $\CosmoConst = 0$ and with $k$ in the LW case also equalling $+1$, $0$, or $-1$ according to whether the universe is open, flat, or closed respectively.  Moreover, we find that the density $\rho$ in the LW case equals a cell's Schwarzschild mass divided by the spherical volume $4 \pi r_b^3/3$ where $r_b$ is the radius of a cell; thus, the density behaves as if the cell were a Euclidean sphere.  Thirdly, the CF global co-ordinate system was found unsuitable for studying closed lattice universes.  We shall show that for closed universes, the CF co-ordinates do not actually cover the interior of the cell completely but leave a gap region instead.  For this reason, we must adopt an alternative co-ordinate system if we wish to propagate photons through closed universes, and in fact, a suitable system was proposed by Lindquist and Wheeler themselves at the end of their original paper \citep{LW, *LW-err}.  Fourthly, though our method for propagating photons across the universe is strongly influenced by Clifton and Ferreira's approach, we have adopted several modifications to the set of conditions used to propagate photons across cell boundaries; these modifications will be justified in the paper.  Finally, we shall examine the redshifts of photons travelling along a range of trajectories in closed, flat, and open universes.  So far, redshifts have only been studied in the flat universe using both a perturbative approach \citep{BrunetonLarena2} and the LW formalism \citep{CF, CFO}; redshifts for the closed and open universes currently do not exist anywhere in the literature, so this part of our results is entirely new.  We shall show that for all three types of universes, the redshifts behave broadly in the same way as their FLRW counterparts though with certain differences that can be attributed to the inhomogeneities of the lattice.  One of our most striking results is that LW redshifts can differ from their FLRW counterparts by as much as 30\%, which has implications on estimating the age of the photon's source, and that an LW universe can give rise to a non-zero integrated Sachs-Wolfe effect without needing any cosmological constant.

This paper is organised as follows.  The second section will explain the origins of the LW approximation and detail its construction; the CF generalisation to flat and open universes will also be reviewed.  This section will also compare the CF and LW global co-ordinate system and demonstrate the existence of a gap region if the CF system is applied to closed universes.  The third section will derive the LW analogues of the FLRW scale factor and the modified CF Friedmann equation for all three types of lattice universes.  The latter half of this paper will examine the behaviour of redshifts in the lattice universe.  The fourth section will briefly explain the manner by which redshifts are computed, which follows the method of Clifton and Ferreira.  The fifth section will explain the boundary conditions that are used to propagate photons across the boundary between two contiguous cells.  The sixth section will explain the numerical method used to simulate the propagation of photons through the lattice.  We shall apply Williams and Ellis' Regge calculus formalism for Schwarzschild space-time to each cell \citep{WilliamsEllis1, WilliamsEllis2}; however we have derived an improvement to their rules for tracing geodesics through the Williams-Ellis Schwarzschild space-time which significantly enhances numerical accuracy.  Our improvement will be derived in this section with supporting numerical evidence presented in \appendref{BV}; the improvement is not specific to the models being considered in this paper but is applicable to any situation where the Williams-Ellis scheme is being used to numerically simulate geodesics through Schwarzschild space-time.  The penultimate section presents and analyses the redshift results of our simulations.  The final section concludes with a discussion of possible directions in which this work might be extended.

In this paper, we shall use geometric units where $G = c = 1$.

\section{Constructing the LW approximation}
The LW approximation of lattice universes was inspired by Wigner and Seitz's \citep{WignerSeitz} method for approximating electronic wavefunctions in crystal lattices.  This method approximates the polyhedral cell of a crystal lattice by a sphere of the same volume; any conditions that the wavefunction must satisfy on the original cell boundary get imposed on the spherical boundary instead.  For instance in the original lattice, reflection symmetry at the cell boundary means that the wavefunction $\Psi$ of the lowest energy free electron must satisfy $\Tensb{n} \cdot \nabla\Psi = 0$ at the boundary, where the vector $\Tensb{n}$ is orthogonal to the boundary; in the Wigner-Seitz approximation, this same vanishing-derivative condition gets imposed instead on the spherical boundary; that is, $\Psi$ must now satisfy $\partial \Psi / \partial r = 0$ at the boundary; and this effectively assumes the electron potential within a cell to be spherically symmetric.  The higher the symmetries of the original polyhedral cell, the closer the cell resembles a sphere, and the more accurate the results obtained from the Wigner-Seitz models.  Indeed when applied to crystals where exact solutions are known, the Wigner-Seitz construction yields very accurate results \citep{Shockley, LageBethe}.

In analogy with the Wigner-Seitz construction, Lindquist and Wheeler approximate each elementary cell of the lattice universe by a spherical cell and the metric inside each cell by the Schwarzschild metric
\begin{equation}
d\LinEl^2 = - \left(1 - \frac{2 m}{r} \right) dt^2 + \frac{dr^2}{\left(1 - \frac{2 m}{r} \right)} + r^2 \left(d\theta^2 + \sin^2\theta d\phi^2 \right).
\label{Schwarz}
\end{equation}
They have called this spherical cell the \emph{Schwarzschild-cell}.  Without knowing the true metric, one cannot directly assess how well is a polyhedral cell in the lattice universe approximated by a spherical one.  However Lindquist and Wheeler have shown that polyhedra in a closed or flat 3-space of constant curvature are reasonably approximated by spheres of equal volume \citep{LW, *LW-err, Houches}; hence when such polyhedra are combined into lattices in such 3-spaces, the polyhedral lattice cells would be reasonably well-approximated by spherical cells.  They therefore provide this as partial evidence that the approximation would probably be reasonable as well for the polyhedral cells in the lattice universe.

\begin{figure}[htb]
\input{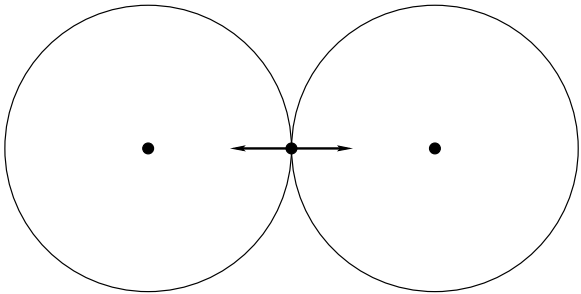}
\caption{\label{fig:boundary_particle}A test particle co-moving with the boundary between two cells must be simultaneously falling towards both central masses.  Thus the boundary itself must expand and contract accordingly.}
\end{figure}

In contrast to the Wigner-Seitz lattice, the LW lattice is itself dynamical.  A test particle sitting at the boundary between two Schwarzschild-cells will by symmetry always remain at the boundary.  Yet like any other test particle in a Schwarzschild geometry, this test particle must also be radially falling towards the centre of one of the Schwarzschild-cells.  And by the same reasoning, this particle is also radially falling towards the centre of the other Schwarzschild-cell.  Therefore this test particle, and hence the cell boundary itself, is radially falling towards both cell centres simultaneously, as depicted in \FigRef{fig:boundary_particle}.
The boundary's motion is hence given by the equation of motion for a radial time-like geodesic,
\begin{equation}
\left(\frac{dr}{d\LattTime}\right)^2 = E - 1 + \frac{2m}{r},
\label{rad_geod}
\end{equation}
where $\LattTime$ is the proper time of a test particle following the geodesic and $E$ is a positive constant of motion; it can be shown that $\sqrt{E}$ is the particle's energy per unit mass at radial infinity.  However this simultaneous free-fall of the boundary towards the two masses is actually the result of the two masses themselves falling towards each other under their mutual gravitational attraction.  This mutual attraction of all the point-masses thereby gives rise to the expansion and contraction of the lattice itself, which manifests as the expansion and contraction of the Schwarzschild-cell boundary.

We can see from \eqref{rad_geod} that the value of $E=E_b$ at the boundary determines whether the cell boundaries will expand indefinitely or eventually re-collapse, and hence whether the underlying lattice universe is open, flat, or closed.  For $E_b < 1$, the boundaries will expand until reaching a maximum radius of
\begin{equation}
r_{max} = \frac{2m}{1-E_b}
\label{r_max}
\end{equation}
before re-collapsing; this corresponds to a closed universe.  For $E_b > 1$, the boundaries will expand indefinitely; this corresponds to an open universe.  And for $E_b = 1$, the boundaries travel at the escape velocity and just reach radial infinity; this corresponds to a flat universe.  Thus it is through this constant $E_b$ that Clifton and Ferreira generalise Lindquist and Wheeler's work to flat and open universes in a natural manner.  

\begin{figure}[tbhp]
\input{coord-mesh1.pspdftex}
\caption{\label{fig:coord_mesh1}Hypersurfaces of constant Schwarzschild time $t$ intercept each other rather than mesh together at the cell boundaries.  The normals from different cells point in different directions at the boundary.  Therefore, Schwarzschild $t$ cannot serve as a global cosmological time co-ordinate for our model.}
\vspace{10pt}
\input{coord-mesh2.pspdftex}
\caption{\label{fig:coord_mesh2}Hypersurfaces of constant $\LattTime_{\scriptscriptstyle CF}$ and $\LattTime_{\scriptscriptstyle LW}$ mesh at the boundaries, and the normals from the two cells coincide.  Therefore either can serve as cosmological time.}
\end{figure}
In order to `glue' the individual cells together into a lattice, we require that the 3-space of constant time in one Schwarzschild-cell mesh at the boundary with the corresponding 3-space of the neighbouring cell.  As Lindquist and Wheeler have pointed out, two 3-spaces will mesh together if and only if they intersect their common boundary orthogonally.  Surfaces of constant Schwarzschild time $t$ do not satisfy this meshing condition; instead, they intersect each other when they meet, as illustrated in \FigRef{fig:coord_mesh1}, and some other time co-ordinate must be found.  

Both Lindquist and Wheeler as well as Clifton and Ferreira devised new time co-ordinates that satisfied the meshing condition.  The LW time co-ordinate is defined for closed universes only while the CF time co-ordinate can only be applied to flat and open universes.  Both time co-ordinates are defined to equal the proper times of a congruence of radial time-like geodesics, which must include the geodesics of test particles co-moving with the boundary.  Test particles following these geodesics will travel at the same velocity if at the same radius, thus forming freely falling shells.  All geodesics' clocks are calibrated to read identical proper time on some initial space-like hypersurface that intersects the geodesics orthogonally.  Then if the congruence was well-chosen, all other constant proper time hypersurfaces will intersect the congruence orthogonally as well.  In particular, such hypersurfaces would always intersect the boundary orthogonally, thus satisfying the meshing condition, as illustrated in \FigRef{fig:coord_mesh2}.  Such a definition of time is well-defined globally and can therefore be used as a `cosmological time' parametrising the lattice universe.  Such a construction of cosmological time is natural because we expect co-moving cosmological observers to be stationary with respect to constant time hypersurfaces and also to follow freely falling geodesics.  However there is no unique choice of congruence; any choice that satisfies our constraints above is acceptable.  Thus herein lies the difference between LW time and CF time.  Note that \eqref{rad_geod} actually describes any radial time-like geodesic in Schwarzschild space-time, including boundary geodesics, for which $r = r_b$.  Clifton and Ferreira's congruence consists of geodesics all having the same constant $E$, namely that of the boundary $E_b$.  Lindquist and Wheeler's congruence uses different $E<1$ for different geodesics.  Indeed, it may be possible to find an alternative choice of congruence based on some stronger physical motivation, but we shall not consider this problem here.  Instead, we shall use LW time for the closed universe and CF time for the flat and open universes.

For boundaries following out-going trajectories, the CF time co-ordinate $\LattTime_{\scriptscriptstyle CF}$ is given by
\begin{equation}
d\LattTime_{\scriptscriptstyle CF} = \sqrt{E_b}\, dt - \frac{\sqrt{E_b - 1 + \frac{2m}{r}}}{\left(1 - \frac{2m}{r}\right)} dr,
\label{tau_CF}
\end{equation}
where $E_b$ is the same positive constant as $E$ in \eqref{rad_geod} for the boundary geodesic.  The Schwarzschild metric \eqref{Schwarz} now becomes
\begin{equation}
\begin{aligned}
d\LinEl^2 =& - \frac{1}{E_b} \! \left(1 - \frac{2 m}{r} \right) \! d\LattTime_{\scriptscriptstyle CF}^2 - \frac{2}{E_b}\sqrt{E_b - 1 + \frac{2 m}{r}} d\LattTime_{\scriptscriptstyle CF} dr \\
&{}+ \frac{dr^2}{E_b} + r^2 d\Omega^2.
\end{aligned}
\label{Schwarz_CF}
\end{equation}
The boundary's equation of motion is still given by \eqref{rad_geod} but with the proper time $\LattTime$ replaced by $\LattTime_{\scriptscriptstyle CF}$, as it can be shown that $\LattTime_{\scriptscriptstyle CF}$ is identical to the boundary's proper time.  In CF co-ordinates, the 4-vector tangent to any trajectory satisfying \eqref{rad_geod} for $E=E_b$ is
\begin{equation}
u^a = \left( 1, \sqrt{E_b - 1 + \frac{2m}{r}}, 0, 0 \right).
\label{boundary_vec}
\end{equation}
For an arbitrary vector $\Tensb{n}$ tangent to a constant $\LattTime_{\scriptscriptstyle CF}$ surface, 
\begin{equation}
n^a = \left( 0, n^r, n^\theta, n^\phi \right),
\label{tau_vec}
\end{equation}
it can be shown that $\Tensb{u} \cdot \Tensb{n} = 0$ and hence that surfaces of constant $\LattTime_{\scriptscriptstyle CF}$ are orthogonal to all geodesics satisfying \eqref{rad_geod} for $E=E_b$, including particularly the boundary geodesics.  Thus the meshing condition is satisfied, and thus $\LattTime_{\scriptscriptstyle CF}$ can serve as a cosmological time.

\begin{figure}[htb]
\hspace{-0.25cm}\input{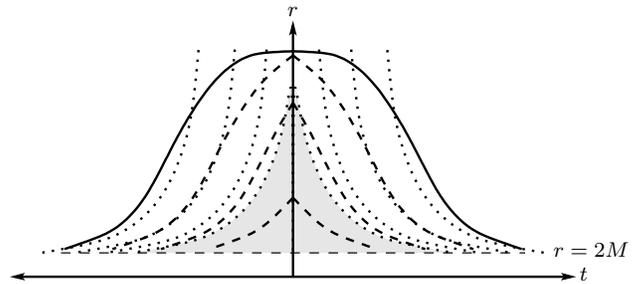}
\caption{\label{fig:closed-cell}A qualitative plot of various co-moving test-particle trajectories (dashed line) inside a closed Schwarzschild-cell and of various constant $\LattTime_{\scriptscriptstyle CF}$ surfaces (dotted line); the plot is of Schwarzschild co-ordinates $r$ versus $t$.  The cell boundary's trajectory has been drawn with a solid line.  The $r$ axis passes through the boundary's moment of maximum expansion, which happens when $\LattTime_{\scriptscriptstyle CF} = \LattTime_{max}$.  Since the radial velocity must change sign abruptly when the boundary starts closing, there is a noticeable discontinuity in the gradients of the test particle graphs at the $r$ axis.  The constant $\LattTime_{\scriptscriptstyle CF}$ surfaces for $\LattTime_{\scriptscriptstyle CF}$ just greater than and less than $\LattTime_{max}$ are not identical.  Between the two surfaces is a `gap' region, shaded in the figure, where space-time points do not have a well-defined cosmological time $\LattTime_{\scriptscriptstyle CF}$.}
\end{figure}

If we attempt to apply the CF co-ordinate system to closed universes though, we encounter problems.  When the boundary is contracting, its radial velocity is given by the negative root of \eqref{rad_geod} instead, and as a result, the square-roots in \eqref{tau_CF}--\eqref{boundary_vec} must change sign.  However the combined co-ordinate patches do not correctly cover the interior of the Schwarzschild-cell.  Rather, they leave an uncovered region centred on the boundary's moment of maximum expansion, as illustrated in \FigRef{fig:closed-cell}.

We shall denote by $\LattTime_{max}$ the boundary's proper time at maximum expansion.  For $\LattTime_{\scriptscriptstyle CF} < \LattTime_{max}$, surfaces of constant $\LattTime_{\scriptscriptstyle CF}$ are generated by the integral curves of \eqref{tau_CF} for $d\LattTime_{\scriptscriptstyle CF}=0$, that is, by
\begin{equation}
\frac{d r}{d t} = \frac{\sqrt{E_b}\left(1-\frac{2m}{r}\right)}{\sqrt{E_b-1+\frac{2m}{r}}} > 0\qquad \forall \, r_b \geq r > 2m.
\label{drdt}
\end{equation}
This has a solution of the form $t = f(r) + t_0$ where $t_0$ is the Schwarzschild time co-ordinate of the surface when it passes through $r = r_0$ and $r_0$ is the Schwarzschild radial co-ordinate satisfying $f(r_0)=0$.  Thus on the $r$-$t$ plane, like the one shown in \FigRef{fig:closed-cell}, all of these surfaces are identical apart from a horizontal shift corresponding to different $t_0$ constants.  Increasing $t_0$ shifts the surface to the right.  This implies the surface would intercept the boundary later in the boundary's trajectory, and therefore $\LattTime_{\scriptscriptstyle CF}$ would increase as well.  As we keep increasing $t_0$, we shall eventually reach a limiting surface $\LattTime_{\scriptscriptstyle CF} \to \LattTime_{max}$ from the right, and this surface we denote by $\LattTime_{max}^-$.

For $\LattTime_{\scriptscriptstyle CF} > \LattTime_{max}$, surfaces of constant $\LattTime_{\scriptscriptstyle CF}$ are generated instead by the negative of \eqref{drdt}.  This has the solution $t = -f(r) + t_0$, and again, $t_0$ is the Schwarzschild time co-ordinate of the surface when it passes through $r = r_0$.  On the $r$-$t$ plane, all $\LattTime_{\scriptscriptstyle CF} > \LattTime_{max}$ surfaces are also identical to each other apart from a horizontal shift.  Decreasing $t_0$ shifts the surface to the left and decreases $\LattTime_{\scriptscriptstyle CF}$.  As we decrease $t_0$, we shall eventually reach a limiting surface $\LattTime_{\scriptscriptstyle CF} \to \LattTime_{max}$ from the left, and this surface we denote by $\LattTime_{max}^+$.

The complete set of constant $\LattTime_{\scriptscriptstyle CF}$ surfaces will completely cover the interior of the cell without any overlaps or gaps if and only if surfaces $\LattTime_{max}^-$ and $\LattTime_{max}^+$ are identical.  This implies that
\begin{align*}
f(r) + t_0^- &= -f(r) + t_0^+\\
\implies f(r) & = constant,
\end{align*}
where $t_0^-$ and $t_0^+$ are the $t_0$ constants for $\LattTime_{max}^-$ and $\LattTime_{max}^+$, respectively.  However we know that function $f(r)$ is not constant, so we have a contradiction.  

Instead, we actually have a `gap' between $\LattTime_{max}^-$ and $\LattTime_{max}^+$.  Both surfaces meet at $\left(\LattTime_{max}, r_b(\LattTime_{max})\right) = (\LattTime_{max}, r_{max})$.  If we move along $\LattTime_{max}^-$ into the cell, then decreasing $r$ would decrease $t$ because $dr/dt$ is always positive along this surface.  If we move along $\LattTime_{max}^+$ into the cell, then decreasing $r$ would increase $t$ because $dr/dt$ is always negative along this surface.  On the $r$-$t$ plane, $\LattTime_{max}^-$ moves to the left and $\LattTime_{max}^+$ to the right as $r$ decreases from $r_{max}$.  Since $\LattTime_{max}^-$ is the rightmost of the $\LattTime_{\scriptscriptstyle CF} < \LattTime_{max}$ surfaces and $\LattTime_{max}^+$ the leftmost of the $\LattTime_{\scriptscriptstyle CF} > \LattTime_{max}$ surfaces, the region between $\LattTime_{max}^-$ and $\LattTime_{max}^+$ is not covered by any $\LattTime_{\scriptscriptstyle CF}$ surface and is hence a gap.

We shall therefore use the LW co-ordinate system, which covers closed cells correctly.  This system is constructed from a congruence of closed geodesics that attain maximum radii at the same Schwarzschild time.  Being a congruence, the geodesics have maximum radii spanning the entire range of $2m < r \leq r_b$, and from a generalised form of \eqref{r_max}, the geodesics must therefore have different constants $E$.  The geodesics' clocks are calibrated to have identical proper time at maximum radii.  The LW time co-ordinate $\LattTime_{\scriptscriptstyle LW}$ at any point is then defined to be the proper time of the geodesic passing through it.  Lindquist and Wheeler also defined a new radial co-ordinate $\tilde{r}_{\scriptscriptstyle LW}$ that is constant along each geodesic and equals the geodesic's maximum radius in Schwarzschild co-ordinates.

The co-ordinate transformation to $\left(\LattTime_{\scriptscriptstyle LW}, \tilde{r}_{\scriptscriptstyle LW} \right)$ is then given implicitly by
\begin{widetext}
\begin{align}
\begin{split}
t ={}& t_0 \pm \left\{\left[\left(\frac{\tilde{r}_{\scriptscriptstyle LW}}{2m}-1 \right)(\tilde{r}_{\scriptscriptstyle LW} - r) r \right]^{1/2} + 2m \left(\frac{\tilde{r}_{\scriptscriptstyle LW}}{2m}-1 \right)^{1/2} \left(\frac{\tilde{r}_{\scriptscriptstyle LW}}{2m}+2 \right) \cos^{-1}\left(\frac{r}{\tilde{r}_{\scriptscriptstyle LW}}\right)^{1/2} \right.\\
& {} \hphantom{t_0} + \left.  2m \ln \frac{r^{1/2}\left(\tilde{r}_{\scriptscriptstyle LW}/2m-1 \right)^{1/2} + \left(\tilde{r}_{\scriptscriptstyle LW}-r \right)^{1/2}}{r^{1/2}\left(\tilde{r}_{\scriptscriptstyle LW}/2m-1 \right)^{1/2} - \left(\tilde{r}_{\scriptscriptstyle LW}-r \right)^{1/2}}\right\},
\end{split}\\
\LattTime_{\scriptscriptstyle LW} ={}& \LattTime_0 \pm \left(\frac{\tilde{r}_{\scriptscriptstyle LW}}{2m}\right)^{1/2}\left[ r^{1/2}\left(\tilde{r}_{\scriptscriptstyle LW}-r \right)^{1/2} + \tilde{r}_{\scriptscriptstyle LW} \cos^{-1}\left(\frac{r}{\tilde{r}_{\scriptscriptstyle LW}}\right)^{1/2} \right], \label{LW_tau}
\end{align}
\end{widetext}
where $t_0$ and $\LattTime_0$ are the time co-ordinates at which the boundary attains its maximum expansion; the positive signs are taken when the boundary is expanding and the negative signs when contracting.  The Schwarzschild metric now becomes
\begin{widetext}
\begin{equation}
d\LinEl^2 = -d\LattTime_{\scriptscriptstyle LW}^2 + \left(\frac{\tilde{r}_{\scriptscriptstyle LW} - r}{4 \, r \, \tilde{r}_{\scriptscriptstyle LW} \left(\tilde{r}_{\scriptscriptstyle LW}/2m -1 \right) } \right) \left[\left(3\, \tilde{r}_{\scriptscriptstyle LW} - r \right) \left(\frac{r}{\tilde{r}_{\scriptscriptstyle LW} - r} \right)^{1/2} + 3\, \tilde{r}_{\scriptscriptstyle LW} \cos^{-1} \left(\frac{r}{\tilde{r}_{\scriptscriptstyle LW}}\right)^{1/2} \right]^2 d\tilde{r}_{\scriptscriptstyle LW}^2 + r^2\, d\Omega^2,
\label{Schwarz_LW}
\end{equation}
\end{widetext}
where $r$ is now a function of $\LattTime_{\scriptscriptstyle LW}$ and $\tilde{r}_{\scriptscriptstyle LW}$.  From this metric, it is clear that surfaces of constant $\LattTime_{\scriptscriptstyle LW}$ are always orthogonal to Lindquist and Wheeler's congruence.  Thus the meshing condition is also satisfied, and $\LattTime_{\scriptscriptstyle LW}$ is also suitable for use as a cosmological time.

We shall henceforth drop any subscript labels from the cosmological time $\LattTime$.  In the case of the closed universe, $\LattTime$ will denote $\LattTime_{\scriptscriptstyle LW}$, while in the flat or open universe, it will denote $\LattTime_{\scriptscriptstyle CF}$.

\section[The cosmological scale factor]{The cosmological scale factor and the Friedmann equations} \label{LWScaleFactor}
There is a clear analogy between the size of the Schwarzschild-cell $r_b(\LattTime)$ and the FLRW scale factor $\FriedScale(t)$, as both provide a scale of their respective universe's size.  Indeed we expect the lattice universe's scale factor $\LattScale(\LattTime)$ should be a function of $r_b$ alone, that is $\LattScale(\LattTime) = \LattScale(r_b(\LattTime))$.\footnote{We shall use $\FriedScale(t)$ when we refer to the FLRW scale factor and $\LattScale(\LattTime)$ when referring to the lattice universe scale factor.  The two functions are completely independent.}  If we expand $r_b$, then each cell of the lattice will expand by some scale $\xi$, and therefore the lattice as a whole will expand by $\xi$.  We shall take $\LattScale(\LattTime)$ to be related linearly to $r_b(\LattTime)$, that is, $\LattScale\big(r_b(\LattTime)\big) = \alpha\, r_b(\LattTime)$ for some constant $\alpha>0$.  As we shall see, $\LattScale(\LattTime)$ then depends on $\LattTime$ in a manner analogous to how $\FriedScale(t)$ depends on $t$ for FLRW universes.
\begin{figure}[htb]
\scalebox{0.77}{\input{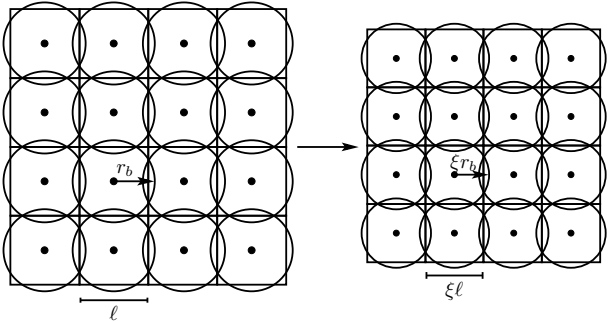}}
\caption{Rescaling the cell size by $\xi$ rescales the entire lattice universe by $\xi$ as well.}
\label{fig:lattice1}
\end{figure}

Lindquist and Wheeler were able to derive a parametric relationship between $\LattScale(\LattTime)$ and $\LattTime$ closely resembling that of \eqref{closed_a}, thus showing that $\LattScale(\LattTime)$ behaves essentially identically to its FLRW counterpart $\FriedScale(t)$; the only difference was the factor of $\FriedScale_0$.  However, their results were limited to the closed lattice universe.  We shall generalise Lindquist and Wheeler's results, showing that this identical behaviour between $\LattScale(\LattTime)$ and $\FriedScale(t)$ holds for lattices of all curvatures.  Equation \eqref{rad_geod} can be integrated to obtain
\begin{IEEEeqnarray}{rCl}
\LattTime &=& \sqrt{\frac{r_{\scriptscriptstyle E}}{2m}}\left(\sqrt{r \left(r_{\scriptscriptstyle E} - r \right)} + r_{\scriptscriptstyle E} \cos^{-1}\sqrt{\frac{r}{r_{\scriptscriptstyle E}}} \;\right) \label{tau_closed}\\[2mm]
&& \qquad\qquad\qquad\qquad\qquad\qquad \text{for } 0 < E < 1, \nonumber \\[3mm]
\LattTime &=& \left(\frac{4}{9\, r_{\scriptscriptstyle E}}\right)^{1/2} r^{3/2} \label{tau_flat}\\[2mm]
&& \qquad\qquad\qquad\qquad\qquad\qquad \text{for } E = 1, \nonumber\\[3mm]
\LattTime &=& \sqrt{\frac{r_{\scriptscriptstyle E}}{2m}}\left(\sqrt{r \left(r_{\scriptscriptstyle E} + r \right)} - r_{\scriptscriptstyle E} \sinh^{-1}\sqrt{\frac{r}{r_{\scriptscriptstyle E}}} \;\right) \label{tau_open}\\[2mm]
&& \qquad\qquad\qquad\qquad\qquad\qquad \text{for } E > 1, \nonumber
\end{IEEEeqnarray}
where 
\begin{equation}
r_{\scriptscriptstyle E} = 
\begin{dcases}
\frac{2m}{\lvert 1-E \rvert} & \text{for } E \neq 1 \text{ and } E>0,\\
2m & \text{for } E=1.
\end{dcases}
\label{r_E}
\end{equation}
In the case of $E<1$, $r_{\scriptscriptstyle E}$ corresponds to the maximum radius attained by the geodesic, and for the boundary geodesic, where $E = E_b$, this maximum radius is identical to \eqref{r_max}.  For boundary geodesics, where $r=r_b$, equations \eqref{tau_closed} to \eqref{tau_open} can be re-cast into the form
\begin{IEEEeqnarray}{ll}
\begin{aligned}
r_b &= \displaystyle \frac{r_{\scriptscriptstyle E_b}}{2}(1 - \cos \eta)\\[3mm]
\LattTime &= \displaystyle \frac{r_{\scriptscriptstyle E_b}}{2}\sqrt{\frac{r_{\scriptscriptstyle E_b}}{2m}}(\eta - \sin \eta)
\end{aligned}
& \qquad \text{for } 0<E_b<1,
\label{closed_a_tau}\\[5mm]
\begin{aligned}
r_b &= \displaystyle \left(\frac{9}{4}r_{\scriptscriptstyle E_b}\right)^{1/3} \LattTime^{2/3}
\end{aligned}
& \qquad \text{for } E_b=1,
\label{flat_a_tau}\\[5mm]
\begin{aligned}
r_b &= \displaystyle \frac{r_{\scriptscriptstyle E_b}}{2}(\cosh \eta - 1)\\[3mm]
\LattTime &= \displaystyle \frac{r_{\scriptscriptstyle E_b}}{2}\sqrt{\frac{r_{\scriptscriptstyle E_b}}{2m}}(\sinh \eta - \eta)
\end{aligned}
& \qquad \text{for } E_b>1,
\label{open_a_tau}
\end{IEEEeqnarray}
where $r_{\scriptscriptstyle E_b}$ simply denotes $r_{\scriptscriptstyle E}$ when $E = E_b$ and $\eta$ is simply a parametrisation.  If we choose $\alpha$ to be
\begin{equation}
\alpha = \displaystyle \sqrt{\frac{r_{\scriptscriptstyle E_b}}{2m}},
\label{alph_LW}
\end{equation}
then the scale factor $\LattScale(\LattTime)$ is given by
\begin{IEEEeqnarray}{rCll}
\LattScale &=& \displaystyle \frac{r_{\scriptscriptstyle E_b}}{2}\sqrt{\frac{r_{\scriptscriptstyle E_b}}{2m}} \, (1 - \cos\eta) & \qquad \text{for } 0<E_b<1, \label{param_a_closed}\\
\LattScale &=& \displaystyle \left(\frac{9}{4}r_{\scriptscriptstyle E_b}\sqrt{\frac{r_{\scriptscriptstyle E_b}}{2m}}\;\right)^{1/3} \LattTime^{2/3} & \qquad \text{for } E_b<1, \label{param_a_flat}\\
\LattScale &=& \displaystyle \frac{r_{\scriptscriptstyle E_b}}{2}\sqrt{\frac{r_{\scriptscriptstyle E_b}}{2m}} \, (\cosh\eta - 1) & \qquad \text{for } E_b>1. \label{param_a_open}
\end{IEEEeqnarray}
Comparing the relations just obtained for $\LattScale(\LattTime)$ and $\LattTime$ with their counterparts in \eqref{closed_a}--\eqref{open_a}, we find that $\LattScale(\LattTime)$ and $\LattTime$ have the same functional form as their FLRW counterparts for all background curvatures.  Moreover, we have deduced an expression for the factor $\alpha$ that is the same for all universes.  The only difference between the lattice and FLRW relations is that $\LattScale_0$ for the lattice universe is
\begin{equation}
\LattScale_0 = \displaystyle r_{\scriptscriptstyle E_b}\sqrt{\frac{r_{\scriptscriptstyle E_b}}{2m}}.
\label{lattice_a0}
\end{equation}
Equivalently, by equating \eqref{lattice_a0} with \eqref{FLRW_a0}, we can say that the density $\tilde{\rho}_0$ for the lattice universe is
\begin{equation}
\tilde{\rho}_0 = \frac{m}{\frac{4}{3}\pi\alpha^{-3}}.
\label{latt_rho0}
\end{equation}
The denominator is simply the `Euclidean volume' of a Schwarzschild-cell with radius $r_b$ corresponding to $\LattScale=1$.

As Clifton and Ferreira have noted, the radius of the cell boundary $r_b(\LattTime)$ satisfies an equation strongly resembling the Friedmann equation \eqref{Friedmann1} for $\Lambda=0$; they also found a similar relation for the closed LW universe's scale factor.  By consistently using the LW scale factors just derived rather than $r_b(\LattTime)$, we can re-express the CF Friedmann equation in a form that makes its resemblance to \eqref{Friedmann1} much more salient; the CF Friedmann equation then becomes
\begin{equation}
\left(\frac{\dot{\LattScale}(\LattTime)}{\LattScale(\LattTime)}\right)^2 = \frac{8\pi\tilde{\rho}}{3} - \frac{k}{\LattScale(\LattTime)^2},
\label{CF_Fried}
\end{equation}
where $k = \alpha^2(1-E_b)$ plays the r{\^o}le of the curvature constant for the lattice universe, and density $\tilde{\rho}$ is given by
\begin{equation}
\tilde{\rho} = \frac{m}{\left(\frac{4}{3}\pi \LattScale(\LattTime)^3\right)/\alpha^3} = \frac{m}{\frac{4}{3}\pi r_b^3}.
\label{rho_CF}
\end{equation}
As noted earlier, $E_b$ determines whether the universe will be open, flat, or closed, and we see that $k$ will take on the correct sign accordingly.  In fact, if we make use of \eqref{r_E} and \eqref{alph_LW}, $k$ simplifies to $-1, 0, +1$ for open, flat, and closed lattice universes, respectively, much like its FLRW analogue.

To complete our model, all that remains is to specify $\alpha$ or $E_b$.  However, Lindquist and Wheeler have actually provided an argument to specify $\alpha$ and an independent argument to effectively\footnote{Recall that the quantity $E_b$ was introduced to the LW construction afterwards by Clifton and Ferreira.  However, Lindquist and Wheeler imposed a tangency condition on the Schwarzschild-cells, which we shall soon describe, that effectively specifies $E_b$.} specify $E_b$.  What is intriguing is that their choices of $\alpha$ and $E_b$ are consistent with the relationship between the two quantities required by \eqref{alph_LW}.  We shall summarise Lindquist and Wheeler's arguments leading to their choices.  Although they dealt only with closed universes, we shall, in this paper, generalise their arguments to flat and open universes as well.

We begin with Lindquist and Wheeler's choice of $\alpha$.  Note that hypersurfaces of constant $t$ in the FLRW metric \eqref{FLRW-metric} correspond to 3-spaces of constant curvature, and through an appropriate choice of scaling, the radius of curvature is given by the scale factor $a(t)$.  We embed the same type of lattice as the lattice universe on such a hypersurface with the appropriate curvature.\footnote{That is, for closed lattice universes, we use hyperspheres; for flat lattice universes, Euclidean space; and for open lattice universes, hyperbolic space.}  We shall call this hypersurface the \emph{comparison hypersurface}.  We then approximate each polyhedral cell by a sphere of the same volume.\footnote{This is actually Clifton and Ferreira's generalisation.  Lindquist and Wheeler's original condition was that the spheres occupy $N^{-1}$ of the comparison hypersphere's total volume, where $N$ is the total number of cells.  Such a condition clearly needs to be modified for flat and open universes, as both $N$ and the hypersurface volume are infinite in these cases.}  For closed hyperspheres, we define $\psi$ to be the angle between the centre of the spherical cell and its boundary as measured from the centre of the hypersphere.  Then $\psi$ is given implicitly by
\begin{equation}
\frac{1}{N} = \frac{2\psi - \sin 2\psi}{2\pi},
\label{psi}
\end{equation}
where $N$ is the total number of cells in the lattice.  For hyperbolic spaces, we define $\psi$ analogously in terms of hyperbolic angles.  If $r_0$ is the radius of one such spherical cell, then we can relate $r_0$ to the comparison hypersurface's radius of curvature $\LattScale_{\scriptscriptstyle LW}$ by 
\begin{equation}
\LattScale_{\scriptscriptstyle LW} = \frac{r_0}{\chi(\psi)},
\label{a_LW}
\end{equation}
where the function $\chi(\psi)$ is given by\footnote{Note that when $k=0$ in the FLRW metric \eqref{FLRW-metric}, we have complete freedom to make a re-scaling of the form $r \rightarrow \xi r$ and $a \rightarrow a/\xi$.  Hence we can always choose a comparison hypersurface such that $a_{\scriptscriptstyle LW} = r_0$.}
\begin{equation}
\chi(\psi) = 
\begin{dcases}
\sin \psi & \text{for closed universes},\\
1 & \text{for flat universes},\\
\sinh \psi & \text{for open universes}.
\end{dcases}
\label{chi_func}
\end{equation}
Lindquist and Wheeler identified the Schwarzschild-cell radius $r_b$ of the lattice universe with $r_0$ and used the corresponding $\LattScale_{\scriptscriptstyle LW}$ as given by \eqref{a_LW} to be the lattice universe's scale factor; that is, 
\begin{equation}
\LattScale_{\scriptscriptstyle LW}(\LattTime) = \frac{r_b(\LattTime)}{\chi (\psi)}.
\label{LW_scalefactor}
\end{equation}
Hence they have chosen $\alpha = 1/\chi (\psi)$.  According to \eqref{alph_LW}, $E_b$ would therefore correspond to
\begin{equation}
E_b = 
\begin{dcases}
\cos^2 \psi & \text{for closed universes},\\[1mm]
1 & \text{for flat universes},\\[1mm]
\cosh^2 \psi & \text{for open universes}.
\end{dcases}
\label{LW_E}
\end{equation}

However, Lindquist and Wheeler prescribed $E_b$ independently of \eqref{alph_LW}.  They instead embedded the Schwarzschild-cell in the comparison hypersurface and required it to be in some sense tangent to the hypersurface.  Their prescription for the embedding is as follows.  Suppose we replaced each of the original spherical cells in the hypersurface with a Schwarzschild-cell from the lattice universe.  Then we want to choose an $E_b$ to make the Schwarzschild-cell tangent to the hypersurface.  Lindquist and Wheeler formulate this tangency condition as follows.  Take any great circle on the boundary of the original sphere and compare its circumference with that of the corresponding great circle on an infinitesimally smaller sphere.  Depending on which hypersurface the cell is embedded in, the circumferences will obey the relation
\begin{align*}
\lefteqn{\frac{1}{2\pi} \frac{d(\text{circumference})}{d(\text{radial distance})}}\\[1mm]
&\qquad =
\begin{dcases}
\frac{1}{2\pi} \frac{d(2\pi a_{\scriptscriptstyle LW} \sin \psi)}{a_{\scriptscriptstyle LW} d\psi} & \text{for closed hyperspheres},\\[1mm]
1 & \text{for flat hypersurfaces},\\[1mm]
\frac{1}{2\pi} \frac{d(2\pi a_{\scriptscriptstyle LW} \sinh \psi)}{a_{\scriptscriptstyle LW} d\psi} & \text{for open hypersurfaces},
\end{dcases}
\\[1mm]
&\qquad = 
\begin{dcases}
\cos \psi, &\\[1mm]
1, &\\[1mm]
\cosh \psi, &
\end{dcases}
\end{align*}
as depicted by \FigRef{fig:3sphere} to \FigRef{fig:hyperbolic}.  Then for the Schwarzschild-cell to be tangent to the hypersphere, we also require that at its boundary,
\begin{equation}
\frac{1}{2\pi} \frac{d(\text{circumference})}{d(\text{radial distance})} = 
\begin{dcases}
\cos \psi & \text{for closed universes},\\[1mm]
1 & \text{for flat universes},\\[1mm]
\cosh \psi & \text{for open universes}.
\end{dcases}
\end{equation}
From the Schwarzschild metric expressed in CF co-ordinates \eqref{Schwarz_CF}, we can see that $d(\text{circumference}) = 2\pi d r_b$ and that $d(\text{radial distance}) = d r_b / \sqrt{E_b}$.  Thus,
$$\frac{1}{2\pi} \frac{d(\text{circumference})}{d(\text{radial distance})} = \sqrt{E_b}.$$
From the Schwarzschild metric in LW co-ordinates \eqref{Schwarz_LW}, it can also be shown, by making use of \eqref{LW_tau}, that $(2\pi)^{-1} d(\text{circumference}) / d(\text{radial distance})$ is the same.  Solving for $E_b$, we then obtain the same result as in \eqref{LW_E}.

\begin{figure}[htbp]
\input{3-sphere.pspdftex}
\caption{\label{fig:3sphere}A spherical cell of radius $\LattScale_{\scriptscriptstyle LW} \sin \psi$ and an infinitesimally smaller shell, indicated by the dashed line, have been embedded in a three-dimensional hypersphere of radius $\LattScale_{\scriptscriptstyle LW}$, with one of the angular dimensions suppressed.  The radial distance between the two shells, as measured along the 3-sphere, is $\LattScale_{\scriptscriptstyle LW} d\psi$.  Because of the curvature of the underlying 3-sphere, we have that $d(\text{circumference})/d(\text{radial distance}) = 2\pi \cos \psi$.}
\vspace{10pt}
\input{flat.pspdftex}
\caption{\label{fig:flat}An analogous embedding in a flat hypersurface.  In this case, the circumference of the boundary is $2\pi \LattScale_{\scriptscriptstyle LW}$; the radial distance between the two shells is $d r$; and $d(\text{circumference})/d(\text{radial distance}) = 2\pi$.}
\vspace{10pt}
\input{hyperbolic.pspdftex}
\caption{\label{fig:hyperbolic}An analogous embedding in a hyperbolic hypersurface.  The circumference of the boundary is $2\pi \LattScale_{\scriptscriptstyle LW} \sinh \psi$; the radial distance between the two shells is $\LattScale_{\scriptscriptstyle LW} d\psi$; and $d(\text{circumference})/d(\text{radial distance}) = 2\pi \cosh \psi$.}
\end{figure}

We close this section with a few remarks about the large-$N$ behaviour of the closed LW model.  By taking $N \to \infty$, we can deduce what limiting behaviour the model would approach as the number of masses increases while the universe's total mass is held constant.  It can be shown from \eqref{psi} that as $N \to \infty$, the angle $\psi \to \left(\frac{3\pi}{2N}\right)^{1/3}$.  If $M = Nm$ is the total mass of the universe, then
$$\lim_{N \to \infty} \alpha^3\, m = \lim_{N \to \infty} \frac{M}{N \sin^3 \psi} = \frac{2M}{3\pi},$$
and therefore the density $\tilde{\rho}_0$ for the lattice universe, as given by \eqref{latt_rho0}, becomes
\begin{equation}
\lim_{N \to \infty} \tilde{\rho}_0 = \frac{M}{2\pi^2}.
\end{equation}
In closed FLRW space-time, a hypersphere of constant $t$ has a volume of $2\pi^2 \FriedScale(t)^3$.  Since the FLRW $\rho_0$ is defined to be the density when $\FriedScale(t)=1$, then $\rho_0$ also equals $M/2\pi^2$.  Therefore as $N \to \infty$, the lattice universe density $\tilde{\rho}_0$ approaches its FLRW equivalent $\rho_0$, and hence the lattice universe factor $\LattScale_0$ approaches its FLRW equivalent as well.  Consequently, $\LattScale(\LattTime)$ in \eqref{param_a_closed} approaches $\FriedScale(t)$ in \eqref{closed_a}, and $\LattTime$ in \eqref{closed_a_tau} becomes identical to $t$ in \eqref{closed_a}.  We also note that $\tilde{\rho}$ in \eqref{rho_CF} becomes 
\begin{equation}
\lim_{N \to \infty} \tilde{\rho} = \frac{M}{2\pi^2 \LattScale^3},
\end{equation}
which is identical to the FLRW energy density $\rho$.  This implies that the CF Friedmann equation \eqref{CF_Fried} would be identical to the FLRW Friedmann equation \eqref{Friedmann1} for $k=1$ and $\Lambda=0$.  Therefore, as $N$ increases and the closed lattice universe approaches the continuum limit, it becomes increasingly similar to the closed dust-filled FLRW universe, which is consistent with the matter content itself becoming increasingly similar to homogeneous and isotropic dust.  In fact, Korzy{\'n}ski has proven that this large-$N$ correspondence between the two closed universes is indeed true for the exact solution of the Einstein field equations, at least on the hypersurface of time-symmetry \citep{Korzynski}.  He actually considered universes where there could be an arbitrary number $N$ of identical Schwarzschild-like black holes and demonstrated that as $N \to \infty$, the hypersurface approaches its FLRW counterpart exactly, both in terms of the scale-factor and in terms of the energy density.  Therefore in this regard, the LW approximation agrees with the exact solution. 

\section[Lattice universe redshifts]{Redshifts in the lattice universe}
Cosmological redshifts, $1 + z_{\scriptscriptstyle FLRW}$, in FLRW universes are defined with respect to sources and observers that are co-moving with the universe's expansion.  The redshifts are given by
\begin{equation}
1 + z_{\scriptscriptstyle FLRW} = \frac{\FriedScale_o}{\FriedScale_e},
\label{z_FLRW}
\end{equation}
where $\FriedScale_o$ and $\FriedScale_e$ are the cosmological scale factors at the moments of observation and emission, respectively.

\begin{figure}[htb]
\scalebox{1}{\input{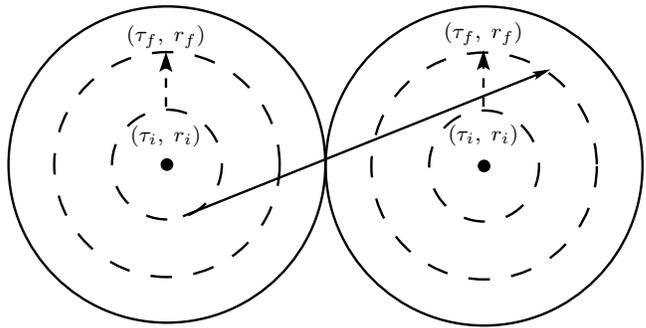}}
\caption{\label{fig:cell_crossing}A photon travels from radius $r_i$ at cosmological time $\LattTime_i$ in one cell to radius $r_f$ at time $\LattTime_f$ in the next, as indicated by the long solid arrow.  We assume that the photon always passes through the boundary in the manner illustrated here: that is, the point of crossing is a point of tangency between the boundaries of the two cells.  A photon is emitted at $(\LattTime_i,\; r_i)$ by a source travelling along a geodesic given by \eqref{rad_geod} and observed by an observer in another cell.  Although in a different cell from the source, the observer is still `co-moving' with the source; that is, for any $\LattTime$, the observer is always at the same radius as the source in their respective cells, and this requires the observer to travel along a geodesic with the same $E$ as the source.}
\end{figure}

We should like to find the lattice universe analogue to $z_{\scriptscriptstyle FLRW}$ so that we can compare redshifts in the two types of universes.  In general, redshifts $1+z$ are defined as the ratio of the photon frequency measured at the source to the frequency measured by the end observer.  The closest analogy to co-moving sources and observers in the lattice universe would be sources and observers that are co-moving with respect to a constant $\LattTime$ surface.  All such observers would be following radial geodesics that obey \eqref{rad_geod}, with $E$ fixed to be $E_b$ in the flat and open universes but geodesic-dependent in the closed universe.  If $\Tensb{u}$ is the 4-velocity of an observer and $\Tensb{k}$ the 4-momentum of a photon as it passes the observer, then the photon frequency measured by the observer would be $-\Tensb{u} \cdot \Tensb{k}$.  For the lattice universe, the cosmological redshift is therefore given by
\begin{equation}
1+z_{\scriptscriptstyle LW} = \frac{\Tensb{u}_s \cdot \Tensb{k}_s}{\Tensb{u}_o \cdot \Tensb{k}_o},
\end{equation}
where the subscripts $s$ and $o$ denote `source' and `observer', respectively.  Clifton and Ferreira have constrained their consideration to observers that are co-moving with the photon's source; that is, at any time $\LattTime$, the observer would be at the same radius as the source in their respective cells' Schwarzschild co-ordinates; this has been illustrated in \FigRef{fig:cell_crossing}.
To facilitate comparison with Clifton and Ferreira's results, we shall compute redshifts for the same set of observers.

Following Clifton and Ferreira's example, we shall also use the lattice universe's scale factor $\LattScale(\LattTime)$ at the moments of emission and observation to compute $z_{\scriptscriptstyle FLRW}$ for comparison; we thus re-express \eqref{z_FLRW} as
\begin{equation}
1 + z_{\scriptscriptstyle FLRW} = \frac{r_b(\LattTime_o)}{r_b(\LattTime_e)},
\end{equation}
where we have made use of the fact that $\LattScale(\LattTime) = \alpha \, r_b(\LattTime)$.

\section{\label{bc}Propagating photons across cell boundaries}
\begin{figure}[htbp]
\input{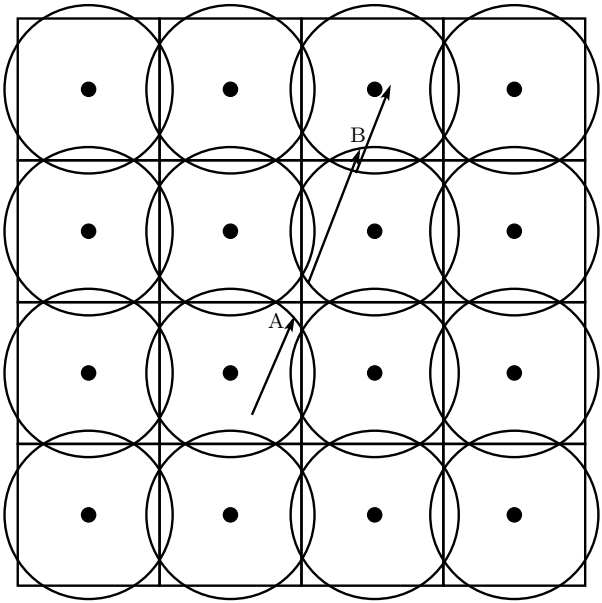}
\caption{\label{fig:lattice2}A flat lattice universe with cubic cells approximated by spheres.  The central masses are depicted by dots.  A photon will on average travel through an equal number of gap and overlap regions.  Because we are using spherical boundaries, at boundary crossing A, the photon will jump over a gap region, but at crossing B, it will pass through the overlap twice.  The cubic boundaries are tangent to each other everywhere along the boundaries while the spherical boundaries are nowhere tangent to each other except where they intersect; though even there, the tangency is only in the entire (3+1)-dimensional space-time, not in any specific 3-dimensional constant-$\LattTime$ hypersurface.  However in the LW approximation, we shall assume that the spherical boundaries are everywhere tangent to each other, both in the entire (3+1)-dimensional space-time and in each 3-dimensional constant-$\LattTime$ hypersurface; this is because we are treating the spherical boundaries as if they were approximately identical to the cubic ones.}
\end{figure}

In order to propagate photons through the lattice universe, we must first specify what boundary conditions trajectories must satisfy whenever they pass from one cell into the next.  Before discussing the boundary conditions though, we first note a difference between Clifton and Ferreira's choice of boundary geometry and ours.  Clifton and Ferreira converted from spherical cell boundaries back to polyhedral ones, deducing the polyhedral boundary velocity from the requirement that the spherical cell always have the same `Euclidean volume' as the polyhedral cell.  We shall instead continue to use spherical boundaries and propagate photons across boundaries in the manner illustrated in \FigRef{fig:cell_crossing}; that is, wherever a photon crosses, we always regard the point of crossing as a point of tangency between the two neighbouring cell boundaries.  We assume that the boundaries are tangent not just in (3+1)-dimensional space-time as a whole, but also in each 3-dimensional constant-$\LattTime$ hypersurface, which would be the case had we been using the true polyhedral lattice cells.  As Lindquist and Wheeler have argued, spherical boundaries should be a good approximation to the shape of polyhedral boundaries, with the approximation improving as the number of symmetries increases.  Therefore any errors due to this approximation would be small to begin with.  We further argue that although spherical cells may tile the lattice universe with gaps and overlaps, an arbitrary photon would on average travel through an equal number of gaps and overlaps such that the overall error approximately cancels out, as illustrated in \FigRef{fig:lattice2}.
Additionally, we note that in Clifton and Ferreira's polyhedral cells, the meshing of constant-$\LattTime$ hypersurfaces at cell boundaries is lost, with the hypersurfaces now meshing in an average manner only.  In contrast, the hypersurfaces in the original spherical cells, by construction, mesh exactly.  Finally, the idea of replacing polyhedral boundaries completely with a spherical approximation seems closer in spirit to the original idea of Wigner and Seitz.  Therefore according to our choice of boundary geometry, if a photon exits its current cell at a radius $r_1 = r_b$, we require it to enter the next cell at radius $r_2 = r_b$.  Because of cell 2's spherical symmetry, we are free to choose any $\theta$ and $\phi$ co-ordinate for the entry point.

We now present the conditions that photon trajectories must satisfy when propagated across boundaries.  These conditions are applied locally at any pair of exit and entry points.  Our conditions are founded upon Clifton and Ferreira's principle that any physical quantity should be independent of which cell's co-ordinate system an observer co-moving with the boundary may choose to use.  In the context of photon trajectories, we require that
\begin{enumerate}[label=\arabic{*}), topsep=2.5mm]
\item the photon frequency match across the boundary,
\begin{equation}
\Tensb{u}_1 \cdot \Tensb{k}_1 = \Tensb{u}_2 \cdot \Tensb{k}_2,
\label{bound_cond1}
\end{equation}

\item and the projection of the photon's 4-momentum onto the vector $\Tensb{n}^r$ orthogonal to the boundary match across the boundary,\footnote{Because of the normalisation condition, this second requirement is equivalent to requiring that the space-like projection of $\Tensb{k}$ onto the boundary be identical in both cells.}
\begin{equation}
\Tensb{n}^r_1 \cdot \Tensb{k}_1 = -\Tensb{n}^r_2 \cdot \Tensb{k}_2;
\label{bound_cond2}
\end{equation}
it can be shown that $(n^r)^a = (0, \sqrt{E_b}, 0, 0)$ for both the LW and CF co-ordinate systems.  There is a negative sign in the above equation because $\Tensb{n}^r$ always points radially out of its respective cell, whereas the radial direction of $\Tensb{k}$ would be out of one cell and into the next.
\end{enumerate}
These conditions along with the normalisation $\Tensb{k} \cdot \Tensb{k}=0$ are sufficient to deduce the components of $\Tensb{k}_2$ in terms of $\Tensb{u}_1$, $\Tensb{u}_2$, and $\Tensb{k}_1$.

The vectors $\Tensb{u}$ and $\Tensb{n}^r$ imply a decomposition of $\Tensb{k}$ into the form
\begin{equation}
\Tensb{k} = - (\Tensb{k} \cdot \Tensb{u})\Tensb{u} + (\Tensb{k} \cdot \Tensb{n}^r)\Tensb{n}^r + \Tensb{k}^\Omega,
\label{k_decomp}
\end{equation}
where it can be shown that $\Tensb{k}^\Omega \cdot \Tensb{u} = \Tensb{k}^\Omega \cdot \Tensb{n}^r = 0$.  We note that because of the cell's spherical symmetry, we can always choose polar co-ordinates such that $\Tensb{k}$ lies in the $\theta = \pi / 2$ plane, thereby allowing us to suppress the $\theta$ co-ordinate.  In this case, $\Tensb{k}^\Omega$ would take the form $(k^\Omega)^a = (0, 0, 0, k^\Omega)$.  The above conditions imply that $(k^\Omega_1)^2 = (k^\Omega_2)^2$, and by a suitable choice of the $\phi$ co-ordinates, we can always make $k^\Omega_1 = k^\Omega_2$.  We also note that there is no physical reason for the photon trajectory to refract when passing through a boundary, and therefore the $\theta = \pi / 2$ planes of the two cells should be aligned.

Using the above decomposition and the two boundary conditions, we can therefore express the photon's 4-momentum $\Tensb{k}_2$ in the new cell as
\begin{equation}
\Tensb{k}_2 = - (\Tensb{k}_1 \cdot \Tensb{u}_1)\Tensb{u}_2 - (\Tensb{k}_1 \cdot \Tensb{n}_1^r)\Tensb{n}_2^r + \Tensb{k}_1^\Omega.
\end{equation}

\section{\label{implem_sect}Numerical implementation of the LW model}
To numerically simulate photons propagating through the lattice universe, we have chosen to implement Williams and Ellis' Regge calculus scheme \citep{WilliamsEllis1} for Schwarzschild space-times.  We shall use their method to discretise the space-time of each Schwarzschild-cell into Regge calculus blocks, and then employ their geodesic-tracing method to propagate photons and test particles through a cell, though with some modifications.  In general, the discrete space-times of Regge calculus are expected to converge at second order in the lattice edge-lengths to the continuum space-times of general relativity \citep{BrewinGentle}.

\begin{figure*}[tb]
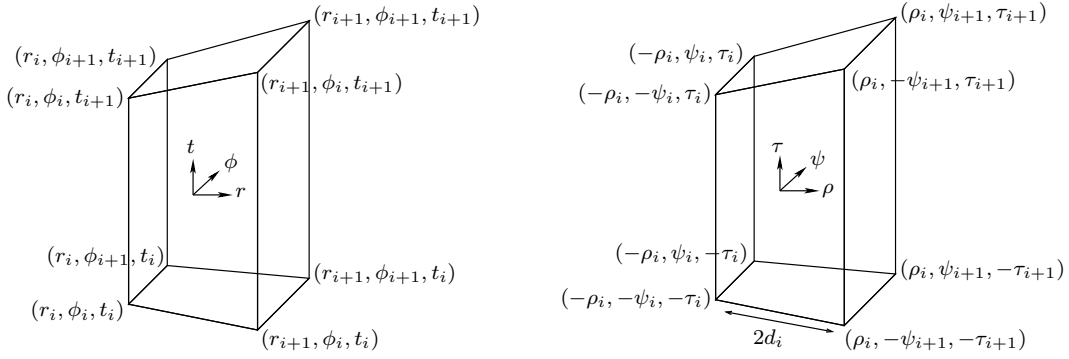

\hspace{-1cm}
\input{Regge_block1.pspdftex}
\hspace{5cm}
\input{Regge_block2.pspdftex}
\caption{\label{fig:RegBlock}On the left is an example of the original Schwarzschild block with one angular co-ordinate suppressed, and on the right is the Regge block to which it is mapped.  $(\tau, \rho, \psi)$ is a Minkowski co-ordinate system within the Regge block.}
\end{figure*}

Under Willliams and Ellis' scheme, Schwarzschild space-time is discretised as follows.  A grid is constructed in the Schwarzschild space-time such that along any particular gridline, only one Schwarzschild co-ordinate $(t, r, \theta, \phi)$ changes while the other three are held constant.  The lines intersect at constant $\Delta t$, $\Delta r$, $\Delta \theta$, or $\Delta \phi$ intervals, thus forming the edges of curved rectangular blocks.  Each of these blocks gets mapped to flat rectangular Regge blocks such that the straight edges of the Regge blocks have the same lengths as the curved edges of the original blocks; \Figuref{fig:RegBlock} illustrates an example of a curved Schwarzschild block with its corresponding Regge block.
The Schwarzschild co-ordinates of the original blocks' vertices $(t_i, r_j, \theta_k, \phi_l)$ are now taken over as labels for the Regge blocks' vertices.  As the Regge blocks are flat, the metric inside is simply the Minkowski metric.

Suppose a Regge block had vertices $(t_i, r_j, \theta_k, \phi_l)$ and $(t_{i+1}, r_{j+1}, \theta_{k+1}, \phi_{l+1})$; then we shall use the label $(t_i, r_j, \theta_k, \phi_l)$ to refer to this block as a whole.  The edge-lengths for this block are
\begin{align*}
d[(t_i, r_j, \theta_k, \phi_l), (t_{i+1}, r_j, \theta_k, \phi_l)] &= \sqrt{1 - \frac{2m}{r_j}} \Big (t_{i+1} - t_i \Big),\\
d[(t_i, r_j, \theta_k, \phi_l), (t_i, r_{j+1}, \theta_k, \phi_l)] &= \bigintss_{r_j}^{r_{j+1}} \mkern-18mu \frac{dr}{\sqrt{1 - \frac{2m}{r}}},\\
d[(t_i, r_j, \theta_k, \phi_l), (t_i, r_j, \theta_{k+1}, \phi_l)] &= r_j (\theta_{k+1} - \theta_k),\\
d[(t_i, r_j, \theta_k, \phi_l), (t_i, r_j, \theta_k, \phi_{l+1})] &= r_j \sin \theta_k (\phi_{l+1} - \phi_l).
\end{align*}

Since photons always move in planes, we can always suppress one of the angular co-ordinates by setting $\theta$ to be $\pi/2$.  We then set-up a Minkowski co-ordinate system $(\tau, \rho, \psi)$ in the block, with block vertices located at $(\pm\, \tau_i, -\rho_i, \pm\, \psi_i)$ and $(\pm\, \tau_{i+1}, \rho_i, \pm\, \psi_{i+1})$, where $\tau_i$, $\rho_i$, and $\psi_i$ are defined by\footnote{Note that an arbitrary index $j$ appears in the expressions for $\tau_i$ and $\psi_i$.  However, because both $(t_{j+1} - t_j)$ and $(\phi_{j+1} - \phi_j)$ are actually independent of $j$, it does not matter what value of $j$ is used to calculate $\tau_i$ and $\psi_i$.}
\begin{equation}
\begin{aligned}
2 \tau_i &= \displaystyle \sqrt{1 - \frac{2m}{r_i}}\, \Big (t_{j+1} - t_j \Big),\\
2 \psi_i &= \displaystyle  r_i (\phi_{j+1} - \phi_j),\\
\rho_i &= \displaystyle \left[d_i^2 + \frac{(\tau_{i+1} - \tau_i)^2}{4} - \frac{(\psi_{i+1} - \psi_i)^2}{4} \right]^{1/2},
\end{aligned}
\end{equation}
and where\footnote{In \citep{WilliamsEllis1}, there was a missing factor of $2m$ in front of the logarithm for this equation; this factor has been restored here.}
\begin{equation}
\begin{aligned}
2 d_i &= \bigintss_{r_i}^{r_{i+1}} \mkern-18mu \frac{dr}{\sqrt{1 - \frac{2m}{r}}}\\
&= \left\{r \sqrt{1 - \frac{2m}{r}} - 2m \ln \left[ \sqrt{\frac{r}{2m}} - \sqrt{\frac{r}{2m} - 1}\; \right] \right\}^{r_{i+1}}_{r_i}. \label{rad_dist}
\end{aligned}
\end{equation}

Williams and Ellis propagate geodesics through Regge blocks on the principle that geodesics should follow straight lines both within a block and on crossing from one block into the next.  There is an apparent refraction of the geodesic in crossing into a new block because the $(\tau, \rho, \psi)$ co-ordinate systems of the two blocks are not aligned; therefore the same tangent vector of a geodesic would be represented differently in different blocks' co-ordinate systems.  Williams and Ellis have demonstrated that their scheme successfully reproduces the orbits and redshifts of particles travelling in Schwarzschild space-time \citep{WilliamsEllis1, WilliamsEllis2}.

Referring to \FigRef{fig:RegBlock}, we now summarise the rules for propagating geo\-desics from one block into the next.  Suppose the particle is at position $(\tau_0, \rho_0, \psi_0)$ and travelling in direction $k^a = (k_1, k_2, k_3)$.  It exits the block at $(\tau^\prime, \rho^\prime, \psi^\prime) = (\tau_0, \rho_0, \psi_0) + \lambda\, \Tensb{k}$, where the value of $\lambda$ depends on the face exited.
\begin{enumerate}[label=\arabic{*}), topsep=2.5mm]
\item If the particle exits the block by the top face, then
\begin{equation}
\lambda = \frac{\tilde{\tau} - \tau_0 + \rho_0 \tanh \beta_i}{k_1 - k_2 \tanh \beta_i},
\end{equation}
where $\tilde{\tau} = (\tau_{i+1} + \tau_i)/2$ and $\tanh \beta_i = (\tau_{i+1} - \tau_i)/2\rho_i$.  In the new block, the particle's new trajectory is given by applying to $\Tensb{k}$ the matrix
\begin{equation}
\begin{pmatrix}
\cosh 2\beta_i & - \sinh 2\beta_i & 0 \\
- \sinh 2\beta_i & \cosh 2\beta_i & 0 \\
0 & 0 & 1
\end{pmatrix},
\end{equation}
and the particle's starting position is given by $(-\tau^\prime, \rho^\prime, \psi^\prime)$.

\item If the particle exits by the back face, then
\begin{equation}
\lambda = \frac{\tilde{\psi} - \psi_0 + \rho_0 \tan \alpha_i}{k_3 - k_2 \tan \alpha_i},
\end{equation}
where $\tilde{\psi} = (\psi_{i+1} + \psi_i)/2$ and $\tan \alpha_i = (\psi_{i+1} - \psi_i)/2\rho_i$.  In the new block, the particle's trajectory $\Tensb{k}$ gets transformed by
\begin{equation}
\begin{pmatrix}
1 & 0 & 0 \\
0 & \cos 2\alpha_i & \sin 2\alpha_i \\
0 & - \sin 2\alpha_i & \cos 2\alpha_i
\end{pmatrix},
\label{rotn}
\end{equation}
and the particle's starting position is $(\tau^\prime, \rho^\prime, -\psi^\prime)$.

\item If the particle exits by the front face, then\footnote{In \citep{WilliamsEllis1}, there were a few sign errors for this equation, which have been corrected here.}
\begin{equation}
\lambda = \frac{-\tilde{\psi} - \psi_0 - \rho_0 \tan \alpha_i}{k_3 + k_2 \tan \alpha_i},
\end{equation}
where $\tilde{\psi}$ and $\tan \alpha_i$ are the same as above.  In the new block, the particle's trajectory $\Tensb{k}$ gets transformed by the same rotation matrix as \eqref{rotn} but with angle $-\alpha_i$ instead.  The particle's starting position in the new block is again $(\tau^\prime, \rho^\prime, -\psi^\prime)$, since $\psi^\prime < 0$ at the front face of the original block.

\item If the particle exits by the right/left face, then
\begin{equation}
\lambda = \frac{\pm \rho_i - \rho_0}{k_2},
\end{equation}
and $\rho^\prime$ is simply $\rho^\prime = \pm\rho_i$.  There is no refraction as the particle enters the next block.  Its new starting position is $(\tau^\prime, -\rho_{i+1}, \psi^\prime)$ if entering the block to the right and $(\tau^\prime, \rho_{i-1}, \psi^\prime)$ if entering the block to the left.
\end{enumerate}

However from numerical simulations, we have discovered an empirical relation for the tangent 4-vectors of radially out-going time-like geodesics.  This has led us to introduce a correction to the above propagation rules.  We have found that as a function of $r_i$, these 4-vectors obey
\begin{equation}
\begin{split}
u^a_{\scriptscriptstyle Regge} &= \left(\dot{\tau}, \dot{\rho}, \dot{\psi}\right) \\
&= \left(\left(\frac{1 - \frac{2m}{\tilde{r}_{max}}}{1 - \frac{2m}{r_i}}\right)^{1/2},\; \left(\frac{\frac{2m}{r_i} - \frac{2m}{\tilde{r}_{max}}}{1 - \frac{2m}{r_i}}\right)^{1/2},\; 0 \right),
\end{split}
\label{Reg_bound_vec}
\end{equation}
where $\tilde{r}_{max}$ is a constant of motion.  Our numerical results supporting this have been presented in \appendref{BV}.

This relation can be derived analytically by comparing $\Tensb{u}_{\scriptscriptstyle Regge}$ with its counterpart 4-vector in continuum Schwarzschild space-time.  The tangent 4-vector of a radially out-going time-like geodesic in Schwarzschild space-time is
\begin{equation}
\begin{split}
u^a_{\scriptscriptstyle Schwarz} &= \left(\dot{t}, \dot{r}, \dot{\Omega}\right) \\
&= \left(\frac{\sqrt{1 - \frac{2m}{r_{max}}}}{1 - \frac{2m}{r}},\; \left(\frac{2m}{r} - \frac{2m}{r_{max}}\right)^{1/2},\; 0 \right),
\end{split}
\label{cont_bound_vec}
\end{equation}
where $r_{max}$ is also a constant of motion and is related to $E$ in \eqref{rad_geod} by $r_{max} = 2m/(1-E)$.  For in-going geodesics, the radial components of both $\Tensb{u}_{\scriptscriptstyle Regge}$ and $\Tensb{u}_{\scriptscriptstyle Schwarz}$ would have an additional negative sign.  If we take the scalar product of $\Tensb{u}_{\scriptscriptstyle Regge}$ and $\Tensb{u}_{\scriptscriptstyle Schwarz}$ with the unit vector in the time direction, that is, with $\hat{\tau}\,^a = (1,0,0)$ for Regge space-time and $\hat{t}\,^a = \left(\left(1-2m / r \right)^{-1/2},0,0\right)$ for continuum Schwarzschild space-time, we have that
\begin{align*}
\Tensb{\hat{\tau}} \cdot \Tensb{u}_{\scriptscriptstyle Regge} &= \left(\frac{1 - \frac{2m}{\tilde{r}_{max}}}{1 - \frac{2m}{r_i}}\right)^{1/2},\\
\intertext{and that}
\Tensb{\hat{t}} \cdot \Tensb{u}_{\scriptscriptstyle Schwarz} &= \left(\frac{1 - \frac{2m}{r_{max}}}{1 - \frac{2m}{r}}\right)^{1/2}.
\end{align*}
If we identify $r_{max}$ with $\tilde{r}_{max}$, then these two expressions are identical whenever $r = r_i$ in the continuum space-time.  Similarly, if we take the scalar product with the unit vector in the radial direction, that is, with $\hat{\rho}\,^a = (0, 1, 0)$ for Regge space-time and $\tilde{r}\,^a = \left(0, \left(1-2m / r \right)^{1/2}, 0\right)$ for continuum space-time, we have that
\begin{align*}
\Tensb{\hat{\rho}}  \cdot \Tensb{u}_{\scriptscriptstyle Regge} &= -\left(\frac{\frac{2m}{r_i} - \frac{2m}{\tilde{r}_{max}}}{1 - \frac{2m}{r_i}}\right)^{1/2},\\
\intertext{and that}
\Tensb{\hat{r}} \cdot \Tensb{u}_{\scriptscriptstyle Schwarz} &= -\left(\frac{\frac{2m}{r_i} - \frac{2m}{r_{max}}}{1 - \frac{2m}{r}}\right)^{1/2}.
\end{align*}
Again if we identify $r_{max}$ with $\tilde{r}_{max}$, then the two expressions are also identical whenever $r = r_i$ in the continuum space-time.  Therefore provided $\tilde{r}_{max} = r_{max}$, we see that \eqref{Reg_bound_vec} is indeed the Regge analogue of \eqref{cont_bound_vec}.  Furthermore, the choice of $r_{max}$ and $\tilde{r}_{max}$ determines whether the resulting particle orbit will be closed or open in the corresponding space-time.  If $2m/r_{max} > 0$, then the orbit in the continuum space-time will be closed and $r_{max}$ would indeed be the maximum radius of the orbit.  Similarly, we found that if $2m/\tilde{r}_{max} > 0$, then the orbit in Regge space-time will also be closed, and the maximum radius would be $\tilde{r}_{max}$.  If $2m/r_{max} = 0$, then the geodesic will just reach spatial infinity in the continuum space-time, and \eqref{cont_bound_vec} gives the particle's escape velocity as a function of its radial position $r > 2m$.  Similarly, we show in \appendref{BV} that when $2m/\tilde{r}_{max} = 0$, then \eqref{Reg_bound_vec} gives the escape velocity for a test particle in the Regge space-time.  Finally if $2m/r_{max} < 0$, then the orbit in continuum space-time will be open.  And if $2m/\tilde{r}_{max} < 0$, then the orbit in Regge space-time will also be open, as we show in \appendref{BV}.  We shall henceforth make the identification of $\tilde{r}_{max} = r_{max}$.

Inspired by this, we can generalise the expression for $\Tensb{u}_{\scriptscriptstyle Regge}$ to any geodesic in Regge Schwarzschild space-time.  The Lagrangian for particles moving in continuum Schwarzschild space-time can be written as
$$
L = - \left(1 - \frac{2 m}{r} \right) \dot{t}^2 + \frac{\dot{r}^2}{\left(1 - \frac{2 m}{r} \right)} + r^2 \dot{\Omega}^2,
$$
where the dot denotes differentiation with respect to some parameter $\lambda$.  Since $0 = \partial L / \partial t$, we have a constant of motion $E$, which we define by the relation
\begin{equation}
E := \left(1 - \frac{2 m}{r} \right)^2 \dot{t}^2.
\end{equation}
Similarly, since $0 = \partial L / \partial \Omega$, we have another constant of motion $J$ defined by
\begin{equation}
J := r^2 \dot{\Omega}.
\end{equation}
These two constants correspond to the square of the particle's energy per unit mass at radial infinity and to the angular momentum.  For radial time-like geodesics, $E$ here is the same as $E$ in \eqref{rad_geod}.  In terms of these constants, the tangent 4-vector $\Tensb{v}_{Schwarz}$ to the particle's geodesic can be expressed as
\begin{equation}
v^a_{\scriptscriptstyle Schwarz} = \left( \dot{t},\, \dot{r},\, \dot{\Omega} \right) = \left( \frac{\sqrt{E}}{\left(1 - \frac{2 m}{r} \right)},\, \dot{r},\, \frac{J}{r^2} \right),
\end{equation}
which is clearly a function of $r$ alone, since $\dot{r}$ can be deduced from $\dot{t}$ and $\dot{\Omega}$ through the normalisation of $\Tensb{v}_{\scriptscriptstyle Schwarz}$.

Let $\Tensb{v}_{\scriptscriptstyle Regge}$ denote the Regge analogue of $\Tensb{v}_{\scriptscriptstyle Schwarz}$.  If we assume that $\Tensb{v}_{\scriptscriptstyle Regge}$ and $\Tensb{v}_{\scriptscriptstyle Schwarz}$ are related in a manner analogous to how $\Tensb{u}_{\scriptscriptstyle Regge}$ and $\Tensb{u}_{\scriptscriptstyle Schwarz}$ are related, then we can deduce the components of $\Tensb{v}_{\scriptscriptstyle Regge}$ from $\Tensb{v}_{\scriptscriptstyle Schwarz}$.  Specifically, by equating $\Tensb{\hat{\tau}} \cdot \Tensb{v}_{\scriptscriptstyle Regge}$ to $\Tensb{\hat{t}} \cdot \Tensb{u}_{\scriptscriptstyle Schwarz}$, we deduce $\dot{\tau}$ to be
$$
\dot{\tau} = \frac{\sqrt{E}}{\left(1 - \frac{2 m}{r} \right)^{1/2}}.
$$
Similarly by equating $\Tensb{\hat{\psi}} \cdot \Tensb{v}_{\scriptscriptstyle Regge}$ to $\Tensb{\hat{\Omega}} \cdot \Tensb{u}_{\scriptscriptstyle Schwarz}$, where $\hat{\psi}^a = (0,\, 0,\, 1)$ and $\hat{\Omega}^a = (0,\, 0,\, r^{-1})$, we deduce $\dot{\psi}$ to be
$$
\dot{\psi} = \frac{J}{r}.
$$
Thus the components of $\Tensb{v}_{\scriptscriptstyle Regge}$ are given by
\begin{equation}
v^a_{\scriptscriptstyle Regge} = \left(\frac{\sqrt{E}}{\left(1 - \frac{2 m}{r} \right)^{1/2}},\, \dot{\rho},\, \frac{J}{r} \right),
\label{GenRegSchwarzVel}
\end{equation}
with $\dot{\rho}$ deducible from the normalisation of $\Tensb{v}_{\scriptscriptstyle Regge}$.  Finally, it is easy to verify that $\Tensb{\hat{\rho}} \cdot \Tensb{v}_{\scriptscriptstyle Regge}$ and $\Tensb{\hat{r}} \cdot \Tensb{v}_{\scriptscriptstyle Schwarz}$ are then consistent.

Our generalised expression for tangent 4-vectors can be used for both null and time-like geodesics following both radial and non-radial trajectories.  For time-like radial geodesics, where $J=0$, it can be shown that $\Tensb{v}_{\scriptscriptstyle Regge}$ reduces to $\Tensb{u}_{\scriptscriptstyle Regge}$.  As we wish to simulate geodesics in continuum rather than Regge Schwarzschild-cells, we have introduced a correction to the Williams-Ellis scheme: whenever a geodesic crosses a right or left face, its tangent 4-vector is changed back to \eqref{GenRegSchwarzVel}.

Having established the modified Williams-Ellis scheme, we now discuss its application to the Schwarzschild-cell.  The boundary is simulated by propagating a test particle that is co-moving with the boundary.  Whenever the photon reaches the same radial and time co-ordinates as the boundary particle, then at that point the photon exits one cell and enters the next, and we need to apply conditions \eqref{bound_cond1} and \eqref{bound_cond2} to determine the photon's new trajectory.  However vector components will differ from previously as we are now working in Regge space-time.

Let $u^a = (u^\tau, u^\rho, 0)$ be the boundary particle velocity and $k^a = (k^\tau, k^\rho, k^\psi)$ be the photon's 4-momentum at the boundary.  We shall use subscripts $1$ and $2$ to indicate the cell being exited and the cell being entered, respectively.  As in \eqref{k_decomp}, we decompose the Regge vector $\Tensb{k}$ into
\begin{equation}
\Tensb{k} = \nu\Tensb{u} + \Tensb{n}^\rho + \Tensb{n}^\psi,
\label{Regge_k_decomp}
\end{equation}
where $\nu = -\Tensb{u} \cdot \Tensb{k}$, $(n^\psi)^a = (0, 0, n^\psi)$, and $\Tensb{n}^\rho$ satisfies $\Tensb{n}^\rho \cdot \Tensb{u} = \Tensb{n}^\rho \cdot \Tensb{n}^\psi = 0$.  If $\Tensb{\hat{n}}^\rho$ is the normalised vector of $\Tensb{n}^\rho$, that is $\Tensb{n}^\rho = n\, \Tensb{\hat{n}}^\rho$, then the components of $\Tensb{\hat{n}}^\rho$ are
$$
(\hat{n}^\rho)^a = (u^\rho, u^\tau, 0),
$$
as this satisfies all orthogonality relations required of $\Tensb{n}^\rho$.  We then deduce $n$ to be
\begin{equation}
n = \Tensb{\hat{n}}^\rho \cdot \Tensb{k} = \left(u^\tau k^\rho - u^\rho k^\tau \right).
\label{n_val}
\end{equation}
Condition \eqref{bound_cond1} implies that
\begin{equation}
\nu_1 = \nu_2.
\label{Reg_bc1}
\end{equation}
Condition \eqref{bound_cond2} in this context is equivalent to
\begin{align}
\Tensb{k}_1 \cdot \Tensb{\hat{n}}^\rho_1 &= -\Tensb{k}_2 \cdot \Tensb{\hat{n}}^\rho_2,\nonumber\\
\shortintertext{which implies that}
n_1 &= -n_2.
\label{Reg_bc2}
\end{align}
As at the end of \secref{bc}, these conditions imply that $(n_1^\psi)^2 = (n_2^\psi)^2$, and we again have freedom to choose polar co-ordinates such that 
\begin{equation}
n_1^\psi = n_2^\psi,
\label{Reg_bc3}
\end{equation}
which we shall henceforth assume to be the case.  Using relation \eqref{n_val}, conditions \eqref{Reg_bc1} and \eqref{Reg_bc2}, relation \eqref{Reg_bc3}, and decomposition \eqref{Regge_k_decomp}, we can express the components of $\Tensb{k}_2$ as
\begin{equation}
k_2^a = (\nu u_2^\tau + (u_1^\rho k_1^\tau - u_1^\tau k_1^\rho)u_2^\rho, \; \nu u_2^\rho + (u_1^\rho k_1^\tau - u_1^\tau k_1^\rho)u_2^\tau,\; k_1^\psi).
\end{equation}

We note that constant $E$ in \eqref{GenRegSchwarzVel} will differ between $\Tensb{k}_1$ and $\Tensb{k}_2$, but $J$ will remain the same since $k^\psi = J / r$ is identical on both sides of the boundary.

\section[Redshift results]{Redshifts from lattice universe simulations}
\begin{figure*}[p]
{\fontsize{8pt}{9.6pt}\input{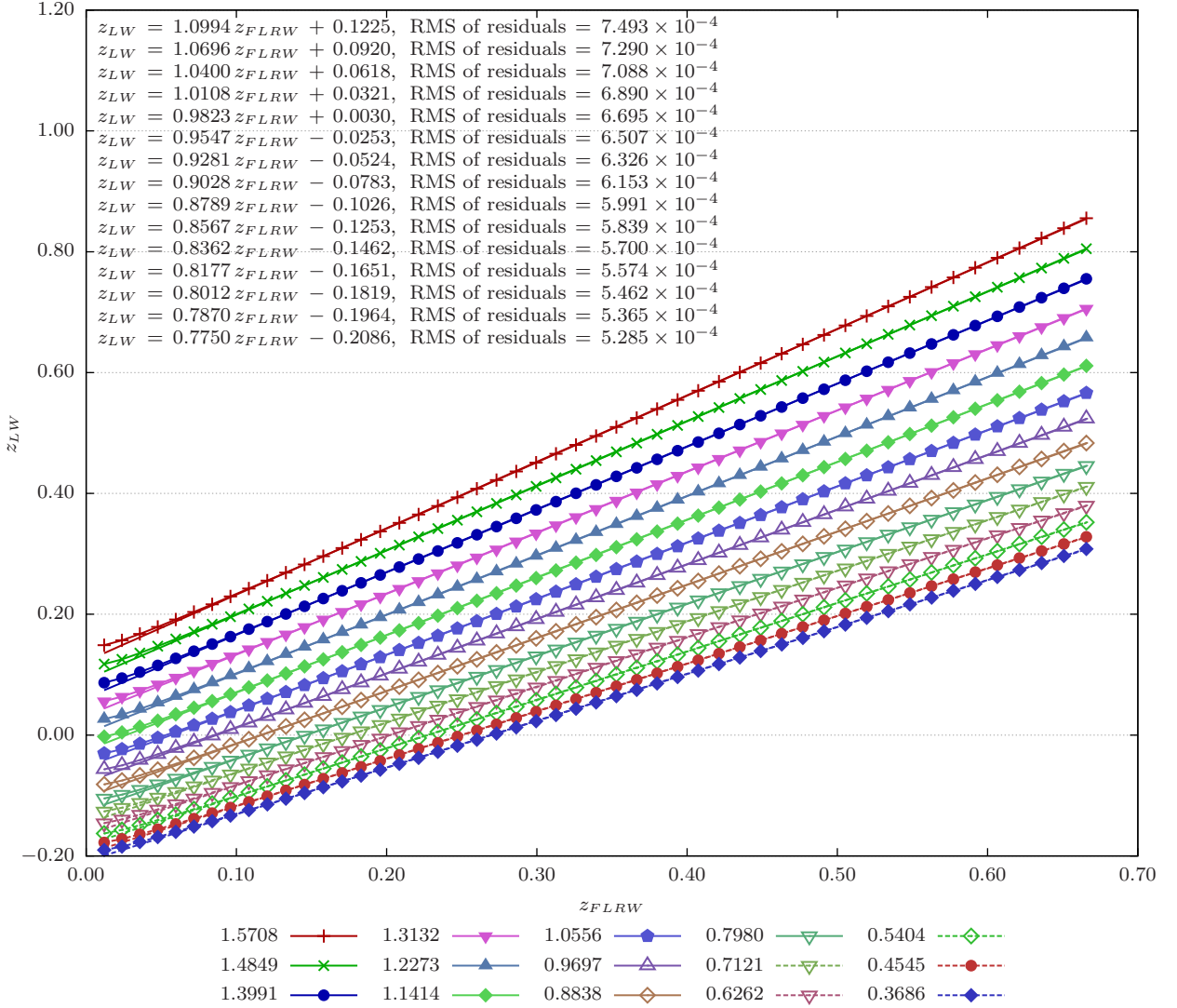}}
\caption{\label{fig:complete_3E4}A plot of $z_{\scriptscriptstyle LW}$ against $z_{\scriptscriptstyle FLRW}$ for 15 trajectories in a flat universe with an initial cell size of $r_{b_0} = 3\times 10^4\, R$.  The initial angle $\theta_n$ of the trajectory is given in the legend.  Each trajectory was traced across 50 cells.  A linear regression has been performed for each graph, with the first five data points excluded so as to focus only on the linear regime.  The regression equations and corresponding root-mean-squares of the residuals are listed above in order of decreasing $\theta_n$.}
\end{figure*}
We have simulated the propagation of photons through multiple Schwarzschild-cells for closed, flat, and open LW universes.  Each time, a photon was propagated outwards in various directions from an initial radius of $10\, R$, $R$ being the Schwarzschild radius.  The initial direction of travel was given, in terms of block co-ordinates, by
$$
k^a = (1,\, \cos \theta_n,\, \sin \theta_n),
$$
with the range of $\theta_n$ starting from the purely tangential direction of $\theta_0 = \pi / 2$ and decreasing until the direction was almost completely radial.  Both LW and FLRW redshift factors were computed whenever the photon, while travelling outwards again, passed an observer co-moving with the source.

We can in principle re-scale the initial $\Tensb{k}$ by a constant factor $\lambda$, but this should not affect the redshifts: such a re-scaling would not alter the corresponding null geodesic, so the space-time points at which the observer and photon meet would remain unchanged; at these points, the only change in the photon's frequencies would be a re-scaling by $\lambda$, but this factor would then cancel out of the redshifts.  We verified this invariance of the redshifts by simulating a photon with different scalings of $\Tensb{k}$, and the results, not shown, were indeed invariant for at least 11 significant figures; any discrepancies can be attributed to numerical error.

All length-scales in our simulations have been specified in terms of $R$.  By simply re-scaling $R$ in one set of results, we can readily obtain the results for an equivalent simulation where the only difference is the magnitude of $R$.  In particular, because redshifts are dimensionless quantities, they would not depend on the choice of $R$, so we therefore made the arbitrary choice of setting $R=1$ for all our simulations.

\begin{figure*}[htb]
{\fontsize{8pt}{9.6pt}\input{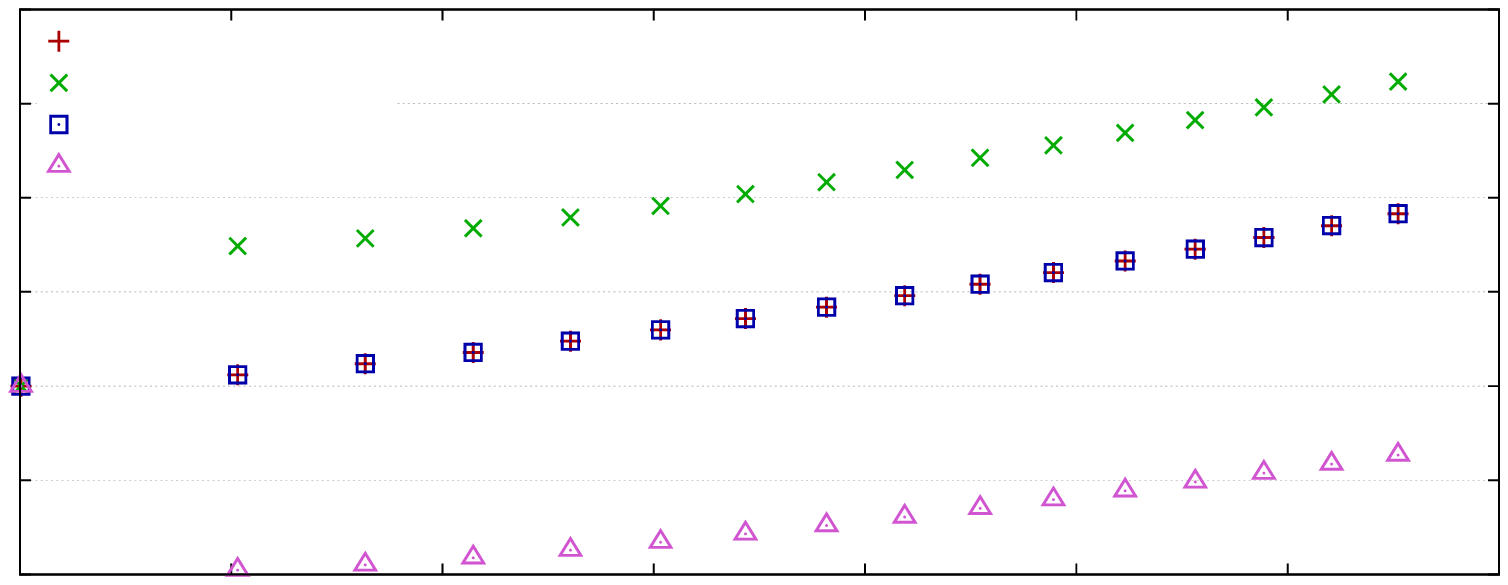}}
\caption{\label{fig:z_vs_r_obs}A plot of redshifts $z$ against the radius of observation $r_{obs}$ for photons travelling in the flat universe with an initial cell size of $r_{b_0} = 3\times 10^4\, R$.  The photons were travelling initially in the directions of $\theta_0 = \pi / 2$ and $\theta_{29} = 311 \pi / 3000$.  Plots for both LW and FLRW redshifts are shown.  All four series start at the same data point, and there is a clear jump in the LW graphs between the first and second data points.}
\end{figure*}
\begin{figure}[htb]
{\fontsize{8pt}{9.6pt}\input{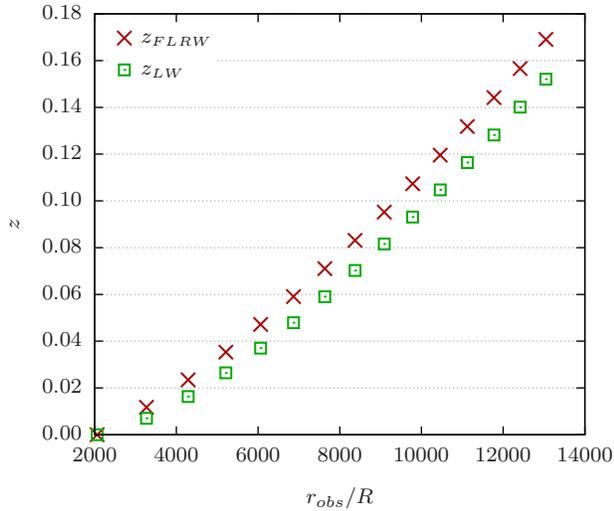}}
\caption{\label{fig:z_vs_r_obs-rescaled}A plot of redshifts $z$ against the radius of observation $r_{obs}$ for a photon starting instead with a position and trajectory corresponding to the second data point of the $\theta_0 = \pi / 2$ trajectory in \FigRef{fig:z_vs_r_obs}.  Plots for both LW and FLRW redshifts are shown.  This time, both graphs extend smoothly out from the zero-redshift data point.}
\end{figure}
\begin{figure*}[p]
{\fontsize{8pt}{9.6pt}\input{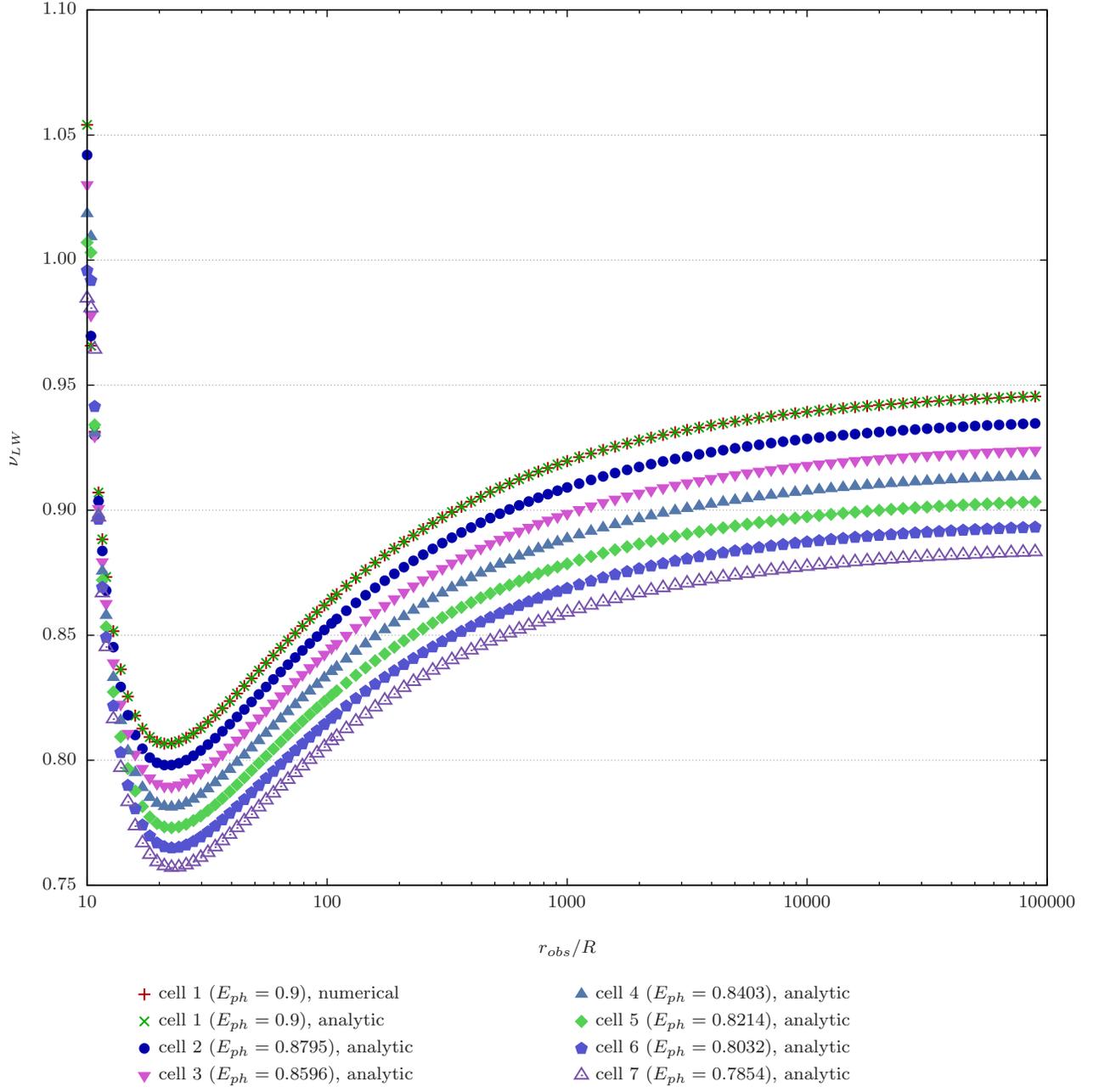}}
\caption{\label{fig:single-cell-freq-1_5707}A plot of photon frequencies $\nu_{\scriptscriptstyle LW}$ against the radius of observation $r_{obs}$ for a photon travelling initially in the direction of $\theta_0 = \pi / 2$; the initial cell size is $r_{b_0} = 3\times 10^4\, R$.  Each graph represents the photon's frequencies within a single cell; these are the frequencies that would be seen by a co-moving observer if the photon intercepted the observer at $r_{obs}$.  The analytic frequencies are given by \eqref{analytic-single-cell-freq}, and graphs for the first seven cells traversed are shown.  Frequencies were also computed numerically by simulating the propagation of a photon across the first cell only; the corresponding graph is shown and completely overlaps with its analytic counterpart.}
\end{figure*}
\begin{figure*}[p]
{\fontsize{8pt}{9.6pt}\input{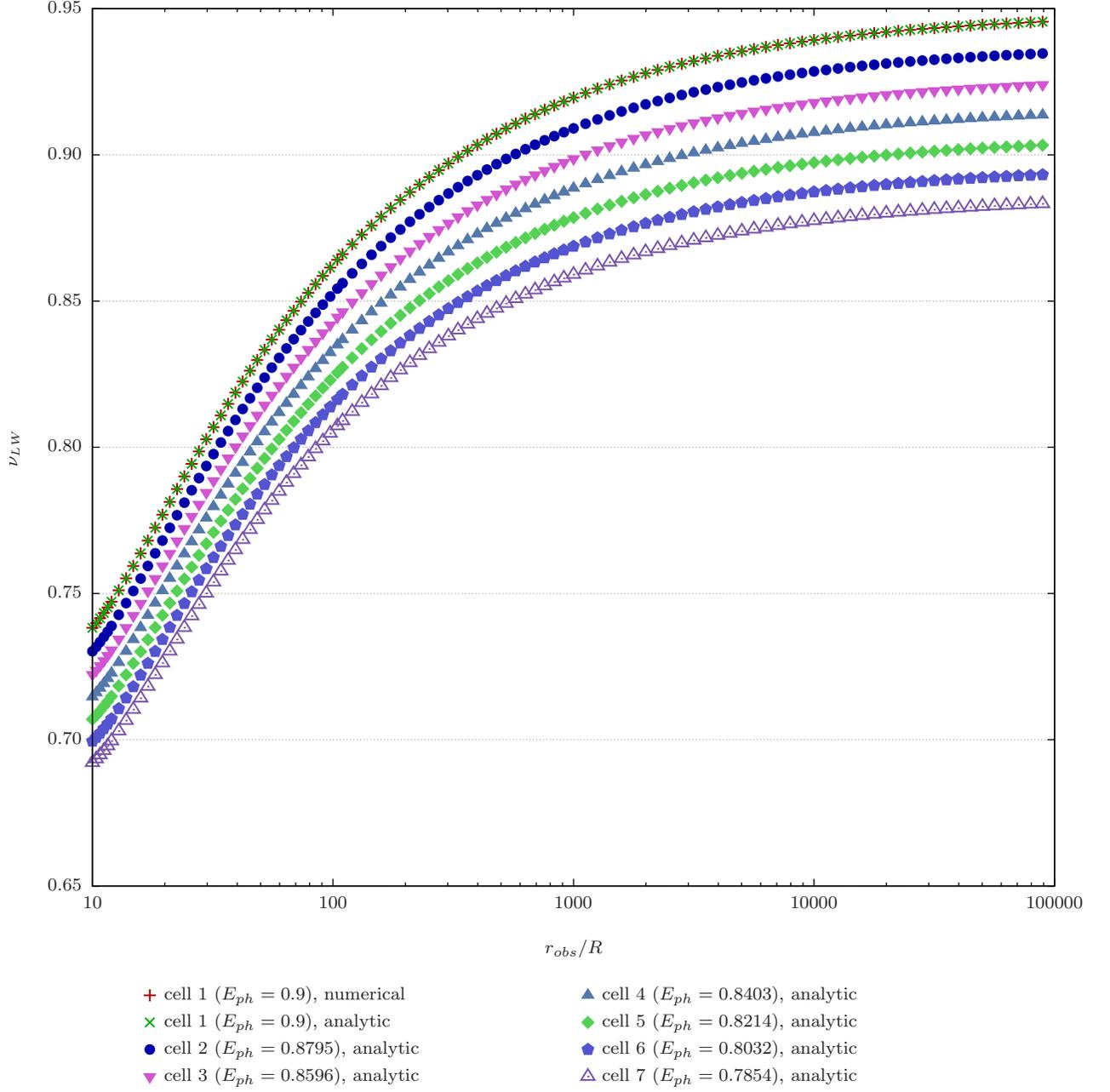}}
\caption{\label{fig:single-cell-freq-0_3257}The equivalent plot to \FigRef{fig:single-cell-freq-1_5707} but for a photon travelling initially in the direction of $\theta_{29} = 311 \pi / 3000$.  Once again, the numerical and analytic graphs for cell 1 overlap completely.}
\end{figure*}
\begin{figure*}[tb]
{\fontsize{8pt}{9.6pt}\input{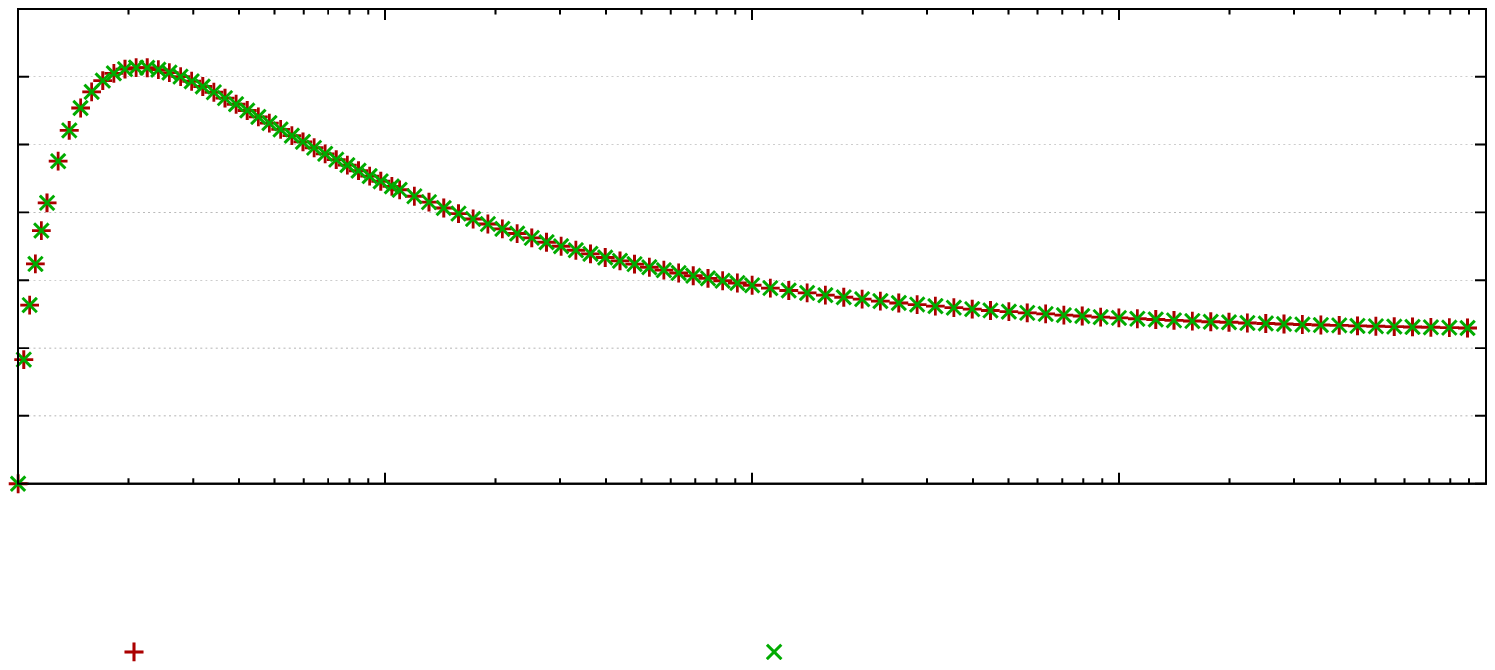}}
\caption{\label{fig:single-cell-redshifts-1_5707}The equivalent plot to \FigRef{fig:single-cell-freq-1_5707} but showing the redshifts $z_{\scriptscriptstyle LW}$ of cell 1 only against the radius of observation $r_{obs}$.  The two graphs overlap each other completely.}
\end{figure*}

We have chosen the dimensions of our Regge block as follows.  The angular length $2\psi_i$ was chosen so that $\Delta \phi = 2\pi / (3\times 10^{10})$.  For the closed universe, the radial length $2 d_i$ was chosen to represent a fixed interval of $\Delta r = 10^{-7}\, R$.  For the flat and open universes, it was instead chosen to lengthen for blocks further away from the cell centre.  This lengthening was implemented so as to increase computation speed with only a marginal expense to the accuracy: the underlying continuum Schwarzschild space-time becomes flatter as one moves further away from the centre, so a flat Regge block would approximate the region more accurately.  Our exact method for determining the block's length has been described in \appendref{RadLen}.  Although $\Delta r$ is no longer constant, it is constrained to be an integral multiple of a minimum interval $\Delta r_0$, a parameter we can freely specify, and the Regge grid is fixed so that its first set of blocks covers the region between $R + \Delta r_0$ and $R + 2\Delta r_0$.\footnote{As an informal check on the accuracy, we simulated photons propagating across 100 cells in the flat, $E=1$ universe using both fixed and increasing block-lengths.  The specific simulation parameters were an initial Schwarzschild-cell radius of $r_{b_0} = 3\times 10^4\, R$ with an initial photon radius of $10R$; the Regge blocks' non-radial dimensions were $\Delta t = 10\, \Delta r$ and $\Delta \phi = 2\pi / (3\times 10^{7})$; the radial dimension $\Delta r$ was fixed at $10^{-4} R$ for the fixed-length simulation and was set to be at least that length for the increasing-length simulation.  The two simulations' redshift results, not shown, agreed for at least five significant figures, even at the largest radii; however the fixed block-length simulation required about 6.5 hours of computation time, while the increasing block-length simulation required only 20 minutes.}  As we shall see below, $\Delta r_0$ was chosen according to which universe was being simulated.  Finally, the temporal interval was chosen to be $\Delta t = 10\, \Delta r_0$ for the flat and open universes and simply $\Delta t = 10\, \Delta r$ for the closed universe.

In this section we shall present the redshift results of our simulations.  We begin with the flat universe, for which $E=1$.  To investigate the effect of the initial cell size $r_{b_0}$ on the redshifts, we have simulated flat universes for a range of initial sizes from $r_{b_0} = 3\times 10^4\, R$ to $r_{b_0} = 10^8\, R$.  Following the example of Clifton and Ferreira, we have chosen the largest initial size to approximate cells with Milky Way-like masses at their centres, as this is thought to best represent our actual universe.  Depending on the initial cell size, $\Delta r_0$ ranged from $\Delta r_0=10^{-5}\, R$ for $r_{b_0} = 3\times 10^4\, R$ to $\Delta r_0=10^{-3}\, R$ for $r_{b_0} = 10^8\, R$.

\Figuref{fig:complete_3E4} plots $z_{\scriptscriptstyle LW}$ against $z_{\scriptscriptstyle FLRW}$ for the universe where $r_{b_0} = 3\times 10^4\, R$.  Each trajectory was traced across 50 cells; only results for 15 trajectories are shown, although we simulated trajectories for 30 angles $\theta_n$ ranging from $\theta_0 = \pi / 2$ to $\theta_{29} = 311 \pi / 3000$ in decrements of $41 \pi / 3000$.  Apart from a brief curve at the start, each graph clearly demonstrates a strong linear relationship between $z_{\scriptscriptstyle LW}$ and $z_{\scriptscriptstyle FLRW}$, and the gradient of the line is different for each angle.  There are several differences, however, between our results and those of Clifton and Ferreira.  Clifton and Ferreira's graphs showed more initial scatter, which varied depending on the trajectory's initial angle, but their graphs would eventually converge upon a common mean graph as the photon passed through an increasing number of cells.  Our graphs do not display such scatter nor any common mean.  However, their common mean agreed rather closely with FLRW redshifts, as it was given by $1 + z_{\scriptscriptstyle LW} \approx (1 + z_{\scriptscriptstyle FLRW})^{0.98}$ \citep{CFO}; thus in their model, $z_{\scriptscriptstyle LW}$ and $z_{\scriptscriptstyle FLRW}$ are almost completely linear.  Our graphs, apart from the initial curve, are completely linear, but the gradients depend on the trajectory followed.

However Clifton and Ferreira's relation has the desirable feature of passing through the origin, since there must be zero redshift at the start of any trajectory; that is, both $z_{\scriptscriptstyle LW}$ and $z_{\scriptscriptstyle FLRW}$ must be zero.  All our graphs instead show a jump between the origin and the next data point, and this was true of our simulations of all other universes as well.  When we plotted the redshifts against the radius of observation for the $r_{b_0} = 3\times 10^4\, R$ flat universe, as shown in \FigRef{fig:z_vs_r_obs}, we found that the $z_{\scriptscriptstyle FLRW}$ graph would extend naturally outwards from the zero-redshift point at the starting radius of $10\, R$, but that the $z_{\scriptscriptstyle LW}$ graph would jump suddenly from the zero-redshift point to the next data point.  Suppose we ignored our current zero-redshift data point and assumed the photon actually began its trajectory at the next data point; then when we re-calculated all subsequent redshifts based on our new initial frequency for the photon, we found that the resulting $z_{\scriptscriptstyle LW}$ graph progresses naturally from the new zero-redshift point to the next data point without any jumps, as shown in \FigRef{fig:z_vs_r_obs-rescaled}.\footnote{We did check this result by simulating a photon starting at the same radius and direction as that of the original photon when it generated the second data point; the resulting graph was indeed identical to that of \FigRef{fig:z_vs_r_obs-rescaled}.}

To investigate this initial jump further, we note that in universes using CF co-ordinates, the frequency $\nu_{\scriptscriptstyle LW}$ measured by a co-moving observer can be expressed as a function of the radius at which the measurement is made, that is, the radius $r_{obs}$ at which the photon and observer intercept.  This implies the redshifts $z_{\scriptscriptstyle LW}$ can be expressed as a function of $r_{obs}$.  Recall that the 4-velocity of a co-moving observer is given by \eqref{GenRegSchwarzVel} in Regge Schwarzschild space-time, with $E=E_b$ and $J=0$; this vector is a function of the observer's radial position alone.  The 4-momentum of the photon is given by the same relation as well but with $E=E_{ph}$ and $J$ an arbitrary constant; as long as the photon does not cross into the next cell, $E_{ph}$ will be constant, so this vector is also a function of the photon's radial position alone.  When the observer and photon intercept, they will have the same radial position, that is, the same value for $r$, and we have denoted above this common $r$ by $r_{obs}$.  By taking the scalar product of these two vectors, we obtain the measured frequency as a function depending only on the value of $r_{obs}$,
\begin{equation}
\begin{split}
\nu_{\scriptscriptstyle LW} ={}& \frac{\sqrt{E_b\, E_{ph}}}{1-\frac{2m}{r_{obs}}} \\
& {}- \left[\left(\frac{E_b}{1-\frac{2m}{r_{obs}}} - 1 \right) \left(\frac{E_{ph}}{1-\frac{2m}{r_{obs}}} - \frac{J^2}{r^2_{obs}} \right) \right]^{1/2}.
\end{split}
\label{analytic-single-cell-freq}
\end{equation}

We note that if the photon were to cross into the next cell, this relation would still hold, but from the boundary conditions, $E_{ph}$ only would change.\footnote{Although we have an analytic function for $z_{\scriptscriptstyle LW}$ based on $r_{obs}$, we must rely on the simulation to tell us where $r_{obs}$ is, that is, where the photon intercepts the observer.  And although we can also determine $E_{ph}$ analytically from the boundary condition, to do this, we still need to know the radius at which the photon and cell boundary intercept, and we must rely on the simulation to tell us this radius as well.}  In \FigRef{fig:single-cell-freq-1_5707} and \FigRef{fig:single-cell-freq-0_3257}, we have plotted this function for photons travelling at initial angles of $\pi / 2$ and $311 \pi / 3000$, respectively, in the $r_{b_0} = 3\times 10^4\, R$ flat universe.  We have included plots for different $E_{ph}$ corresponding to the first few cells traversed.  We have also simulated a photon's propagation across the first cell for both trajectories, and for a selection of radii, we have computed the frequencies that a co-moving observer would measure if at those radii.  These numerical results are also included in the figures, and we see that they agree very closely with their analytic counterparts.  In \FigRef{fig:single-cell-redshifts-1_5707}, we show redshifts $z_{\scriptscriptstyle LW}$ instead against $r_{obs}$ for the $\theta_0= \pi / 2$ photon travelling through cell 1 only.

\begin{figure*}[p]
\subfloat[\hspace{-1.2cm}\label{fig:3E4-no_jump-linear}]{{\fontsize{8pt}{9.6pt}\input{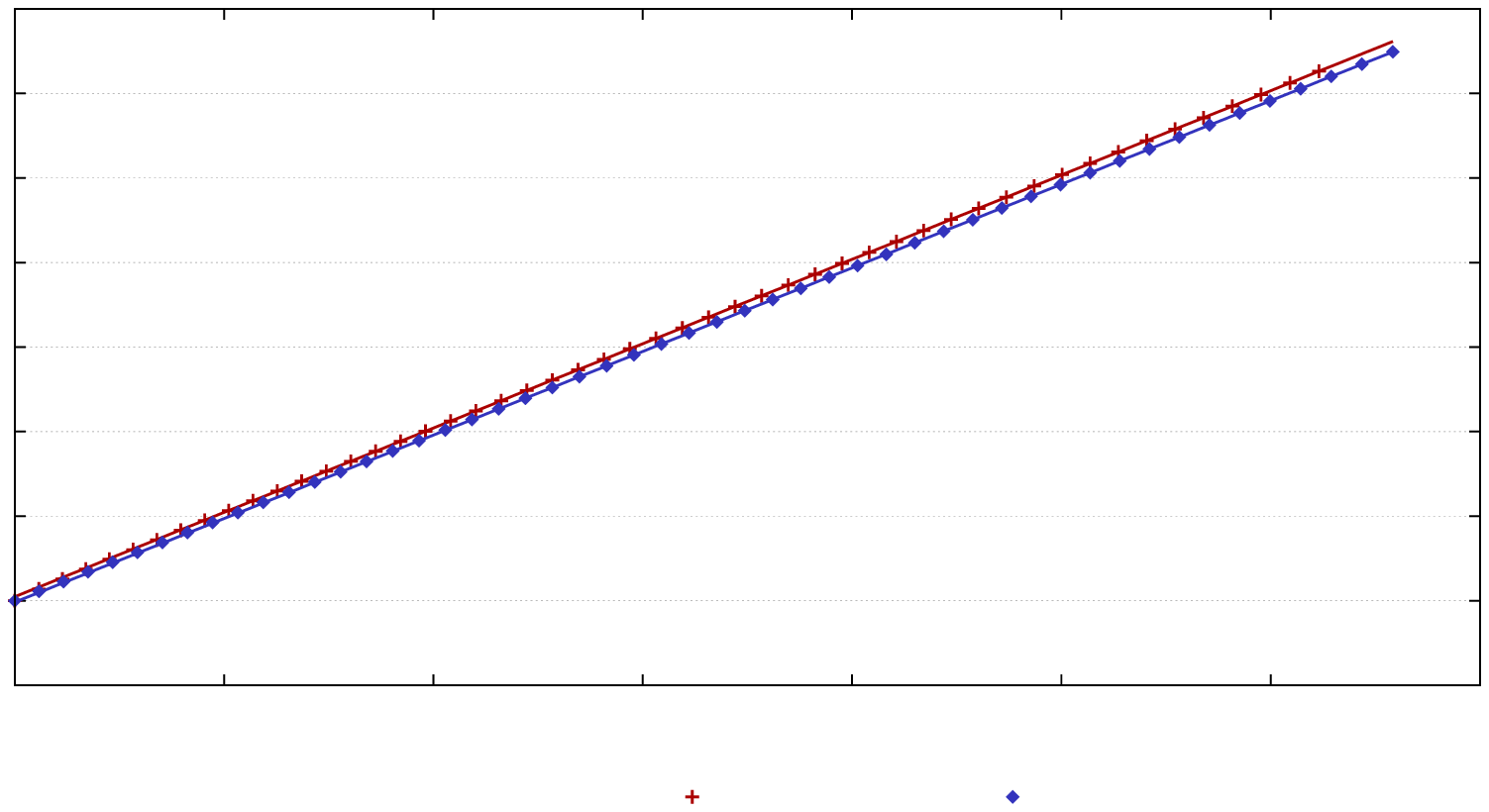}}}\\
\subfloat[\hspace{-1.2cm}\label{fig:3E4-no_jump-diff}]{{\fontsize{8pt}{9.6pt}\input{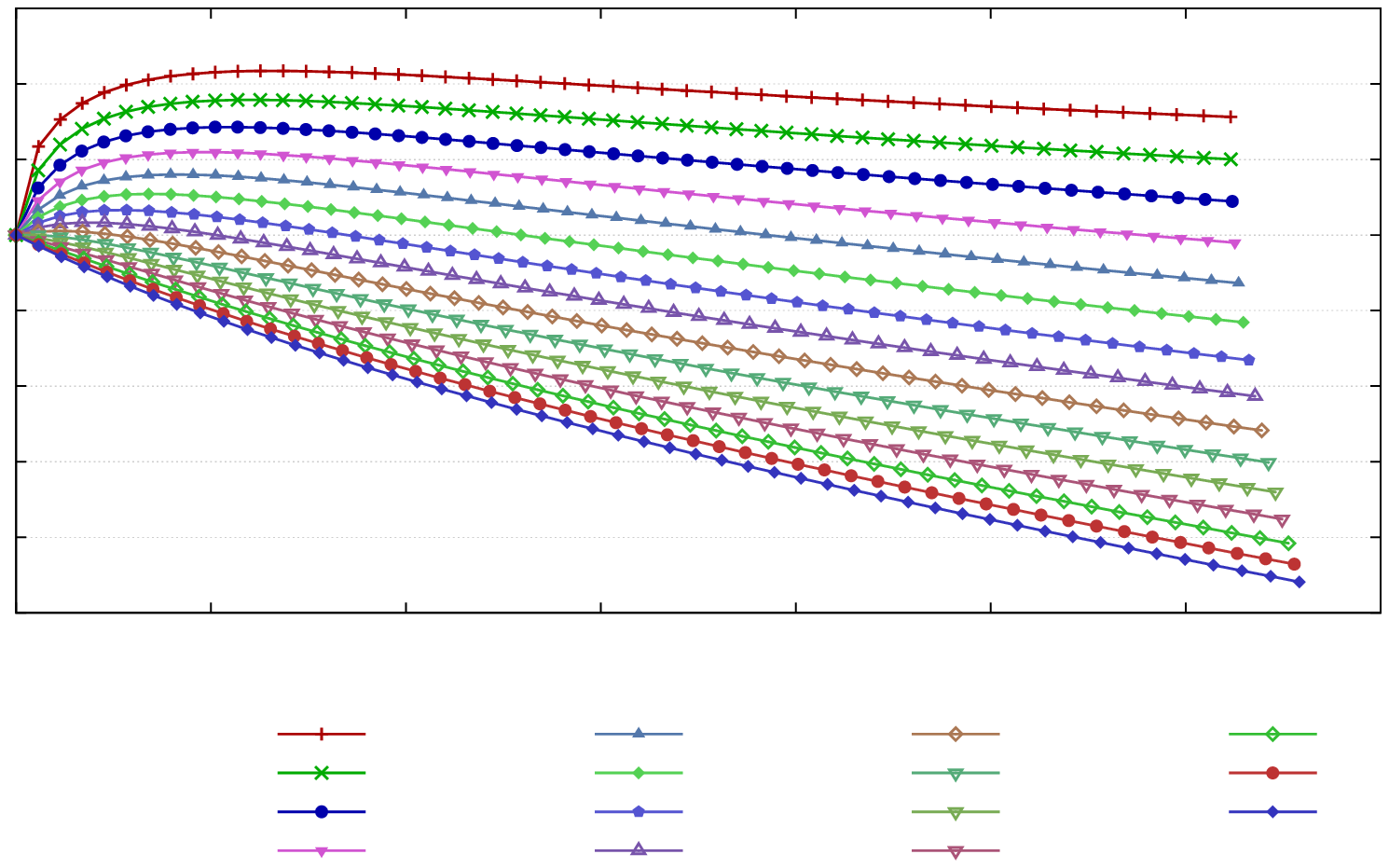}}}
\caption{\label{fig:3E4-no_jump}(a) A plot of $z_{\scriptscriptstyle LW}$ against $z_{\scriptscriptstyle FLRW}$ for a tangential and a nearly radial trajectory in the flat universe with initial cell size of $r_{b_0} = 3\times 10^4\, R$.  This is the equivalent plot to that of \FigRef{fig:complete_3E4} but with the photon starting at $10^4\, R$ instead.  The regression lines now pass very close to the origin, and the initial curve is nearly absent; we also note that the gradients are nearly unity.  Other trajectories, not shown, showed similar behaviour.  (b) A plot of $z_{\scriptscriptstyle LW} - z_{\scriptscriptstyle FLRW}$ against $z_{\scriptscriptstyle FLRW}$ for 15 trajectories in the same universe, including the two trajectories in (a).  Each graph still shows an initial curve, but it is small and short-lived, and linearity is soon established.}
\end{figure*}
\begin{table*} [htb]
\centering
\caption[Regression equations for $z_{\scriptscriptstyle LW}$ versus $z_{\scriptscriptstyle FLRW}$ graphs for photon redshifts in the $r_{b_0} = 3 \times 10^4\, R$ flat universe]{Linear regression equations and the root-mean-squares of the corresponding residuals for $z_{\scriptscriptstyle LW}$ versus $z_{\scriptscriptstyle FLRW}$ graphs from simulations of photons in the flat universe.  The initial cell size was $r_{b_0} = 3\times 10^4\, R$.  This is the same simulation as for \FigRef{fig:3E4-no_jump}.  The constants in the regression equations here are much smaller than those in the regression equations in \FigRef{fig:complete_3E4}.}
\begin{tabular*}{14cm}{>{\centering\arraybackslash}m{3cm} >{\centering\arraybackslash}m{7cm} >{\centering\arraybackslash}m{4cm}}
\hline\hline
\textbf{Initial angle} & \textbf{Regression equation} & \textbf{RMS of residuals}\\
\hline
1.5708 & $z_{\scriptscriptstyle LW} \, = \, 0.9973\, z_{\scriptscriptstyle FLRW}\, +\, 4.75 \times 10^{-3}$ & $1.84 \times 10^{-5}$ \\
1.4849 & $z_{\scriptscriptstyle LW} \, = \, 0.9966\, z_{\scriptscriptstyle FLRW}\, +\, 4.07 \times 10^{-3}$ & $1.88 \times 10^{-5}$ \\
1.3991 & $z_{\scriptscriptstyle LW} \, = \, 0.9959\, z_{\scriptscriptstyle FLRW}\, +\, 3.41 \times 10^{-3}$ & $2.03 \times 10^{-5}$ \\
1.3132 & $z_{\scriptscriptstyle LW} \, = \, 0.9952\, z_{\scriptscriptstyle FLRW}\, +\, 2.78 \times 10^{-3}$ & $2.29 \times 10^{-5}$ \\
1.2273 & $z_{\scriptscriptstyle LW} \, = \, 0.9944\, z_{\scriptscriptstyle FLRW}\, +\, 2.18 \times 10^{-3}$ & $2.63 \times 10^{-5}$ \\
1.1414 & $z_{\scriptscriptstyle LW} \, = \, 0.9937\, z_{\scriptscriptstyle FLRW}\, +\, 1.62 \times 10^{-3}$ & $3.05 \times 10^{-5}$ \\ 
1.0556 & $z_{\scriptscriptstyle LW} \, = \, 0.9929\, z_{\scriptscriptstyle FLRW}\, +\, 1.09 \times 10^{-3}$ & $3.53 \times 10^{-5}$ \\
0.9697 & $z_{\scriptscriptstyle LW} \, = \, 0.9922\, z_{\scriptscriptstyle FLRW}\, +\, 6.06 \times 10^{-4}$ & $4.04 \times 10^{-5}$ \\
0.8838 & $z_{\scriptscriptstyle LW} \, = \, 0.9915\, z_{\scriptscriptstyle FLRW}\, +\, 1.75 \times 10^{-4}$ & $4.89 \times 10^{-5}$ \\
0.7980 & $z_{\scriptscriptstyle LW} \, = \, 0.9908\, z_{\scriptscriptstyle FLRW}\, -\, 2.35 \times 10^{-4}$ & $5.49 \times 10^{-5}$ \\
0.7121 & $z_{\scriptscriptstyle LW} \, = \, 0.9902\, z_{\scriptscriptstyle FLRW}\, -\,6.06 \times 10^{-4}$ & $6.07 \times 10^{-5}$ \\
0.6262 & $z_{\scriptscriptstyle LW} \, = \, 0.9897\, z_{\scriptscriptstyle FLRW}\, -\,9.37 \times 10^{-4}$ & $6.62 \times 10^{-5}$ \\
0.5404 & $z_{\scriptscriptstyle LW} \, = \, 0.9892\, z_{\scriptscriptstyle FLRW}\, -\,1.23 \times 10^{-3}$ & $7.12 \times 10^{-5}$ \\
0.4545 & $z_{\scriptscriptstyle LW} \, = \, 0.9888\, z_{\scriptscriptstyle FLRW}\, -\,1.48 \times 10^{-3}$ & $7.57 \times 10^{-5}$ \\
0.3686 & $z_{\scriptscriptstyle LW} \, = \, 0.9884\, z_{\scriptscriptstyle FLRW}\, -\,1.69 \times 10^{-3}$ & $7.94 \times 10^{-5}$ \\
\hline\hline
\end{tabular*}
\label{tab:3E4-no_jump-regressions}
\end{table*}
\begin{figure*}[p!tbh]
\subfloat[\hspace{-1.2cm}\label{flat-1E5}]{{\fontsize{8pt}{9.6pt}\input{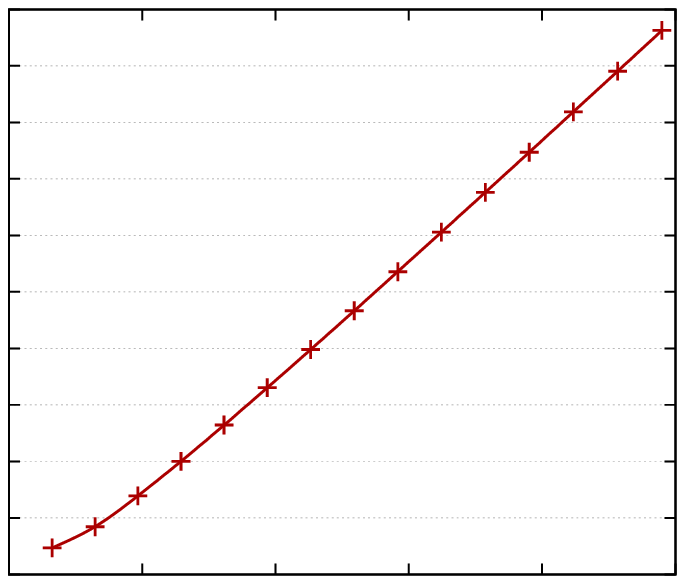}}}
\subfloat[\hspace{-1.2cm}\label{flat-1E6}]{{\fontsize{8pt}{9.6pt}\input{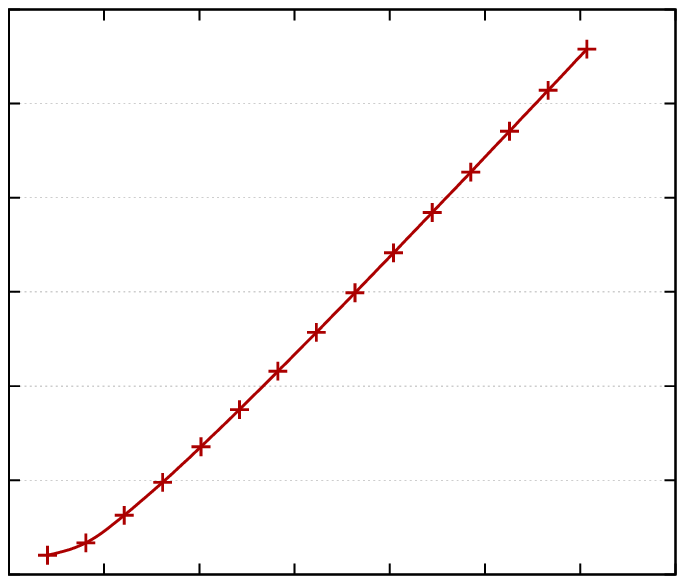}}}\\
\subfloat[\hspace{-1.2cm}\label{flat-1E7}]{{\fontsize{8pt}{9.6pt}\input{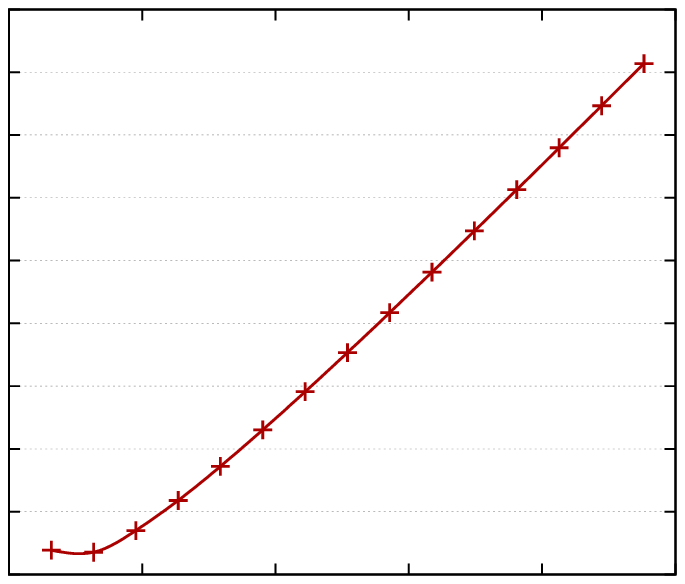}}}
\subfloat[\hspace{-1.2cm}\label{flat-1E8}]{{\fontsize{8pt}{9.6pt}\input{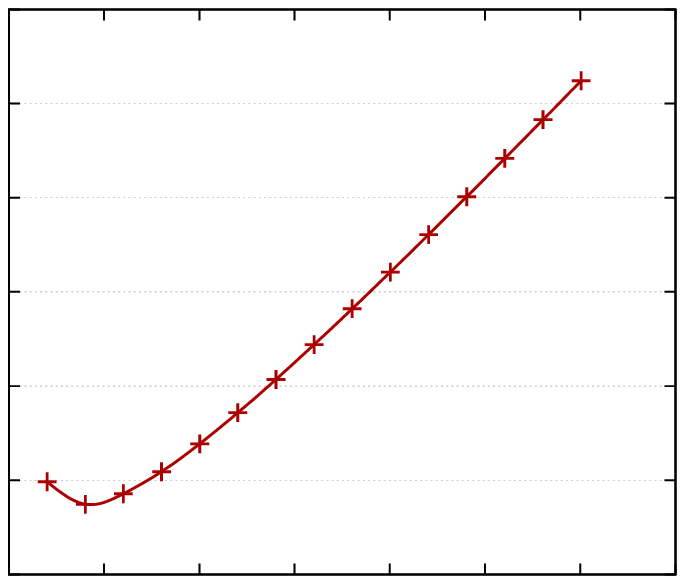}}}
\caption{\label{fig:1E5}A plot of $z_{\scriptscriptstyle LW}$ against $z_{\scriptscriptstyle FLRW}$ for the $\theta_0=\pi / 2$ photon trajectory in the (a) $r_{b_0} = 10^5\, R$ universe, (b) $r_{b_0} = 10^6\, R$ universe, (c) $r_{b_0} = 10^7\, R$ universe, and (d) $r_{b_0} = 10^8\, R$ universe.  $\Delta r_0$ was $10^{-5}\, R$ for (a) and (b), $10^{-4}\, R$ for (c), and $10^{-3}\, R$ for (d).}
\end{figure*}

\begin{figure*}[p]
\subfloat[\hspace{-1.2cm}\label{fig:1E8-no_jump-linear}]{{\fontsize{8pt}{9.6pt}\input{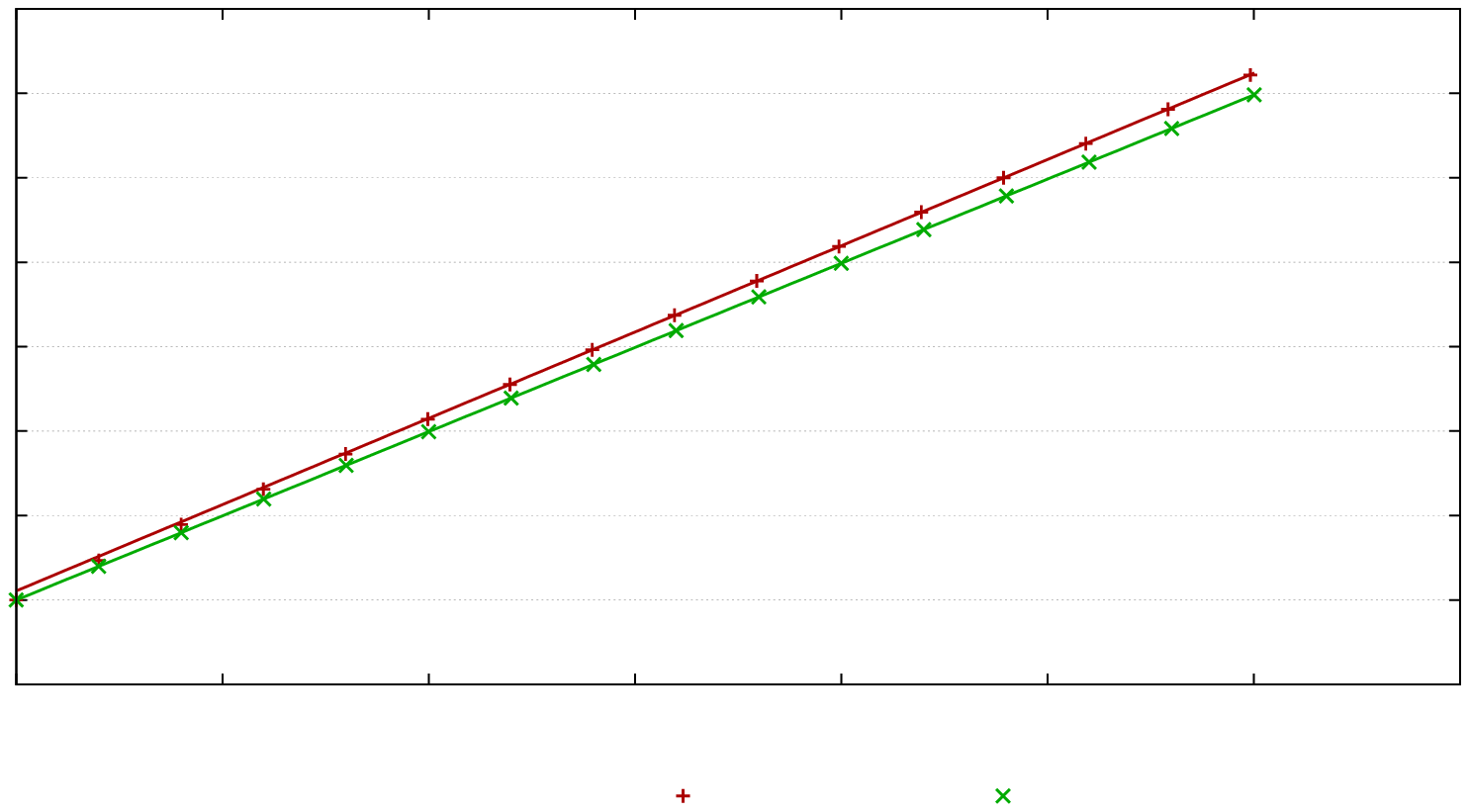}}}\\
\subfloat[\hspace{-1.2cm}\label{fig:1E8-no_jump-diff}]{{\fontsize{8pt}{9.6pt}\input{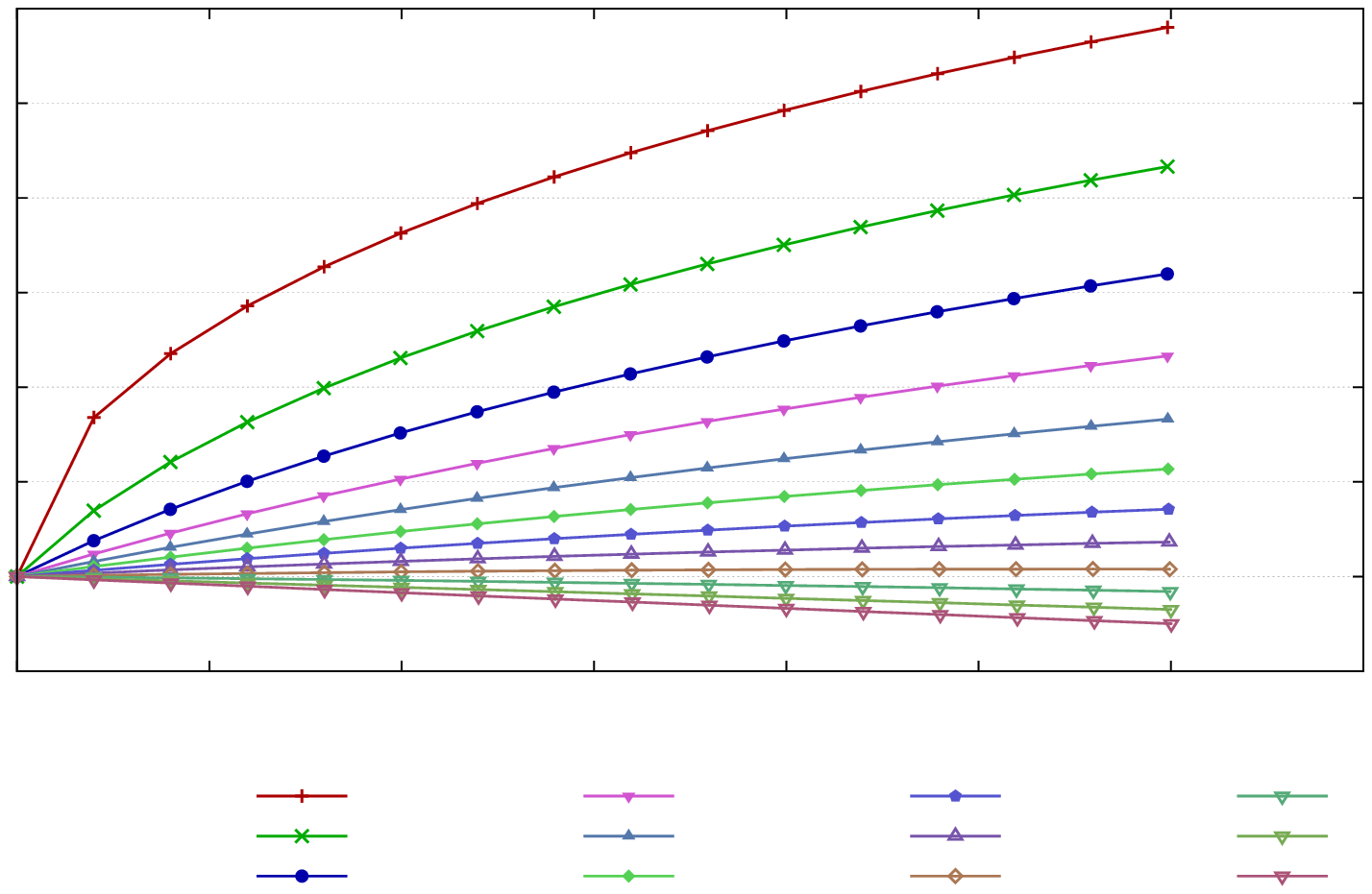}}}
\caption{\label{fig:1E8-no_jump}(a) A plot of $z_{\scriptscriptstyle LW}$ against $z_{\scriptscriptstyle FLRW}$ for a tangential and a nearly radial trajectory starting at radius $10^7\, R$ in the $r_{b_0} = 10^8\, R$ flat universe.  Each trajectory was traced across 15 cells.  Compared to \FigRef{fig:3E4-no_jump-linear}, the regression lines here pass even closer to the origin; the initial curve is even more suppressed; and the gradients are even closer to unity.  Other trajectories, not shown, showed similar behaviour.  (b) A plot of $z_{\scriptscriptstyle LW} - z_{\scriptscriptstyle FLRW}$ against $z_{\scriptscriptstyle FLRW}$ for 15 trajectories in the same universe, including the two trajectories in (a).  Compared to \FigRef{fig:3E4-no_jump-diff}, the initial curve here is even smaller and more short-lived.}
\end{figure*}
\begin{table*} [htb]
\centering
\caption{Linear regression equations and the root-mean-squares of the corresponding residuals for $z_{\scriptscriptstyle LW}$ versus $z_{\scriptscriptstyle FLRW}$ graphs from simulations of photons in the $r_{b_0} = 10^8\, R$ flat universe.  This is the same simulation as for \FigRef{fig:1E8-no_jump-linear} and \FigRef{fig:1E8-no_jump-diff}.  Regression was performed on the $z_{\scriptscriptstyle FLRW} > 0.001$ region.  Both the regression equation constants and the residuals are even smaller here than in \tabref{tab:3E4-no_jump-regressions}.}
\begin{tabular*}{14cm}{>{\centering\arraybackslash}m{3cm} >{\centering\arraybackslash}m{7cm} >{\centering\arraybackslash}m{4cm}}
\hline\hline
\textbf{Initial angle} & \textbf{Regression equation} & \textbf{RMS of residuals}\\
\hline
1.5708 & $z_{\scriptscriptstyle LW} \, = \, 1.0215\, z_{\scriptscriptstyle FLRW}\, +\, 5.4 \times 10^{-5}$ & $1.66 \times 10^{-6}$ \\
1.4635 & $z_{\scriptscriptstyle LW} \, = \, 1.0200\, z_{\scriptscriptstyle FLRW}\, +\, 2.9 \times 10^{-5}$ & $1.45 \times 10^{-6}$ \\
1.3561 & $z_{\scriptscriptstyle LW} \, = \, 1.0167\, z_{\scriptscriptstyle FLRW}\, +\, 1.5 \times 10^{-5}$ & $1.04 \times 10^{-6}$ \\
1.2488 & $z_{\scriptscriptstyle LW} \, = \, 1.0130\, z_{\scriptscriptstyle FLRW}\, +\, 8.8 \times 10^{-6}$ & $6.97 \times 10^{-7}$ \\
1.1414 & $z_{\scriptscriptstyle LW} \, = \, 1.0095\, z_{\scriptscriptstyle FLRW}\, +\, 5.4 \times 10^{-6}$ & $4.62 \times 10^{-7}$ \\ 
1.0341 & $z_{\scriptscriptstyle LW} \, = \, 1.0066\, z_{\scriptscriptstyle FLRW}\, +\, 3.4 \times 10^{-6}$ & $3.08 \times 10^{-7}$ \\
0.9268 & $z_{\scriptscriptstyle LW} \, = \, 1.0041\, z_{\scriptscriptstyle FLRW}\, +\, 2.2 \times 10^{-6}$ & $2.04 \times 10^{-7}$ \\
0.8194 & $z_{\scriptscriptstyle LW} \, = \, 1.0020\, z_{\scriptscriptstyle FLRW}\, +\, 1.4 \times 10^{-6}$ & $1.31 \times 10^{-7}$ \\
0.7121 & $z_{\scriptscriptstyle LW} \, = \, 1.0002\, z_{\scriptscriptstyle FLRW}\, +\, 9.2 \times 10^{-7}$ & $6.34 \times 10^{-8}$ \\
0.6048 & $z_{\scriptscriptstyle LW} \, = \, 0.9988\, z_{\scriptscriptstyle FLRW}\, +\, 4.5 \times 10^{-7}$ & $3.10 \times 10^{-8}$ \\
0.4974 & $z_{\scriptscriptstyle LW} \, = \, 0.9976\, z_{\scriptscriptstyle FLRW}\, +\, 1.0 \times 10^{-7}$ & $7.08 \times 10^{-9}$ \\
0.3901 & $z_{\scriptscriptstyle LW} \, = \, 0.9967\, z_{\scriptscriptstyle FLRW}\, -\, 1.6 \times 10^{-7}$ & $1.24 \times 10^{-8}$ \\
\hline\hline
\end{tabular*}
\label{tab:1E8-no_jump-regressions}
\end{table*}

\begin{figure*}[tbh]
{\fontsize{8pt}{9.6pt}\input{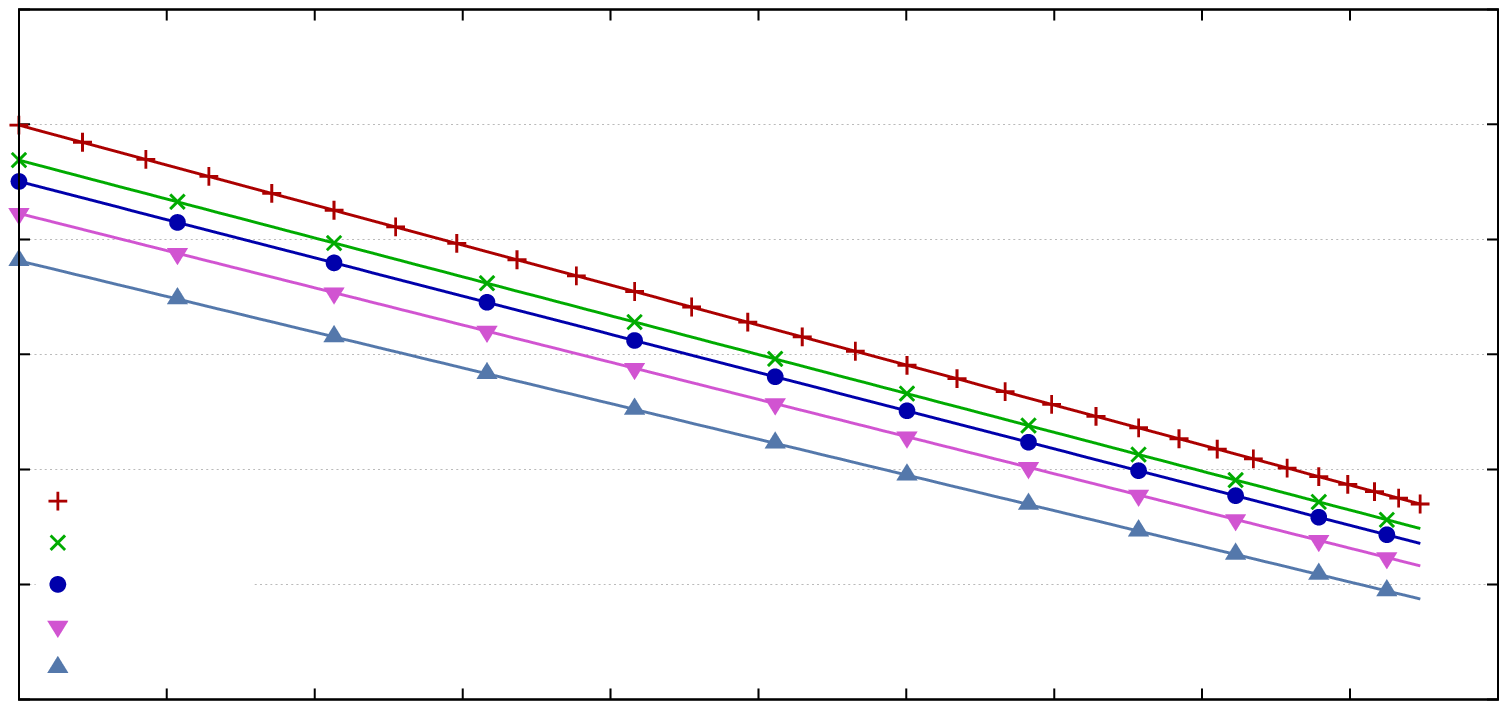}}
\caption{\label{fig:flat-regressions}A plot of the gradients of $z_{\scriptscriptstyle LW}$ vs $z_{\scriptscriptstyle FLRW}$ graphs against $\cos(\theta_n)$ for flat universes of different initial cell size $r_{b_0}$.  A linear regression has been performed on each graph, and the regression equations are displayed in order of increasing $r_{b_0}$.}
\end{figure*}

Based on these graphs, we make some speculations on the origin of the jump as well as the initial curve in the $z_{\scriptscriptstyle LW}$ vs $z_{\scriptscriptstyle FLRW}$ graphs.  We note that for large $r_{obs}$, the frequency asymptotically approaches
\begin{align*}
\nu_{asym} &= \sqrt{E_{ph}} \left(\sqrt{E_b} - \sqrt{E_b-1} \right)\\
& = \sqrt{E_{ph}} \qquad \text{for } E_b=1.
\end{align*}
Equivalently, the redshift asymptotically approaches
\begin{align*}
z_{asym} &= \frac{\nu_e}{\sqrt{E_{ph}}\left(\sqrt{E_b} - \sqrt{E_b-1} \right)}-1\\
& = \frac{\nu_e}{\sqrt{E_{ph}}}-1 \qquad \text{for } E_b=1.
\end{align*}
As the photon traversed ever more cells in our simulations, we found that it would intercept the co-moving observer at ever larger $r_{obs}$.  So after enough cells, $z_{\scriptscriptstyle LW}$ should be very close to the asymptotic value $z_{asym}$.  Also from our simulations, we found that $E_{ph}$ would decrease with each subsequent cell implying that $z_{asym}$ would increase, which is consistent with the $z_{\scriptscriptstyle LW}$ vs $z_{\scriptscriptstyle FLRW}$ graphs' positive gradients.  Thus the asymptotic, large-radius behaviour of the redshifts is responsible for the linear behaviour of the $z_{\scriptscriptstyle LW}$ vs $z_{\scriptscriptstyle FLRW}$ graphs.  The graphs' initial behaviour however must be explained by the behaviour of the redshifts at low $r_{obs}$, where the photon is still sufficiently close to the central mass to feel its influence strongly.  We note that the frequencies always approach $\nu_{asym}$ from below, implying that the redshifts always approach $z_{asym}$ from above.  This is consistent with the initial curve in the $z_{\scriptscriptstyle LW}$ vs $z_{\scriptscriptstyle FLRW}$ graphs, as this curve converges into the linear regime from above as well.  At much smaller $r_{obs}$ however, the redshift may deviate much more significantly from the asymptotic value.  In particular, the deviation of the zero-redshift point, where $r_{obs}$ is smallest, may be much greater than the deviation of the next data point, where $r_{obs}$ may have increased sufficiently for the redshift to be much closer to the asymptotic value.  This would account for the sudden initial jump seen in the $z_{\scriptscriptstyle LW}$ vs $z_{\scriptscriptstyle FLRW}$ graphs.  Therefore, the initial jump and the initial curve in the $z_{\scriptscriptstyle LW}$ vs $z_{\scriptscriptstyle FLRW}$ graphs can be attributed to the photon being closer to the central mass and thus feeling its effects more strongly.

\begin{figure*}[p]
{\fontsize{8pt}{9.6pt}\input{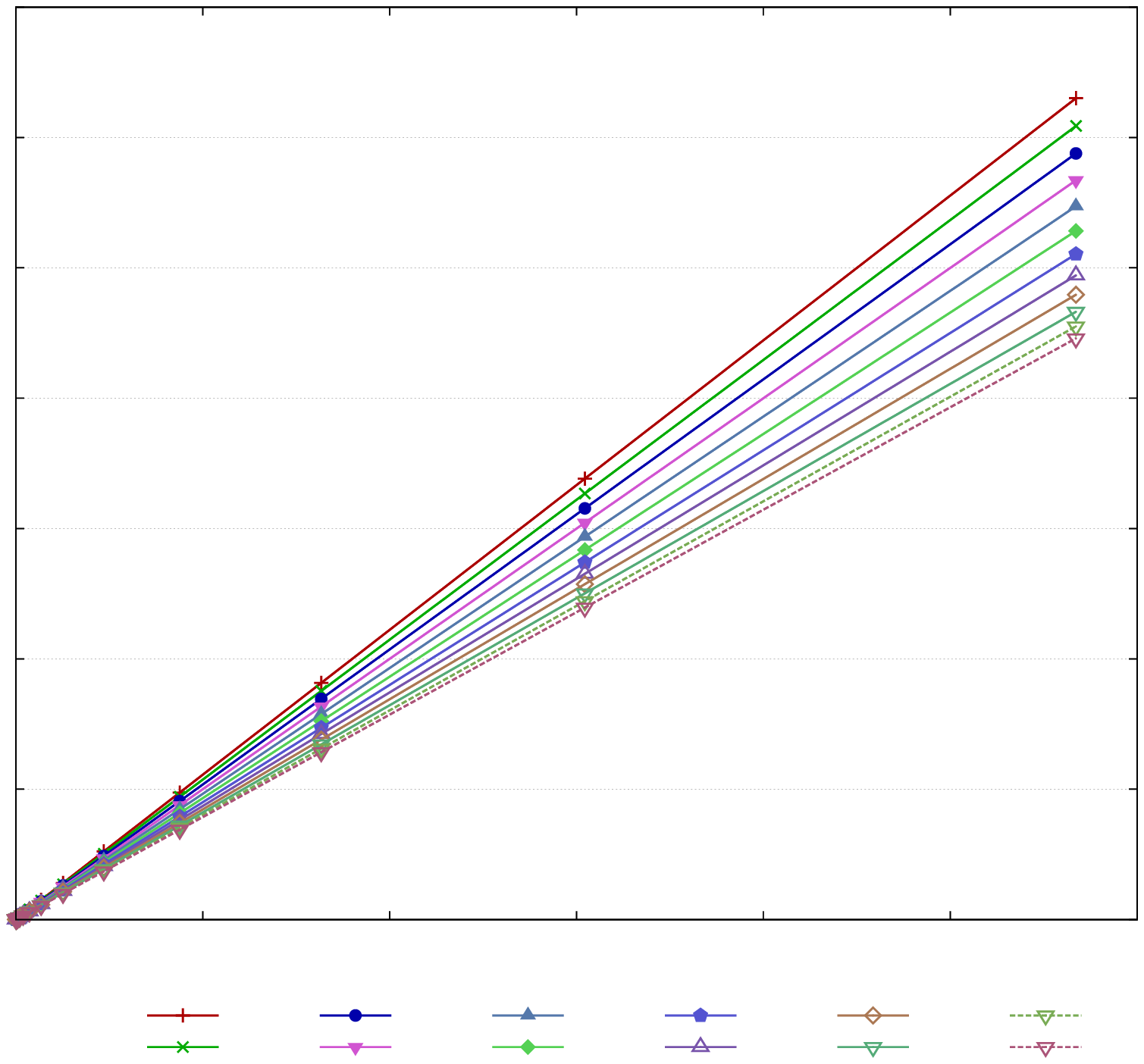}}
\caption{\label{fig:D=-0_10}A plot of $z_{\scriptscriptstyle LW}$ against $z_{\scriptscriptstyle FLRW}$ for 12 trajectories in the $E_b=1.1$ universe.  For this simulation, $\Delta r_0$ was $10^{-3}\, R$.  The initial angle $\theta_n$ of the trajectory is given in the legend.  Each trajectory was traced across 15 cells.  A linear regression has been performed for each graph.  The regression equations and corresponding root-mean-squares of the residuals are listed above in order of decreasing $\theta_n$.}
\end{figure*}

To test our conjecture, we simulated photons in the same universe but starting from a large radius, specifically $r = 10^4\, R$, such that the central mass' influence would be much reduced.  \FigRef{fig:3E4-no_jump-linear} shows the resulting plot of $z_{\scriptscriptstyle LW}$ against $z_{\scriptscriptstyle FLRW}$ for the most radial and the most tangential trajectories simulated.  The graphs now follow completely straight lines passing very close to the origin, with the initial jump completely gone and the initial curve nearly absent.  All other trajectories at intermediate angles displayed similar behaviour, with the graphs' gradients increasing as the initial trajectory became more tangential.  We also note that the gradients are nearly unity, indicating that $z_{\scriptscriptstyle LW}$ is nearly identical to $z_{\scriptscriptstyle FLRW}$.  In \FigRef{fig:3E4-no_jump-diff}, we have plotted the difference between the $z_{\scriptscriptstyle LW}$ graphs and the reference graph of $\tilde{z}_{\scriptscriptstyle LW} = z_{\scriptscriptstyle FLRW}$.  The deviation from linearity at low $r_{obs}$ becomes clearer now, although we also see that it is short-lived.  From this figure, we see that linearity is relatively well established by $z_{\scriptscriptstyle FLRW} \approx 0.2$; thus for each trajectory, we have regressed the corresponding $z_{\scriptscriptstyle LW}$ vs $z_{\scriptscriptstyle FLRW}$ graph for the region $z_{\scriptscriptstyle FLRW} > 0.2$.  All regression equations and corresponding root-mean-squares of the residuals have been listed in \tabref{tab:3E4-no_jump-regressions}, and two of these regressions are shown in \FigRef{fig:3E4-no_jump-linear}.  These equations show that our observations for the graphs in \FigRef{fig:3E4-no_jump-linear} holds generally for all trajectories, specifically that the gradients are nearly unity and that the graphs pass very close to the origin.  Thus, the influence of the masses on redshifts has been significantly suppressed, and the redshifts now behave nearly identically to FLRW, much like Clifton and Ferreira's mean redshift result.

For flat universes with larger initial cell sizes, our simulations yielded results similar to those of the $r_{b_0} = 3\times 10^4\, R$ universe.  \Figuref{fig:1E5} depicts plots of $z_{\scriptscriptstyle LW}$ against $z_{\scriptscriptstyle FLRW}$ for the $\theta_0 = \pi / 2$ trajectory for universes with $r_{b_0}$ ranging from $10^{5}$ to $10^{8}$.  All graphs again converged quickly to a straight line from above, including graphs, not shown, for the other trajectories.  One noticeable development though was that the curve at the start of the graphs became more pronounced as the initial size increased.  However when we simulated photons starting from a large radius again, the initial curve was again suppressed, and the initial jump was again absent, as shown in \FigRef{fig:1E8-no_jump-linear}, \FigRef{fig:1E8-no_jump-diff}, and \tabref{tab:1E8-no_jump-regressions} for photons starting from $r = 10^7\, R$ in the $r_{b_0} = 10^8\, R$ universe; the initial curve is even more short-lived, as seen in \FigRef{fig:1E8-no_jump-diff}, and the regression lines for all trajectories pass even closer to the origin, as demonstrated by the even smaller constants in the regression equations in \tabref{tab:1E8-no_jump-regressions}.

We also notice, in \FigRef{fig:complete_3E4}, that the gradients of $z_{\scriptscriptstyle LW}$ vs $z_{\scriptscriptstyle FLRW}$ graphs decrease with $\theta_n$.  In fact, if we plot the gradients against $\cos(\theta_n)$, we find a strongly linear relationship between the two quantities, as shown in \FigRef{fig:flat-regressions}.  As the figure shows, this is also true of the gradients for all the other flat universes we simulated.  Since $\cos(\theta_n)$ determines the radial component $\dot{\rho}_{init}$ of the photon's initial velocity, this implies that the gradients are linearly related to $\dot{\rho}_{init}/\dot{\tau}_{init}$.  As the photon's initial velocity becomes increasingly radial, the photon's frequency becomes increasingly blueshifted relative to the FLRW redshift.  We believe this is due to the central mass' stronger influence on the photon, as a more radial trajectory would bring the photon closer to the central mass.  Quite remarkably, this influence can alter the lattice universe redshifts $z_{\scriptscriptstyle LW}$ from their FLRW counterparts $z_{\scriptscriptstyle FLRW}$ by as much as 30\%, as shown by the right end of the $r_{b_0} = 10^8$ graph.  Since the lattice universe can generate lower redshift values, an observer fitting these redshifts to an FLRW model could significantly underestimate the age of the photons' source.

\begin{figure*}[tb]
{\fontsize{8pt}{9.6pt}\input{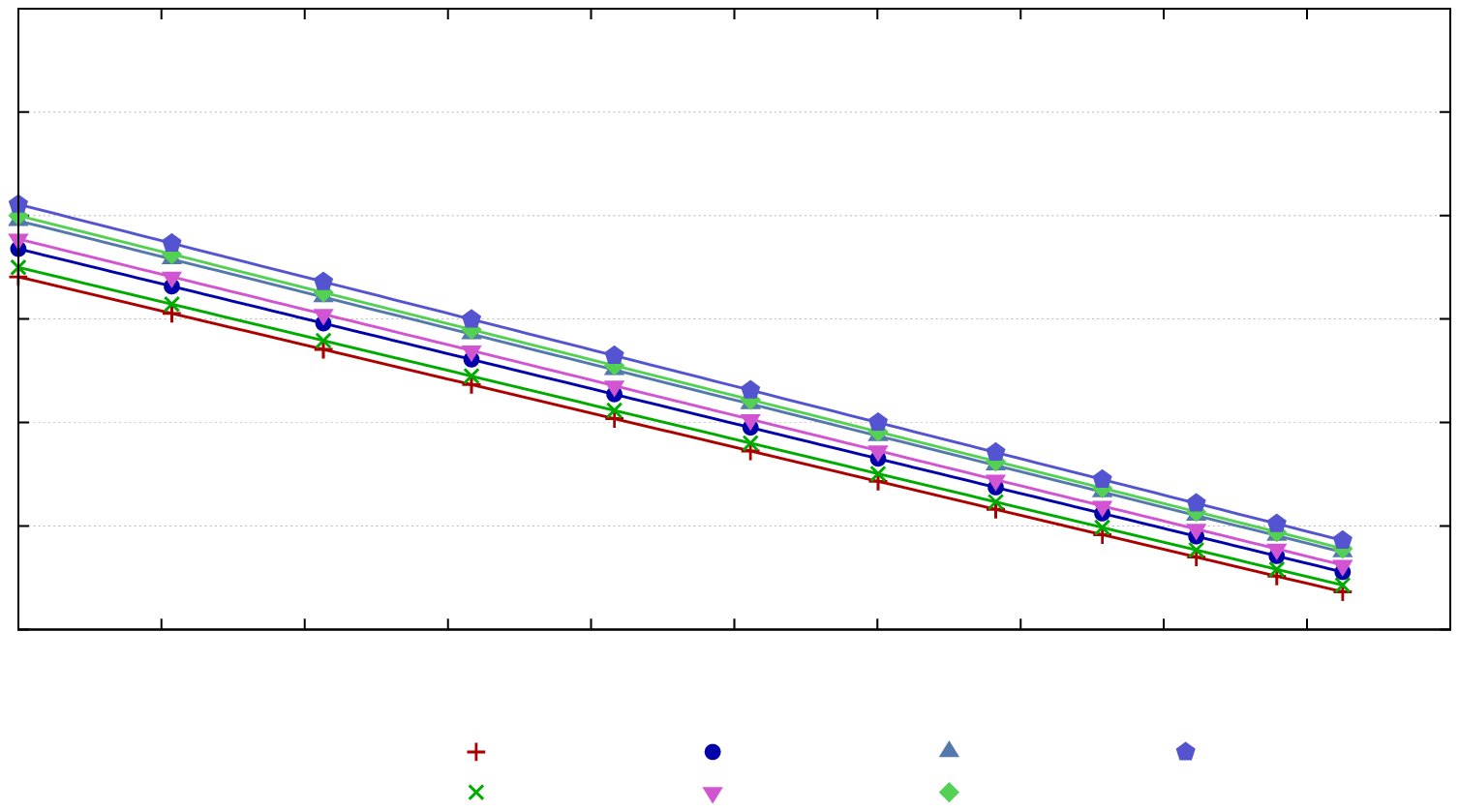}}
\caption{\label{fig:open-gradient_vs_angs}Plot of the gradients of $z_{\scriptscriptstyle LW}$ vs $z_{\scriptscriptstyle FLRW}$ graphs against $\cos(\theta_n)$ for open universes of different values of $E_b$.  Also shown is a plot for the flat universe, corresponding to $E_b=1$.  A linear regression has been performed on each graph, and both the regression equation and the root-mean-square of the residuals are displayed in order of increasing $E_b$.}
\end{figure*}

Everything we noticed about redshifts in the flat universe also applied to open universes.  We simulated open universes for values of $E_b$ in the range of $1.1 \geq E_b > 1$.  In each simulation, the initial cell size was fixed to be $r_{b_0} = 3 \times 10^4$, and photons were propagated along 12 trajectories with initial angles ranging from $\theta_0 = \pi / 2$ to $\theta_{11} = 149 \pi / 1200$ in decrements of $41 \pi / 1200$.  Depending on $E_b$, $\Delta r_0$ ranged from $10^{-5}\,R$ for $E_b = 1$ to $10^{-3}\,R$ for $E_b=1.1$.  \Figuref{fig:D=-0_10} shows a clear linear relationship again between $z_{\scriptscriptstyle LW}$ and $z_{\scriptscriptstyle FLRW}$ for the $E_b=1.1$ universe.  The regression equations' non-zero intercepts indicate the presence again of a jump between the zero-redshift point and the first non-zero redshift point.  Similar behaviour was seen for universes corresponding to other values of $E_b$.  \Figuref{fig:open-gradient_vs_angs} shows that for all $E_b$, there is again a clear negative linear relation between the gradients of $z_{\scriptscriptstyle LW}$ vs $z_{\scriptscriptstyle FLRW}$ graphs and $\cos(\theta_n)$, or equivalently between the gradients and $\dot{\rho}_{init}/\dot{\tau}_{init}$.
 
\begin{figure}[tbh]
{\fontsize{8pt}{9.6pt}\input{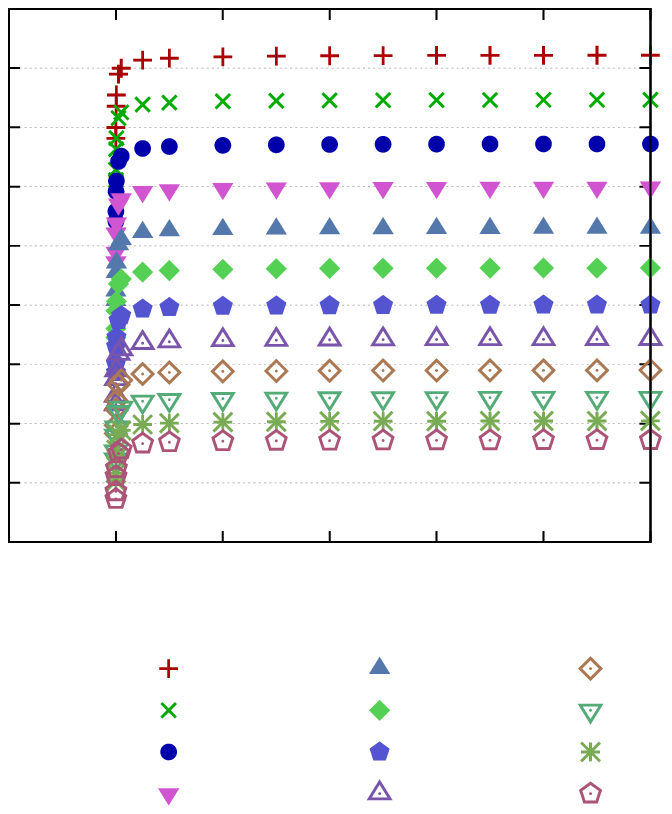}}
\caption{\label{fig:ang}A plot of the $z_{\scriptscriptstyle LW}$ vs $z_{\scriptscriptstyle FLRW}$ gradients against $E_b$ for a selection of initial photon trajectories.  Each graph approaches an asymptotic value as $E_b$ increases.}
\end{figure}
To see how the gradients of $z_{\scriptscriptstyle LW}$ vs $z_{\scriptscriptstyle FLRW}$ change with $E_b$, we have plotted the gradients against $E_b$ in \FigRef{fig:ang} for a selection of trajectories.  The graphs indicate that the gradients approach some asymptotic value as $E_b$ increases and that this asymptotic gradient decreases as the photon trajectory becomes more radial, consistent with the behaviour illustrated in \FigRef{fig:open-gradient_vs_angs}.

Finally, we shall now discuss our simulation of the closed universe.  We have chosen to focus on an LW universe built from the 600-cell Coxeter lattice described in \appendref{App_Latt}, as it is the most finely subdivided closed Coxeter lattice possible.  Using \eqref{psi} and \eqref{LW_E}, we find that $E_b \approx 0.96080152145$ for this universe, and the maximum cell size is therefore approximately $25.5\, R$.  If we embed a Schwarzschild-cell into the comparison hypersphere, as described in \secref{LWScaleFactor}, then the ratio between the angular distance, $\psi_h$, from the cell centre to the horizon, as measured from the hypersphere's centre, and the corresponding distance, $\psi_b$, to the cell boundary is approximately 0.0392;\footnote{This value has been calculated using \eqref{LW_scalefactor}, \eqref{chi_func}, \eqref{r_E}, and \eqref{LW_E}; from these equations, it follows that $\sin \psi_h = \sin^3 \psi_b$, where $\psi_b$ is given by \eqref{psi} for $N=600$.} for comparison, Clifton \emph{et al.}\ \citep{CRT} have shown that on the time-symmetric hypersurface, the exact 3-metric is conformally equivalent to that of a 3-sphere, and on this 3-sphere, the equivalent ratio is 0.0147; our value is about 2.6 times greater than theirs, although their ratio uses the shortest angular distance to the cell boundary, which is not spherical in the exact lattice.  In any case, the 600-cell universe is much smaller than the universes we have heretofore been considering, and the central mass' influence will therefore be much stronger.  We note that in the LW co-ordinates for the closed universe, we do not have freedom to choose the initial size of the cell; once the photon's initial co-ordinates are chosen, then the cell boundary's position is determined, since both the boundary and the source must reach their respective maximum radii at the same time $\tau$.  For our simulation of this universe, we chose to propagate photons along 30 trajectories ranging from $\theta_0 = \frac{1}{2}\pi$ to $\theta_{29} = \frac{17}{150}\pi$ going in decrements of $\frac{2}{150}\pi$.

\begin{figure}[tb]
{\fontsize{8pt}{9.6pt}\input{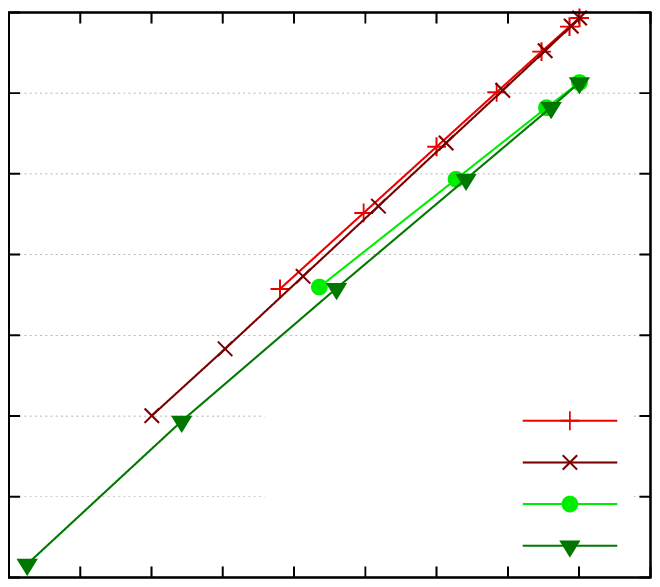}}
\caption{\label{fig:closed_subset}A plot of $z_{\scriptscriptstyle LW}$ against $z_{\scriptscriptstyle FLRW}$ for two trajectories of photons, $\theta_0 = \pi / 2$ and $\theta_{29} = 17 \pi / 150$.  Each trajectory's graph is divided into a subgraph for when the trajectory is outgoing and a separate subgraph for when the same trajectory becomes ingoing; the two subgraphs meet at the redshift data point closest to maximum expansion of the universe.  There is a slight bend in the blueshift regime of the ingoing $\theta_{29} = 17 \pi / 150$ graph.}
\end{figure}
\begin{figure*}[p]
{\fontsize{8pt}{9.6pt}\input{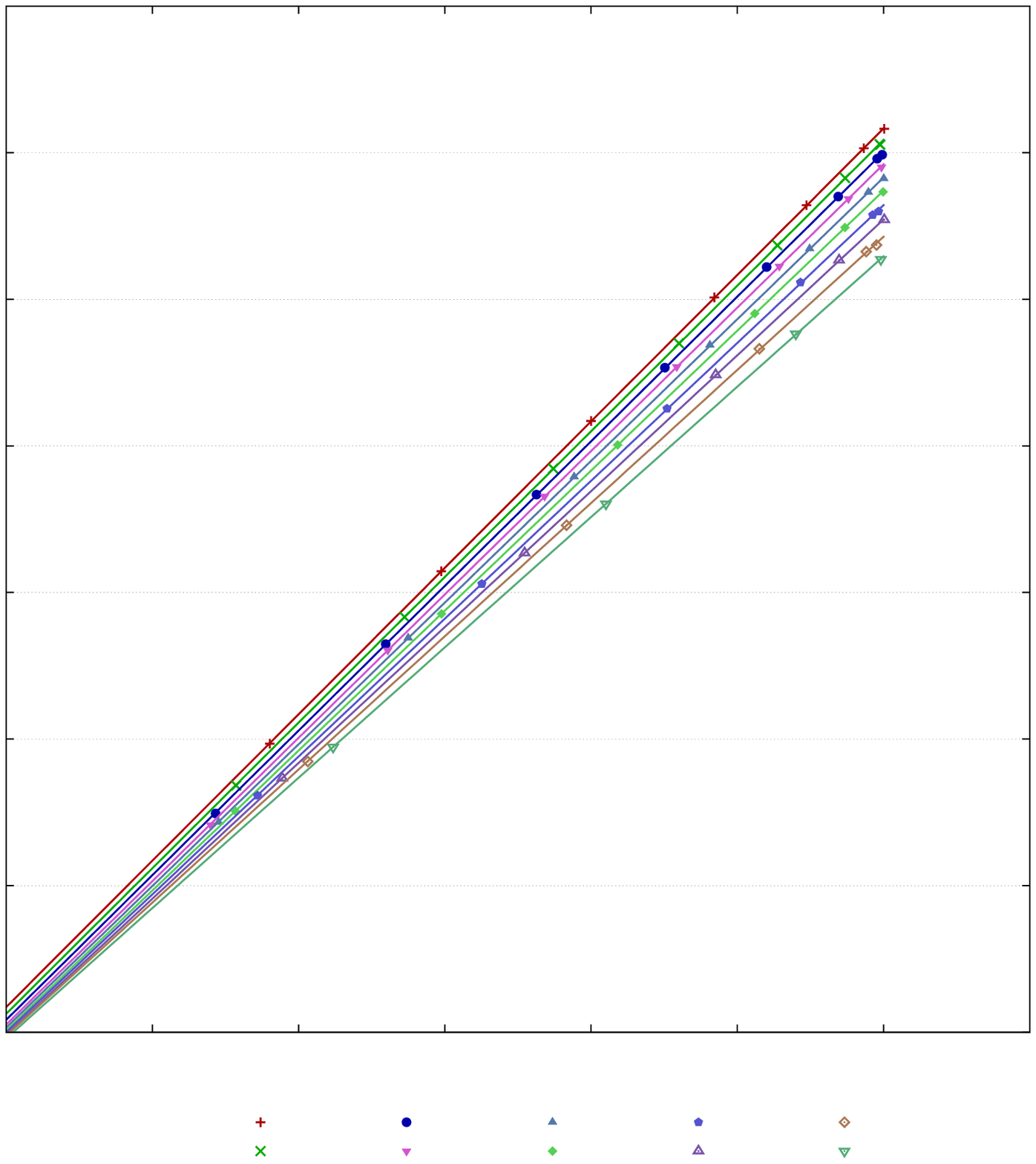}}
\caption{\label{fig:closed_outgoing}A plot of $z_{\scriptscriptstyle LW}$ against $z_{\scriptscriptstyle FLRW}$ while photons are outgoing.  Graphs for a selection of photon trajectories $\theta_n$ are shown.  A linear regression was performed on each graph, and the regression equation is listed in order of decreasing $\theta_n$.}
\end{figure*}
\begin{figure*}[p]
{\fontsize{8pt}{9.6pt}\input{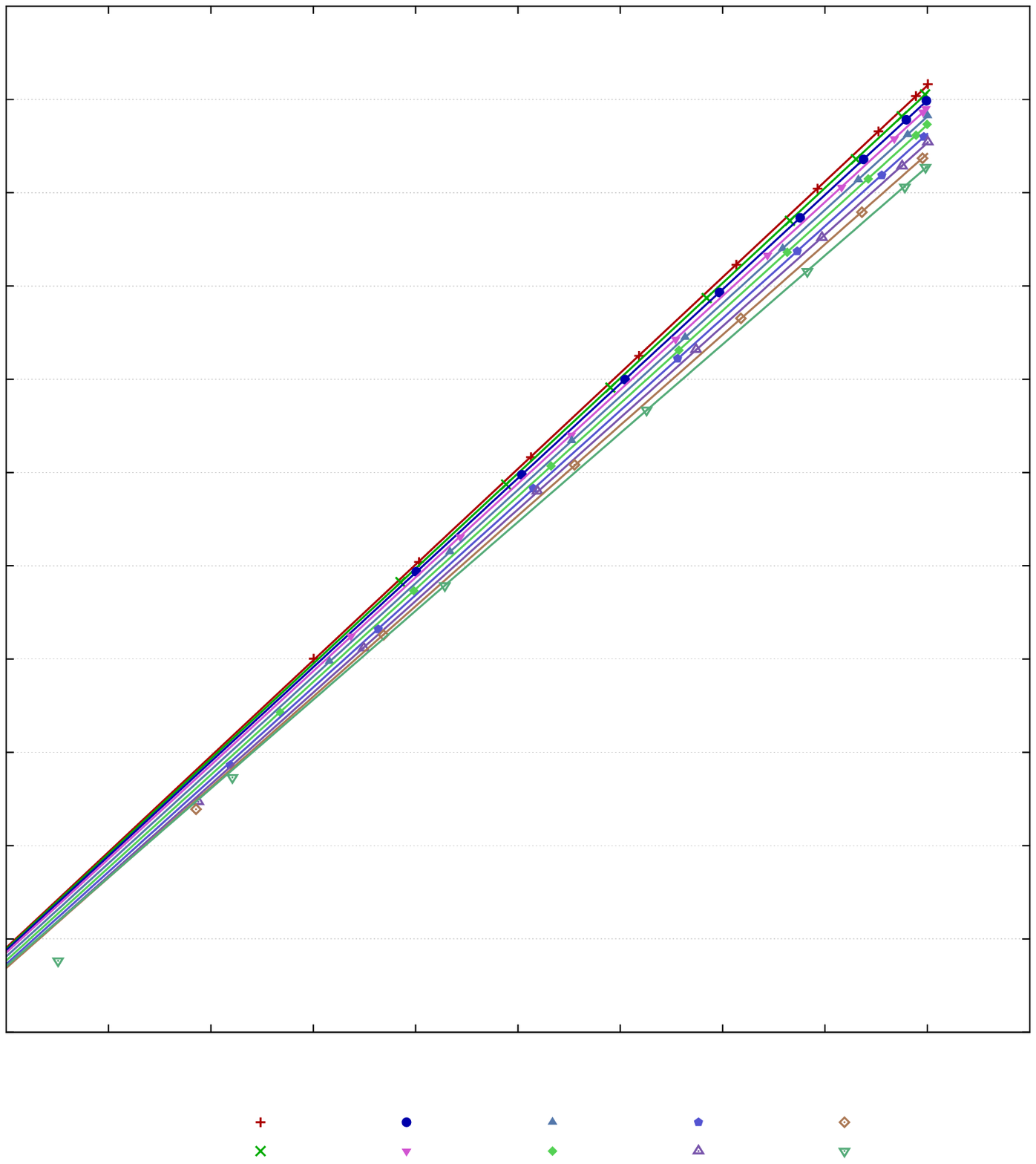}}
\caption{\label{fig:closed_ingoing}A plot of $z_{\scriptscriptstyle LW}$ against $z_{\scriptscriptstyle FLRW}$ while photons are ingoing.  Graphs for the same trajectories as in \FigRef{fig:closed_outgoing} are shown.  A linear regression was performed on each graph, and the regression equation is listed in order of decreasing $\theta_n$.  Though in this case, the regression only includes redshifts measured at $r_{obs}$ equal to at least $10\, R$, the photon's starting radius, so as to exclude points in the initial bend.}
\end{figure*}

The simulation results were again similar to those for the flat and open universes but with several slight differences this time.  The graphs of $z_{\scriptscriptstyle LW}$ against $z_{\scriptscriptstyle FLRW}$ again followed straight lines but the gradient was slightly different for when the photon was outgoing and when it was ingoing, as shown in \FigRef{fig:closed_subset}.  As the figure also shows, the graphs did not necessarily intercept the origin either.  Several of the graphs for ingoing photons also had a subtle bend in their low $z$ end.  We believe this bend to be analogous to the initial curve we saw in the flat universe graphs previously: it corresponds to redshifts measured at low $r_{obs}$ and is more pronounced in the more radial trajectories, that is, the smaller $\theta_n$ trajectories, which pass closer to the central mass; thus, like the initial curve in the flat universe graphs, we believe the bend is caused by the central mass' stronger influence at low $r_{obs}$.\footnote{We note that the earlier analysis of the dependence of $\nu_{\scriptscriptstyle LW}$ on $r_{obs}$, given by function \eqref{analytic-single-cell-freq}, would require some modification in this context.  In particular, as we are now using LW co-ordinates, function \eqref{analytic-single-cell-freq} would change, with parameter $E_b$ replaced by a parameter $E_o$ corresponding to the constant $E$ for the co-moving observer's geodesic.  The new function though would still have the same functional form as the original and therefore the same general behaviour.  However since the radius of the observer's geodesic is bounded, the new function's domain is bounded from above by $r_{obs} \leq 2m/(1-E_o)$, otherwise the square-root in the function would turn imaginary; therefore there cannot be any large-$r_{obs}$ asymptotic regime in the new function, and hence this aspect of our previous analysis is not transferable to the closed universe.  Nevertheless, we believe that the general conclusion still applies, that the initial jump and the bend in $z_{\scriptscriptstyle LW}$ vs $z_{\scriptscriptstyle FLRW}$ graphs is caused by the central mass' stronger influence at low $r_{obs}$.}  In \FigRef{fig:closed_outgoing}, we show graphs of $z_{\scriptscriptstyle LW}$ against $z_{\scriptscriptstyle FLRW}$ for when the photons are outgoing, and in \FigRef{fig:closed_ingoing}, we show the corresponding graphs for when the same photons are ingoing.


\begin{figure}[tb]
{\fontsize{8pt}{9.6pt}\input{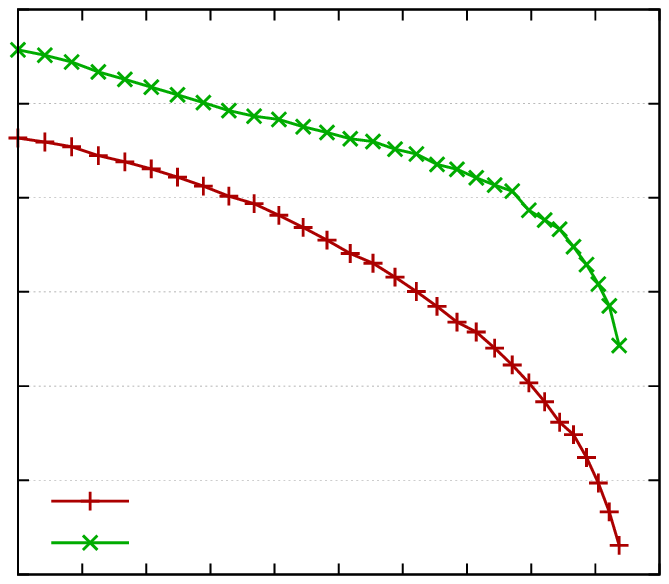}}
\caption{\label{fig:closed_grads}A plot of $z_{\scriptscriptstyle LW}$ vs $z_{\scriptscriptstyle FLRW}$ gradients against $\cos(\theta_n)$, where $\theta_n$ is the initial direction of the photon.   The gradients for when the photons are outgoing and for when they are ingoing are plotted separately, and all gradients are obtained by performing a linear regression on $z_{\scriptscriptstyle LW}$ vs $z_{\scriptscriptstyle FLRW}$ data.  To avoid the initial bend in the $z_{\scriptscriptstyle LW}$ vs $z_{\scriptscriptstyle FLRW}$ graphs for ingoing photons, the ingoing gradients are obtained by regressing over only data points for which $r_{obs}$ is at least equal to the photon's starting radius of $10\, R$.}
\end{figure}

In \FigRef{fig:closed_grads}, we have plotted the gradients of $z_{\scriptscriptstyle LW}$ vs $z_{\scriptscriptstyle FLRW}$ against $\cos(\theta_n)$.
Two graphs are shown, one for the gradients when the photons are outgoing, and the other for when they are ingoing.  To exclude the initial bend in the graphs for ingoing photons, we excluded data points corresponding to $r_{obs}$ less than $10\, R$, the photons' starting radius.\footnote{In fact, including all data points in the regression introduced some `jaggedness' in the graph of ingoing gradients, especially towards the low $\cos(\theta_n)$ regime.  By excluding these points, the `jaggedness' smoothed away, and we were left with a curve much more similar to that for the outgoing gradients.}  The figure shows that, as with the other universes, the gradient decreases as the photon's trajectory becomes more radial, but now, the relationship between the gradients and $\cos(\theta_n)$ is no longer linear.

Thus to summarise, our simulations have demonstrated several features common to redshifts in all LW universes.  The LW redshifts generally increase linearly with their FLRW counterparts in any universe and for any photon trajectory.  When $r_{obs}$ is small such that the central mass has a stronger influence, there is some deviation away from this linear behaviour.  Any influence from the central mass can be suppressed by starting the photon at very large radii; $z_{\scriptscriptstyle LW}$ then becomes completely proportional to $z_{\scriptscriptstyle FLRW}$.  The LW redshifts also generally decrease relative to their FLRW counterparts as the photon takes a more radial trajectory; this can also be attributed to the stronger influence of the central mass since a more radial trajectory would pass closer to it.  The mass' influence on redshifts can sometimes be quite significant, causing LW redshifts to deviate from their FLRW counterparts by as much as 30\%.  This would have significant implications if we attempted to fit redshift data from a lattice universe onto an FLRW model, as quantities such as the age of the universe may be incorrectly estimated.  Thus we see certain significant effects arising from the `lumpiness' of the LW universe which an FLRW-based model could not adequately capture.

This influence of the masses on redshifts can be understood as an integrated Sachs-Wolfe effect (ISW).  The ISW describes the net redshift induced on a photon as it passes through a fluctuating gravitational potential caused by fluctuations in the energy density \citep{SachsWolfe, ReesSciama, MSS, Nishizawa}.  In a flat FLRW universe, a time-varying potential is necessary to generate a non-zero ISW; however when the universe is matter-dominated, the potential is static, and therefore the ISW would be zero.  If there is a non-zero cosmological constant though, the potential in a matter-dominated universe does become time-varying, and hence a non-zero ISW can result \citep{CrittendenTurok}.  Indeed, recent precision observations of the cosmos have not only established that our universe is very nearly flat but that there is definitely a non-zero ISW \citep{BoughnCrittenden, Planck2015}; naturally, this has been interpreted as evidence of a non-zero cosmological constant in our universe.  However, our results suggest that this might be explainable within a matter-dominated universe without needing a cosmological constant; we may simply require a model that better reflects the inhomogeneous matter distribution than FLRW space-times allow.  Because of the significant implications a non-zero ISW may have on the dark energy question, this modelling problem deserves further investigation.

\section{Discussion}
We have investigated the properties of the Lindquist-Wheeler universes, as we hope they might provide some insight into what observable effects the `lumpy' matter distribution of the actual universe might yield.  And although the LW universes are only approximations rather than exact solutions to the Einstein equations, we believe they model enough of the underlying physics to yield at least meaningful qualitative insights into the behaviour of the actual universe.  Much of the LW universes' dynamics bear strong resemblance to those of the matter-dominated FLRW universes.  Additionally, photon redshifts in LW universes behaved roughly similarly to their FLRW counterparts.  Yet there were also certain direction-dependent effects in the redshifts due to the `lumpiness' of the universe's matter distribution.

We noted that there were some differences between Clifton and Ferreira's results and ours for the flat universe.  Most notably, our redshifts displayed the effects of the inhomogeneous matter distribution more.  The effects became stronger the closer the photon passed to a mass and the more radial the photon's trajectory was.  As a result, our redshifts demonstrated a non-zero ISW, which was not present in Clifton and Ferreira's results.  Otherwise, when the photon was far from any mass, its redshift would behave much more similarly to FLRW redshifts as well as to Clifton and Ferreira's mean redshift relation.  The main difference between Clifton and Ferreira's model and ours was the definition of the cell boundary and the choice of conditions for propagating particles across boundaries.  As Clifton and Ferreira have shown, these can significantly alter the redshift relations; for instance, using a different choice of boundary conditions, Clifton and Ferreira obtained the redshift relation of $1 + z_{\scriptscriptstyle LW} \approx (1 + z_{\scriptscriptstyle FLRW})^{7/10}$ instead \citep{CF, CFO}, which clearly agreed far less with FLRW redshifts.  Further investigation is required to determine whether our discrepancy from Clifton and Ferreira's results is indeed significant or merely an artefact of our boundary conditions which somehow exaggerates the effects of the masses.  Nevertheless, neither our inhomogeneous results nor Clifton and Ferreira's more homogeneous one can be ruled out at the moment.

There are several ways in which our investigation can be extended.  It would be interesting to examine the optical properties of the LW universe, as this may have important implications for actual astrophysical observations.  Clifton and Ferreira have already done this for their implementation of the LW universe, but we have used a different implementation from theirs, and this has led to certain differences in the behaviour of redshifts.  However, Clifton and Ferreira have suggested that the optical properties, unlike the redshifts, may actually be insensitive to the choice of boundary conditions \citep{CFO}.  Nevertheless, it would be very interesting to see whether that is indeed the case for our boundary conditions.  It would also be interesting to extend our study to LW universes with a non-zero cosmological constant and attempt to evaluate by how much inhomogeneities reduce the need for such a constant.  Clifton and Ferreira have already constructed an appropriate extension of the LW universe based on the Schwarzschild de-Sitter metric, and they have shown that the corresponding Friedmann-like equation strongly resembles its FLRW counterpart as well.  Moreover, using their boundary conditions, they have also shown that the cosmological constant density ratio $\Omega_\CosmoConst$ can be reduced by about 10\% \citep{CF-Lambda, CFO}, and again this was rather insensitive to which boundary condition they used.  It should be possible to include Clifton and Ferreira's $\CosmoConst$-Schwarzschild-cells in our implementation and to investigate the resulting model.  Finally, our model has a still very idealised distribution of matter.  Each mass is identical and distributed on a perfect lattice which is clearly not the case in the actual universe.  We should like to extend our model to allow for different sized masses and cells.  To this end, we have derived in \appendref{Inhomogeneities} a set of conditions that must be satisfied at the boundary between two neighbouring cells of different mass $m$.  We leave the numerical simulation of such universes and the detailed investigation of their properties to future work.

\begin{acknowledgments}
The author would like to thank Ruth Williams, Timothy Clifton, Pedro Ferreira, and Ulrich Sperhake for much helpful discussion.  Timothy and Ulrich pointed out to the author the resemblance between the paper's redshift results and the integrated Sachs-Wolfe effect.  The numerical simulations of the Lindquist-Wheeler models were performed on the COSMOS Shared Memory system at DAMTP, University of Cambridge, operated on behalf of the STFC DiRAC HPC Facility. This equipment is funded by BIS National E-infrastructure capital grant ST/J005673/1 and STFC grants ST/H008586/1, ST/K00333X/1, and ST/J001341/1.  The author also wishes to thank the COSMOS team, in general, for technical support and James Briggs, in particular, for his kind assistance in parallelising the C++ code with OpenMP.  The author was financed in part by a bursary from the Cambridge Commonwealth Trust.
\end{acknowledgments}

\appendix

\setcounter{section}{0}
\renewcommand{\thesection}{\Alph{section}}
\titleformat{\section}[block]
  {\normalfont\bfseries}{APPENDIX \thesection:}{1em}{\centering \MakeUppercase{#1}}
  

\section{Regular lattices in 3-spaces of constant curvature}
\label{App_Latt}
In this appendix, we shall list all possible lattices that cover 3-spaces of constant curvature with a single regular polyhedral cell.  The cell is tiled to completely cover the 3-space without any gaps or overlaps.  This tessellation problem has been thoroughly studied by Coxeter \citep{Coxeter}.  Clifton and Ferreira \citep{CF} have succinctly summarised Coxeter's results relevant to our discussion, and we have presented their summary in \tabref{tab:Tab_Latt}.
\begin{table} [htbp]
\caption{\label{tab:Tab_Latt}All possible lattices obtained by tessellating 3-spaces of constant curvature with a single regular polyhedron.}
\begin{tabular*}{8.6cm}{@{\extracolsep{\fill} } S N B T }
\hline\hline
\textbf{Elementary cell shape} & \textbf{Number of cells at a lattice edge} & \textbf{Background curvature} & \textbf{Total cells in lattice} \\
\hline
tetrahedron & 3 & + & 5 \\
cube & 3 & + & 8 \\
tetrahedron & 4 & + & 16 \\
octahedron & 3 & + & 24 \\
dodecahedron & 3 & + & 120 \\
tetrahedron & 5 & + & 600 \\
cube & 4 & 0 & $\infty$ \\
cube & 5 & - & $\infty$ \\
dodecahedron & 4 & - & $\infty$ \\
dodecahedron & 5 & - & $\infty$ \\
icosahedron & 3 & - & $\infty$ \\
\hline\hline
\end{tabular*}
\end{table}

Column 1 gives the lattice cell's shape.  Column 2 effectively determines the lattice's structure: it indicates how many of Column 1's elementary cells meet at any lattice edge.  Column 3 indicates whether the background 3-space has positive, flat, or negative curvature.  Column 4 gives the number of cells needed to cover the 3-space; only lattices on positively curved space can have a finite number of cells.  To construct a closed lattice universe, one has six choices of lattices; for example, one can choose a 600-cell lattice where cells are equilateral tetrahedra and where five tetrahedra meet at any edge.  To construct a flat lattice, one has but a single choice.  To construct an open lattice, one has four.

\section{Radial velocities in Regge space-time}
\label{BV}

\begin{figure*}[tb]
{\fontsize{8pt}{9.6pt}\input{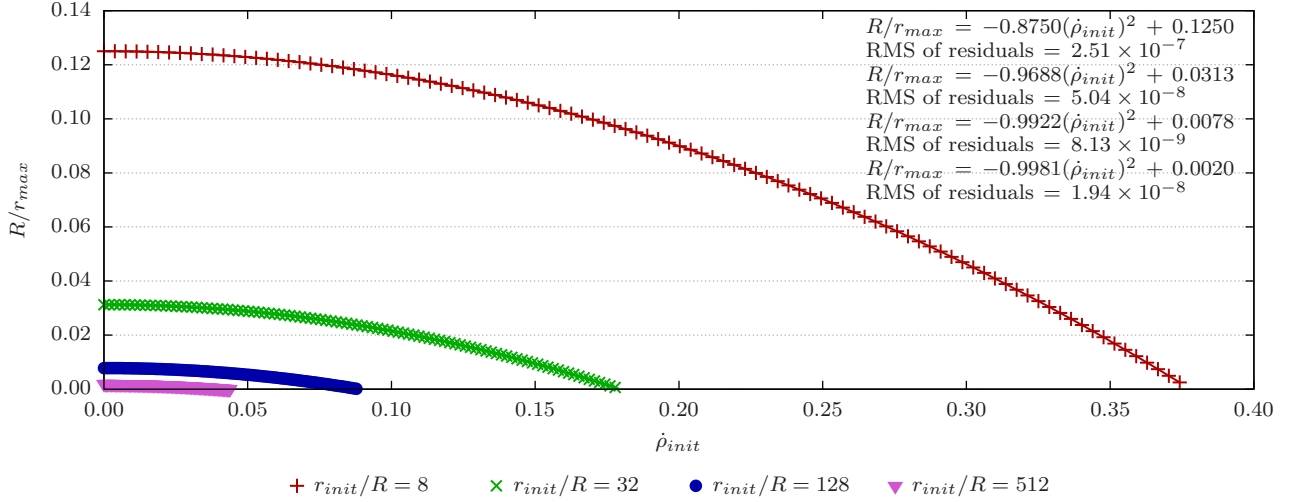}}
\caption{\label{fig:esc_var_r}A plot of $R/r_{max}$ versus $\dot{\rho}_{init}$ for various $r_{init}$.  Each graph has 100 data points.  A quadratic regression has been performed on each graph, and the regression equations are ordered from $r_{init} = 8\, R$ to $512\, R$.  The regressions were performed without a linear term; regression was also attempted with a linear term present, but it was found to be many orders of magnitude smaller than the other two terms.  The simulation's block parameters were $R=31$, $\Delta t=10\Delta r$, $\Delta r = R/10^5$, and $\Delta \phi = 2\pi / (3\times 10^7)$.}
\end{figure*}

\begin{figure*}[tb]
{\fontsize{8pt}{9.6pt}\input{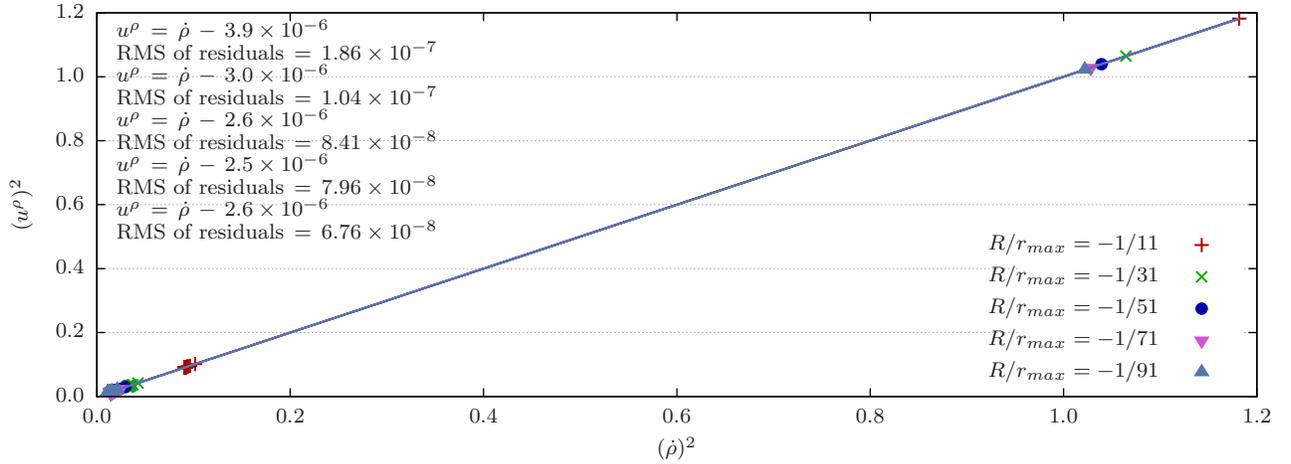}}
\caption{\label{fig:esc_open}A plot of $(u^\rho)^2$ against $\dot{\tilde{\rho}}^{\,2}$ for various $r_{max}<0$, where $u^\rho$ is given by \eqref{Reg_bound_vec}.  Each plot consists of velocities computed at 100 different radii along the course of the test particle's trajectory.  A linear regression was performed on each graph, and the corresponding equations are ordered from $R/r_{max} = -1/11$ to $-1/91$.  The graphs of the regressions completely overlap each other.  The gradients only begin deviating from unity at the $10^{-5}$ order of magnitude.  The simulation's block parameters were $r_{init} = 2\, R$, $R=21$, $\Delta t=\Delta r = R/10^5$, and $\Delta \phi = 2\pi / (3\times 10^7)$.}
\end{figure*}

In this appendix, we shall present our numerical results supporting \eqref{Reg_bound_vec} to be the 4-velocity tangent to radial time-like geodesics in Regge Schwarzschild space-time.  These geodesics include, most importantly, those of test particles co-moving with a Schwarzschild-cell boundary.  We began by determining numerically the escape velocity for a test particle.  We propagated a test particle radially outwards from a series of initial radii $r_{init}$ and with a series of initial velocities $\dot{\rho}_{init}$.  We started $\dot{\rho}_{init}$ from 0 and increased it until the maximum radius $r_{max}$ attained by the particle was very large.  For each $r_{init}$, we then plotted $R/r_{max}$ against its corresponding $\dot{\rho}_{init}$, where $R$ is the Schwarzschild radius.  Both $r_{init}$ and $r_{max}$ refer to the lower Schwarzschild label of the block in which the particle is found.  Our plots are shown in \FigRef{fig:esc_var_r}.  Each graph is very well-fitted by a quadratic curve of the form
$$
\frac{R}{r_{max}} = -A\, \dot{\rho}_{init}^2 + B.
$$
Moreover, the co-efficients $A$ and $B$ always satisfy the relations $A+B = 1$ and $B = R/r_{init}$.\footnote{We have also looked at graphs for different $R$ and $r_{init}$, not shown, and they also conform to this pattern.}  From these relations, we can therefore infer a relation for $\dot{\rho}$ corresponding to that given in \eqref{Reg_bound_vec}, and the $\dot{\tau}$ component follows from normalisation.

\begin{figure}[tb]
{\fontsize{8pt}{9.6pt}\input{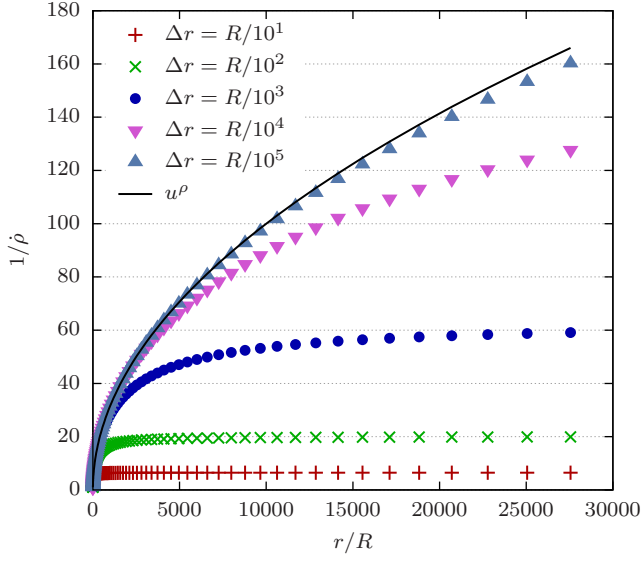}}
\caption{\label{fig:flat_err}A plot of $1/\dot{\rho}$ versus $r/R$ comparing simulated particle velocities with $u^\rho$.  For the simulated particles' graphs $\dot{\rho} = \dot{\tilde{\rho}}$, while for the $u^\rho$ graph, $\dot{\rho} = u^\rho$.  The simulation was run for various sizes $\Delta r$ of the Regge block.  For all simulations, the particle was propagated from $r = 2\, R$ to $3\times 10^4 \, R$.  The simulation's other block parameters were $R=1$, $\Delta t = \Delta r$, and $\Delta \phi = 2\pi / (3\times 10^7)$.}
\end{figure}

We have just provided numerical support for \eqref{Reg_bound_vec} for test particles following geodesics where $r_{max} \geq 0$, corresponding to closed orbits or orbits at the escape velocity.  At this point, we conjectured that for open orbits, \eqref{Reg_bound_vec} would still apply but with $r_{max} < 0$.  To test this, we propagated a test particle outwards with an initial velocity given by \eqref{Reg_bound_vec} but for $r_{max} < 0$, and we examined how the velocity evolved with $r$.  Let us denote the velocity of our simulated particle by $\dot{\tilde{\rho}}$.  Along the particle's trajectory, we compared $\dot{\tilde{\rho}}^{\,2}$ against $(u^\rho)^2$ for the same radius, where $u^\rho$ is given by the radial component of \eqref{Reg_bound_vec}.  Our comparison is presented graphically in \FigRef{fig:esc_open}.  In all cases, the graphs' gradients were effectively unity and the constants effectively zero, thus indicating that the particle's velocity does indeed obey \eqref{Reg_bound_vec}.  In \tabref{tab:esc_open_num}, we provide a partial list of the values of $\dot{\tilde{\rho}}^{\,2}$ and $(u^\rho)^2$ as well as their percentage difference.  The percentage difference is very small in all rows, although there appears to be an increasing trend with increasing $r$.  We find that our conjecture is supported by numerical results, and hence \eqref{Reg_bound_vec} appears valid for all types of radial time-like geodesics.\footnote{Again, we have also simulated the cases of $r_{max} = -R,- 21 \, R,- 41 \, R,- 61 \, R,$ and $-81 \, R$, not shown here, and they all display the same behaviour as in \FigRef{fig:esc_open}.  The percentage difference seen between $\dot{\tilde{\rho}}^{\,2}$ and $(u^\rho)^2$ is also small but slowly increasing in all $r_{max}$ that we looked at.}

\begin{table*}[tbhp]
\caption{\label{tab:esc_open_num}A list of $\dot{\tilde{\rho}}^{\,2}$ and $(u^\rho)^2$ at various radii, and the percentage difference between them.  This data is for $R/r_{max} = -91$.}
\renewcommand{\arraystretch}{1.2}
\begin{tabular*}{14cm}{@{\extracolsep{\fill} } c d{12} d{16} d{17} }
\hline\hline
$r/R$ & \multicolumn{1}{c}{$\dot{\tilde{\rho}}^{\,2}$} & \multicolumn{1}{c}{$(u^\rho)^2$} & \multicolumn{1}{c}{\% difference} \\
\hline
2 & 1.0219794649 & 1.02197802197802 & 0.00014118894047312\\
102 & 0.021001226724 & 0.0209988015595972 & 0.0115477273528382\\
202 & 0.016021230625 & 0.0160188121184031 & 0.0150956356191602\\
302 & 0.014350362849 & 0.0143477802197414 & 0.0179969613713785\\
402 & 0.013512667536 & 0.0135101783683129 & 0.0184209940820187\\
502 & 0.013009455481 & 0.0130069512260618 & 0.0192494985037111\\
602 & 0.012673806084 & 0.0126711894819952 & 0.0206457475160902\\
702 & 0.012433811049 & 0.0124312192608481 & 0.0208446802168433\\
802 & 0.012253604416 & 0.0122511667237755 & 0.0198936748870944\\
$\vdots$ & \multicolumn{1}{c}{$\vdots$} & \multicolumn{1}{c}{$\vdots$} & \multicolumn{1}{c}{$\vdots$} \\
9202 & 0.011101572496 & 0.0110988891274903 & 0.0241710668530401\\
9302 & 0.011100308164 & 0.0110977080370158 & 0.0234239171181774\\
9402 & 0.011099254609 & 0.0110965512181986 & 0.02435650768163\\
9502 & 0.011097990409 & 0.0110954196820738 & 0.023163895728025\\
9602 & 0.011096936964 & 0.0110943114060724 & 0.0236602040377857\\
9702 & 0.011095883569 & 0.0110932263900415 & 0.0239474300718204\\
9802 & 0.011094830224 & 0.0110921626114406 & 0.0240437438476478\\
9902 & 0.011093776929 & 0.011091121081329 & 0.0239399772323361\\
\hline\hline
\end{tabular*}
\end{table*}

\begin{figure}[tbh]
{\fontsize{8pt}{9.6pt}\input{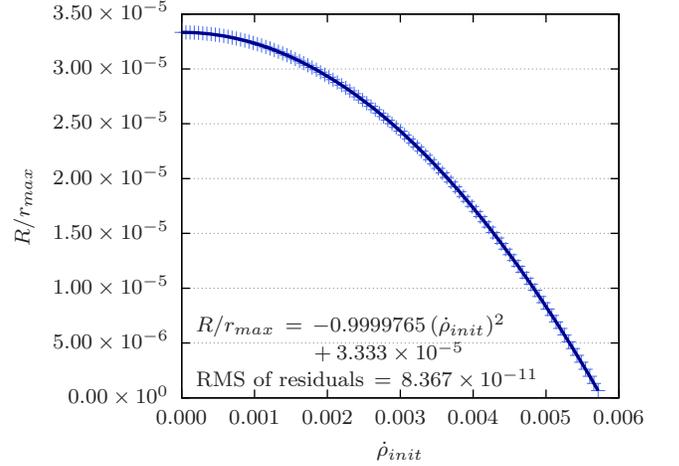}}
\caption{\label{fig:esc_large_r}A plot of $R/r_{max}$ versus $\dot{\rho}_{init}$ for $r_{init} = 3\times 10^4 \, R$.  A quadratic regression has been performed and the corresponding equation shown.  The simulation's block parameters were $R=1$, $\Delta t=10\Delta r$, $\Delta r = R/10$, and $\Delta \phi = 2\pi / (3\times 10^7)$.}
\end{figure}

We have just found that the initial velocity $\dot{\rho}_{init}$ of a simulated particle must equal at least that of \eqref{Reg_bound_vec} in order for the particle to `escape'.  We next examined the long-term behaviour of the particle's velocity $\dot{\tilde{\rho}}$ as the particle propagated outwards from an initial velocity equalling at least the escape velocity.  We found that $\dot{\tilde{\rho}}$ would not stay equal to $u^\rho$ as given by \eqref{Reg_bound_vec} but would instead decrease at a slower rate as a function of the particle's radius.  For example if we started a particle at the escape velocity at $r_{init} = 2\, R$ and propagated it outwards, then by the time it reached $r = 3\times 10^4 \, R$, there was a very significant discrepancy between $\dot{\tilde{\rho}}$ and $u^\rho$, as shown in \FigRef{fig:flat_err}, with $\dot{\tilde{\rho}}$ being consistently larger than $u^\rho$.  This discrepancy reduced if we improved the radial resolution $\Delta r$ of our Regge blocks, but it was still present even for resolutions as high as $\Delta r = R/10^5$.  If instead, using the same method as previously, we numerically determined the escape velocity for a particle starting at $r_{init} = 3\times 10^4 \, R$, we again obtained the same quadratic behaviour between $R/r_{max}$ and $\dot{\tilde{\rho}}$ as that in \FigRef{fig:esc_var_r}, thus indicating the escape velocity should still correspond to \eqref{Reg_bound_vec}.  Indeed, this result was obtained even when the radial resolution was as low as $\Delta r = R/10$, as shown in \FigRef{fig:esc_large_r}, even though there is a huge discrepancy at this resolution between $\dot{\tilde{\rho}}$ and $u^\rho$ in \FigRef{fig:flat_err}.  Therefore even if a particle began at the correct escape velocity for its initial radius, it would gradually travel faster than the correct escape velocity for its subsequent radii as it propagated along its trajectory.

\begin{figure}[tbh]
{\fontsize{8pt}{9.6pt}\input{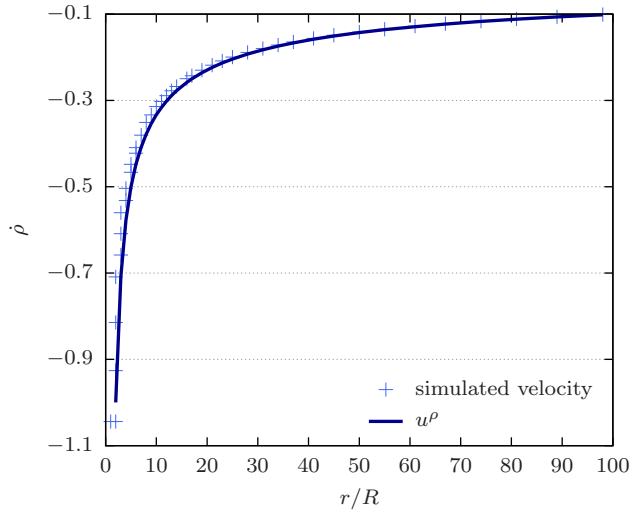}}
\caption{\label{fig:esc_incoming}A plot of $\dot{\tilde{\rho}}$ versus $r/R$ comparing a simulated particle velocity with $u^\rho$.  The simulation began at $r_{init} = 3\times 10^4 \, R$, but only data points from $r=2\, R$ to $100\, R$ are shown.  The simulated velocity is always greater than or equal to $u^\rho$.  The simulation's block parameters were $R=1$, $\Delta t = \Delta r = R$, and $\Delta \phi = 2\pi / (3\times 10^7)$.}
\end{figure}

The discrepancy between $\dot{\tilde{\rho}}$ and $u^\rho$ can be understood by considering the manner by which particles are deflected in the Regge model.  Suppose a particle enters a Regge block of radius $r_i$ by the $-\rho_i$ face and travels at the escape velocity as given by \eqref{Reg_bound_vec}.  If the particle then enters the next block by the $-\rho_{i+1}$ face as well, it will have undergone no deflection and hence continue having the same velocity as at $r_i$.  However, \eqref{Reg_bound_vec} indicates that the escape velocity at $r_{i+1}$ should be smaller than that at $r_i$, and hence we see the beginning of a discrepancy.  If the particle passes through a sizeable number of $\rho$ faces in succession, it would accumulate a significant discrepancy such that the particle's velocity would be noticeably larger than the escape velocity for its radius.  This explains why the graphs for simulated particles in \FigRef{fig:flat_err} are always smaller than the graph for $u^\rho$.  Only when the particle crosses a $\tau$ surface does it undergo a change in velocity, but this change may not be enough to overcome the discrepancy already accumulated.  By a similar argument, we expect the radial velocity of a radially in-going particle to decrease less quickly than $u^\rho$ in \eqref{Reg_bound_vec}, because if the particle crosses a $+\rho$ face with the correct escape velocity, it will cross the neighbouring $+\rho$ face with the same unchanged velocity, which would be less negative than $u^\rho$ for the same radius; and indeed, this is what we see in \FigRef{fig:esc_incoming}.  By improving the resolution of our Regge model, we slow the growth rate of the discrepancy, as we saw in \FigRef{fig:flat_err}, because our model becomes a better approximation to the underlying continuum Schwarzschild space-time and its geodesics therefore become more similar to those in the continuum space-time.

Thus combined with our analytic arguments provided at the end of \secref{implem_sect}, we conclude that \eqref{Reg_bound_vec} gives the correct velocity as a function of $r$ for particles following radial time-like geodesics.

\section{Radial lengths of Schwarzschild Regge blocks}
\label{RadLen}
In our simulations of the flat and open universes, our Regge blocks' radial lengths were not constant but were increased as the blocks were further away from the Schwarschild-cell centre.  This decision was motivated by the fact that the underlying space-time being approximated becomes increasingly flat.  We therefore attempted to maintain higher resolution while the curvature was high but then decrease the resolution with minimal impact on accuracy as the curvature decreased.  This technique allowed us to perform larger-scale simulations within more attainable computation times.  We now describe our method for specifying the block's radial length.

\begin{figure}[htb]
{\fontsize{8pt}{9.6pt}\input{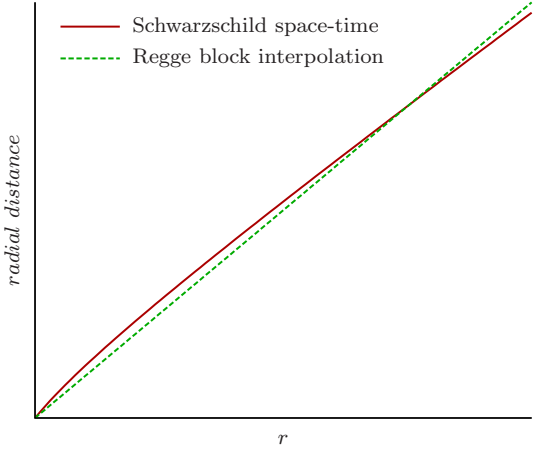}}
\caption{\label{fig:interpolation}The Regge block interpolation intersects the Schwarzschild graph at $r=r_i$ and $r=r_{i+1}$, where $r_i < r_{i+1}$ and where $r_i$ corresponds to the origin of the graph; it interpolates the radial distance for only those values of $r$ lying between these two points of intersection.  Since the Schwarzschild distance function is convex, the interpolation will always underestimate the distance in this regime.}
\end{figure}

In Schwarzschild space-time, the radial distance between radius $r_i > 2m$ and some arbitrary $r > r_i$ is given by
\begin{equation}
d(r) = \left[x \sqrt{1 \! - \! \frac{2m}{x}} - 2m \ln \left( \! \sqrt{\frac{x}{2m}} \! - \! \sqrt{\frac{x}{2m} \! - \! 1}\; \right) \! \right]^{r}_{x\,=\,r_i}.
\end{equation}
When $r=r_{i+1}$, this is identical to \eqref{rad_dist}.  Distances along `radial edges' of Regge blocks approximate radial distances in Schwarzschild space-time by linearly interpolating $d(r)$ between $r_i$ and $r_{i+1}$, as shown in \FigRef{fig:interpolation}.  Since $d(r)$ is convex, the interpolation will always underestimate the true distance, and the error $\varepsilon$ of the approximation is given by $\varepsilon = \max \left[ d(r) - \lin(r) \right]$, where $\lin(r)$ denotes the interpolating function.

Our goal is as follows.  Given some error-tolerance $\tol$, we want our blocks to be as long as possible in the radial direction while still satisfying $\varepsilon \le \tol$.  If we enter a new block from its left or right face, then we would need to re-calculate the block's radial length $h = r_{i+1} - r_i$ in accordance with our goal.  It will always be the case that we know one of $r_i$ or $r_{i+1}$ depending on which face we entered by, and we would need to deduce the other quantity.  However our computation must give consistent results for $h$ regardless of which quantity we used.  In our particular implementation, we have also chosen to impose the constraint that $h$ must be some integer units of a minimal interval $\Delta r_0$, a parameter which we can freely specify.

The most natural approach would be to solve $d^\prime(r) = \lin^\prime(r)$, which maximises $d(r) - \lin(r)$, to obtain $r = r_{max}$, the point at which the error is greatest.  Also through its dependence on $\lin(r)$, $r_{max}$ would be a function of $h$.  We would then set $\varepsilon = \tol$ and solve $d(r_{max}) - \lin(r_{max}) = \tol$ for $h$.  However, this second equation can only be solved numerically, and from a programming point of view, it was easier to implement a different approach instead.

We note that for $r > r_i$, the functions $d(r)$ and $\lin(r)$ are bounded from above by
\begin{equation}
d^+(r) = \sqrt{\frac{r_i}{r_i-1}} (r - r_i) + d(r_i), \label{bound_up}
\end{equation}
since this is just the equation of the tangent to $d(r)$ at $r_i$.  And also, for $r > r_i$, the two functions are bounded from below by
\begin{equation}
d^-(r) = (r - r_i) + d(r_i); \label{bound_down}
\end{equation}
this follows because the gradient of $d(r)$ goes asymptotically to unity from above as $r \to \infty$, which means $d(r)$ will always increase more quickly than $d^-(r)$.  Therefore, after the two functions have intercepted at $r=r_i$, $d(r)$ will always be strictly greater than $d^-(r)$.  The function $\lin(r)$ also intercepts both $d(r)$ and $d^-(r)$ at $r=r_i$.  But it intercepts $d(r)$ again at $r_{i+1} > r_i$, a point where $d(r) > d^-(r)$.  Because $\lin(r)$ is a linear function, this interception at $r_{i+1}$ implies that $\lin(r)$ must have a greater gradient than $d^-(r)$, and therefore $\lin(r)$ must also be strictly greater than $d^-(r)$ in the region $r > r_i$.

Thus we can bound our error by 
$$\varepsilon \leq \max \left[ d^+(r) - d^-(r) \right] = d^+(r_{i+1}) - d^-(r_{i+1}).$$
We shall therefore require that
$$
\tol \geq d^+(r_{i+1}) - d^-(r_{i+1}),
$$
and solve for $r_{i+1}$ from this equation; this yields the solution
$$
h = \frac{\tol}{\sqrt{r_i / (r_i-1)}-1}.
$$

To get $h$ as a function of $r_{i+1}$ instead, we set $r_i$ to be $r_i = r_{i+1} - h$ and solve for $h$ in the preceding equation; this yields
\begin{widetext}
$$
h = \frac{\tol}{2(2\tol - 1)} \left\{ 2 r_{i+1} + \tol-2 \vphantom{\Big]^{1/2}} + \Big[4r_{i+1}(r_{i+1}-1-\tol) + \tol (4+\tol)\Big]^{1/2} \right\}.
$$
\end{widetext}

As mentioned above, we want $h$ to satisfy $h = n\, \Delta r_0$ for some integer $n$.  Therefore our actual $h$ is obtained by taking $n$ to be
\begin{equation}
n = \left\lfloor \frac{\tol}{\Delta r_0 \left(\sqrt{r_i / (r_i-1)}-1\right)} \right\rfloor
\end{equation}
or
\begin{equation}
\begin{split}
n = \left\lfloor \frac{\tol}{2 \Delta r_0 (2\tol - 1)} \right. & \left\{ 2 r_{i+1} + \tol-2 \vphantom{\Big]^{1/2}} \right.\\
& \hphantom{\{} {} + \Big[4r_{i+1}(r_{i+1}-1-\tol) \\
& \left. \left. \hphantom{(2 r} {}+ \tol (4+\tol)\Big]^{1/2} \right\} \right\rfloor.
\end{split}
\end{equation}
as appropriate, where $\lfloor x \rfloor$ gives the greatest integer less than or equal to $x$, and then substituting this $n$ into $h = n\, \Delta r_0$.

Finally, we want to ensure that our algorithm would give the same $h$ regardless of whether it used $r_i$ or $r_{i+1}$.  To satisfy this requirement, whenever the program calculated $r_{i+1}$ from $r_i$, it would check to see if it could recover the same $h$ using the new value for $r_{i+1}$.  If not, then it would decrement $h$ by one unit of $\Delta r_0$ and check again, and it would continue decrementing until obtaining agreement.  The program can only decrement if the error is to remain within tolerance.  We had also required that $h$ be at least $\Delta r_0$; thus the decrementing would stop if $h$ reached this length.

\section{Boundary conditions for Schwarzschild-cells of unequal masses}
\label{Inhomogeneities}
An observer sitting at the interface of two cell boundaries must observe the same physics regardless of which co-ordinate system the observer uses.  In particular, the following two conditions must be satisfied:
\begin{enumerate}[label=\arabic{*}), topsep=2.5mm]
\item the observer must measure the same proper time regardless of which cell's metric is used; \label{bcond1}
\item and the observer must measure the same spatial distances locally along the boundary regardless of which cell's metric is used. \label{bcond2}
\end{enumerate}
However if the cells in the lattice are no longer identical, we would no longer have the same lattice symmetries as before.  In particular, the exact location of the boundary between two unequal cells becomes less transparent, as it is no longer equidistant to the two cell centres.  Recall that in the perfectly symmetric lattice, a test particle sitting on the boundary between two cells would by symmetry always remain at the boundary and yet fall simultaneously towards both cell centres.  Although the symmetry is no longer present when the two cells are no longer identical, we shall still assume that test particles at the boundary fall simultaneously to both centres.  This condition defines the location of the cell boundary for us: it is effectively defined to be where the two cells' gravitational influences are equal.  On one side of the boundary, the gravitational influence of one mass dominates, and we approximate the space-time by a Schwarzschild space-time centred on that mass.  On the other side, the other mass dominates, and we approximate the space-time with a Schwarzschild space-time centred on that mass.  Once again, we can understand this simultaneous free-fall of the boundary towards the two centres as actually the motion of the two masses towards each other under their mutual attraction.  This mutual attraction gives rise to the expansion and contraction of the lattice itself, which manifests as the expansion and contraction of the cell boundary.  Given this definition and our two boundary conditions, we can now derive a set of constraints that $E_b$ and the cell radius $r_b$ must satisfy.

Local spatial distances along the boundary are given by
\begin{equation}
r_b\, d\Omega.
\end{equation}
Since by assumption, particles co-moving with the boundary follow radial geo\-desics, then this distance evolves as
\begin{equation}
\frac{dr_b}{d\tau}\, d\Omega.
\end{equation}
By condition \enumref{bcond2}, we require that 
\begin{equation}
r_{b_1}\, d\Omega_1 = r_{b_2}\, d\Omega_2,
\end{equation}
where the numerical subscripts refer to cells 1 and 2; combined with condition \enumref{bcond1}, we additionally require that
\begin{equation}
\frac{dr_{b_1}}{d\tau}\, d\Omega_1 = \frac{dr_{b_2}}{d\tau}\, d\Omega_2,
\end{equation}
since $d\tau_1 = d\tau_2 = d\tau$.  Combining these two conditions, we obtain
\begin{equation}
\frac{1}{r_{b_1}} \frac{dr_{b_1}}{d\tau} = \frac{1}{r_{b_2}} \frac{dr_{b_2}}{d\tau},
\label{bcond2_a}
\end{equation}
and using \eqref{rad_geod}, we can express this equivalently as
\begin{equation}
\frac{1}{r_{b_1}} \sqrt{E_{b_1} - 1 + \frac{2m_1}{r_{b_1}}} = \frac{1}{r_{b_2}} \sqrt{E_{b_2} - 1 + \frac{2m_2}{r_{b_2}}}.
\label{bcond2_b}
\end{equation}

By condition \enumref{bcond1}, we require that $\Delta \tau_1 = \Delta \tau_2$.  Recall that the proper time of a freely falling particle is given by equations \eqref{tau_closed} to \eqref{r_E}.  However, from the form of equations \eqref{tau_closed} to \eqref{tau_open}, it is clear that we cannot satisfy this condition unless both cells are of the same type, that is, both cells are open, flat, or closed.  This constrains the cases we need to consider to just three, and we shall consider each in turn.

If $E_b=1$ for both cells, then \eqref{tau_flat} and \eqref{r_E} imply that
\begin{align*}
\Delta \tau_1 &= \Delta \tau_2\\
\frac{2}{3}\frac{1}{\sqrt{2m_1}}\, r_{b_1}^{3/2} - \tau_0 &= \frac{2}{3}\frac{1}{\sqrt{2m_2}}\, r_{b_2}^{3/2} - \tau_0,
\end{align*}
where $\tau_0$ is a constant of integration.  From this, we obtain the relation
\begin{equation}
r_{b_2} = \left(\frac{m_2}{m_1}\right)^{1/3}\, r_{b_1}.
\label{r_constraint}
\end{equation}
It can be checked that this relation also satisfies \eqref{bcond2_b} and hence condition \enumref{bcond2} as well.

If $E_b<1$ for both cells, then \eqref{tau_closed} and \eqref{r_E} imply that
\begin{widetext}
\begin{align*}
&\frac{2m_1}{\left(1-E_{b_1}\right)^{3/2}} \left[ \sqrt{\frac{1-E_{b_1}}{2m_1}r_{b_1}}\sqrt{1-\frac{1-E_{b_1}}{2m_1}r_{b_1}} + \cos^{-1}\sqrt{\frac{1-E_{b_1}}{2m_1}r_{b_1}} \right] - \tau_0 \\
&\qquad\qquad = \frac{2m_2}{\left(1-E_{b_2}\right)^{3/2}} \left[ \sqrt{\frac{1-E_{b_2}}{2m_2}r_{b_2}}\sqrt{1-\frac{1-E_{b_2}}{2m_2}r_{b_2}} + \cos^{-1}\sqrt{\frac{1-E_{b_2}}{2m_2}r_{b_2}}\right] - \tau_0.
\end{align*}
\end{widetext}
The two sides of this equation can be made equal if we simultaneously equated
\begin{align*}
\frac{1-E_{b_1}}{2m_1}r_{b_1} = {}& \frac{1-E_{b_2}}{2m_2}r_{b_2}\\
\shortintertext{and}
\frac{2m_1}{\left(1-E_{b_1}\right)^{3/2}} = {}& \frac{2m_2}{\left(1-E_{b_2}\right)^{3/2}}.
\end{align*}
The second equation leads to the constraint
\begin{equation}
\left(E_{b_2} - 1\right) = \left(\frac{m_2}{m_1}\right)^{2/3}\, \left(E_{b_1} - 1 \right),
\label{E_constraint}
\end{equation}
and if we substitute this into the first equation, we recover \eqref{r_constraint}.  Again, it can be checked that \eqref{r_constraint} and \eqref{E_constraint} combined satisfy \eqref{bcond2_b} and hence condition \enumref{bcond2} as well.

Finally if $E_b>1$, then \eqref{tau_open} and \eqref{r_E} imply that
\begin{widetext}
\begin{align*}
&\frac{2m_1}{\left(E_{b_1} - 1\right)^{3/2}} \left[ \sqrt{\frac{E_{b_1} - 1}{2m_1}r_{b_1}}\sqrt{1+\frac{E_{b_1}-1}{2m_1}r_{b_1}} - \sinh^{-1}\sqrt{\frac{E_{b_1}-1}{2m_1}r_{b_1}} \right] - \tau_0 \\
&\qquad\qquad = \frac{2m_2}{\left(E_{b_2}-1\right)^{3/2}} \left[ \sqrt{\frac{E_{b_2}-1}{2m_2}r_{b_2}}\sqrt{1+\frac{E_{b_2}-1}{2m_2}r_{b_2}} - \sinh^{-1}\sqrt{\frac{E_{b_2}-1}{2m_2}r_{b_2}} \right] - \tau_0.
\end{align*}
\end{widetext}
As with the $E_b<1$ case, the two sides of this equation can be made equal if we simultaneously equated
\begin{align*}
\frac{E_{b_1}-1}{2m_1}r_{b_1} = {} & \frac{E_{b_2}-1}{2m_2}r_{b_2}\\
\shortintertext{and}
\frac{2m_1}{\left(E_{b_1}-1\right)^{3/2}} = {} & \frac{2m_2}{\left(E_{b_2}-1\right)^{3/2}}.
\end{align*}
This clearly leads to the same constraints as in the $E_b<1$ case.

We therefore find that for all cases, two neighbouring cells must satisfy constraints \eqref{r_constraint} and \eqref{E_constraint} at the boundary.\footnote{The author wishes to acknowledge that constraints \eqref{r_constraint} and \eqref{E_constraint} were actually first derived by Ruth Williams in an unpublished calculation, where she derived the boundary conditions using the Israeli junction conditions instead.  Equating the two metrics on the boundary gave the condition $dr_{b_1} / d\tau_1 = dr_{b_2} / d\tau_2$, and equating $\Tr{\kappa}$, where $\kappa$ is the extrinsic curvature, gave condition \eqref{bcond2_a}.}  And when $m_1 = m_2$, we recover $E_{b_1} = E_{b_2}$ and $r_{b_1} = r_{b_2}$.

\bibliographystyle{apsrev4-1}
\bibliography{LW}

\end{document}